\documentclass[prx,twocolumn,aps,epsf,showpacs,superscriptaddress,longbibliography,nofootinbib]{revtex4-1}
\usepackage[pdftex]{graphicx}
\usepackage[normalem]{ulem}
\usepackage{xfrac}
\newtheorem{theorem}{Theorem}[section]
\newtheorem{lemma}[theorem]{Lemma}
\usepackage{verbatim, float, complexity}
\usepackage{dcolumn, amsfonts, mathtools}
\usepackage{bm}
\usepackage{latexsym} 
\usepackage{amsmath}
\usepackage{dsfont}
\usepackage{MnSymbol}
\usepackage{color}
\usepackage{array}
\usepackage{bbm}
\usepackage{color}
\usepackage{array}
\usepackage{cancel}
\usepackage[dvipsnames]{xcolor}
\usepackage{braket}
\usepackage{comment}

\usepackage[colorlinks=true,allcolors=blue]{hyperref}%

\newcommand{\<}{\langle}

\renewcommand{\vec}[1]{{\boldsymbol #1}}
\renewcommand{\>}{\rangle}
\renewcommand{\(}{\left(}
\renewcommand{\)}{\right)}
\renewcommand{\[}{\left[}
\renewcommand{\]}{\right]}

\newcommand{\calZ}{\mathcal{Z}}
\newcommand{\N}{\mathcal{N}}
\newcommand{\tr}{\text{tr}}

\newcommand{\kket}[1]{\left| #1\right\rrangle}
\newcommand{\bbra}[1]{\left\llangle #1 \right |}
\definecolor{ultramarine}{RGB}{0,32,96}

\newcommand{\drew}[1]{{\color{BrickRed}{(AP) #1}}}
\newcommand{\sofia}[1]{{\color{WildStrawberry}\footnotesize{(SGG) #1}}}

\newcommand{\romain}[1]{{\color{ForestGreen}{(RV) #1}}}

\usepackage{titlesec}
\titleformat{\paragraph}[runin]
        {\bfseries}
        {}
        {0.0em}
        {}
        [ -- ~]
        
\titlespacing*{\paragraph}{0pt}{4pt}{0pt}

\begin{document}
\title{Random insights into the complexity of two-dimensional tensor network calculations}

\author{Sof\'ia Gonz\'alez-Garc\'ia}
\affiliation{Perimeter Institute for Theoretical Physics, Waterloo, Ontario N2L 2Y5, Canada}
\affiliation{University of Waterloo, Waterloo, Ontario, N2L 3G1, Canada}
\affiliation{Google Quantum AI, Santa Barbara, CA 93111, USA}

\author{Shengqi Sang}
\affiliation{Perimeter Institute for Theoretical Physics, Waterloo, Ontario N2L 2Y5, Canada}
\affiliation{University of Waterloo, Waterloo, Ontario, N2L 3G1, Canada}

\author{Timothy H. Hsieh}
\affiliation{Perimeter Institute for Theoretical Physics, Waterloo, Ontario N2L 2Y5, Canada}

\author{Sergio Boixo}
\affiliation{Google Quantum AI, Santa Barbara, CA 93111, USA}

\author{Guifr\'e Vidal}
\affiliation{Google Quantum AI, Santa Barbara, CA 93111, USA}

\author{Andrew C. Potter}
\affiliation{Department of Physics and Astronomy, and Quantum Matter Institute, University of British Columbia, Vancouver, BC, Canada V6T 1Z1}

\author{Romain Vasseur}
\affiliation{Department of Physics, University of Massachusetts, Amherst, MA 01003, USA}

\begin{abstract}
Projected entangled pair states (PEPS) offer memory-efficient representations of some quantum many-body states that obey an entanglement area law, and are the basis for classical simulations of ground states in two-dimensional (2d) condensed matter systems. However, rigorous results show that exactly computing observables from a 2d PEPS state is generically a computationally hard problem. Yet approximation schemes for computing properties of 2d PEPS are regularly used, and empirically seen to succeed, for a large subclass of (`not too entangled') condensed matter ground states.
Adopting the philosophy of random matrix theory, in this work we analyze the complexity of approximately contracting a 2d random PEPS by exploiting an analytic mapping to an effective replicated statistical mechanics model that permits a controlled analysis at large bond dimension.
Through this statistical-mechanics lens, we argue that: $i)$ although approximately sampling wave-function amplitudes of random PEPS faces a computational-complexity phase transition above a critical bond dimension, $ii)$ one can generically efficiently estimate the norm and correlation functions  for any finite bond dimension.
These results are supported numerically for various bond-dimension regimes. It is an important open question whether the above results for random PEPS apply more generally also to PEPS representing physically relevant ground states.
\end{abstract}

\maketitle
Tensor network states (TNS) provide a compact representation of quantum states with spatially-local entanglement, such as ground states of local Hamiltonians. One-dimensional (1d) TNS, matrix product states (MPS), can be efficiently contracted, enabling dramatic progress in studying 1d many-body ground states~\cite{white_1992, fannes_1992, Verstraete_2008, Schollwock_2011}. Projected entangled pair states (PEPS)~\cite{Verstraete_2004} are a higher dimensional generalization of MPS tensor networks. Contrary to their 1d counterpart, contracting PEPS is a $\#\mathsf{P}$-complete task~\cite{Schuch_2007} (the complexity class of hard counting problems) even for simple square lattice PEPS with constant bond dimension, $D$. 
Moreover, approximately contracting a 2d PEPS with bond dimension $D\sim {\rm poly}(L)$ that scales polynomially in the linear dimension $L$ of the system was shown to be average-case hard even for calculating simple physical quantities like expectation values of local observables~\cite{Haferkamp_2020}.
Yet, in contrast to these complexity results, the practical experience of PEPS practitioners~\cite{Corboz_2014, Corboz_2014_2, Niesen_2017, Zheng_2017, Ponsioen_2019, Chen_2020} is that the standard algorithm~\cite{Verstraete_2004, Jordan_2008} (which we review below) for approximating the physical properties of PEPS seems to work efficiently for finitely-correlated ground states of 2d lattice models in many condensed matter problems.

\begin{figure*}[t!]
  \centering
  {\includegraphics[width=0.67\columnwidth]{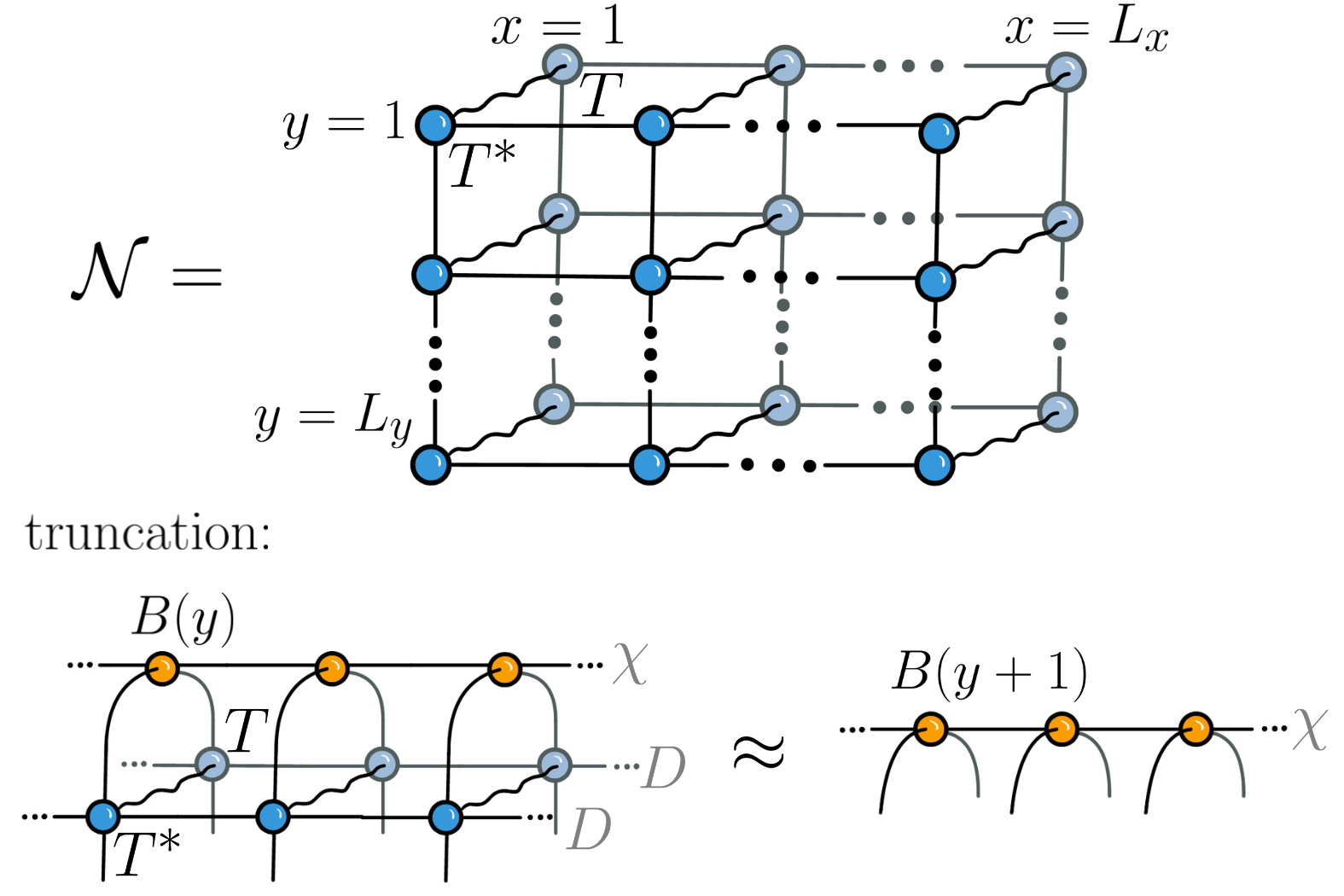}}\hspace{2mm}
  {\includegraphics[width=0.67\columnwidth]{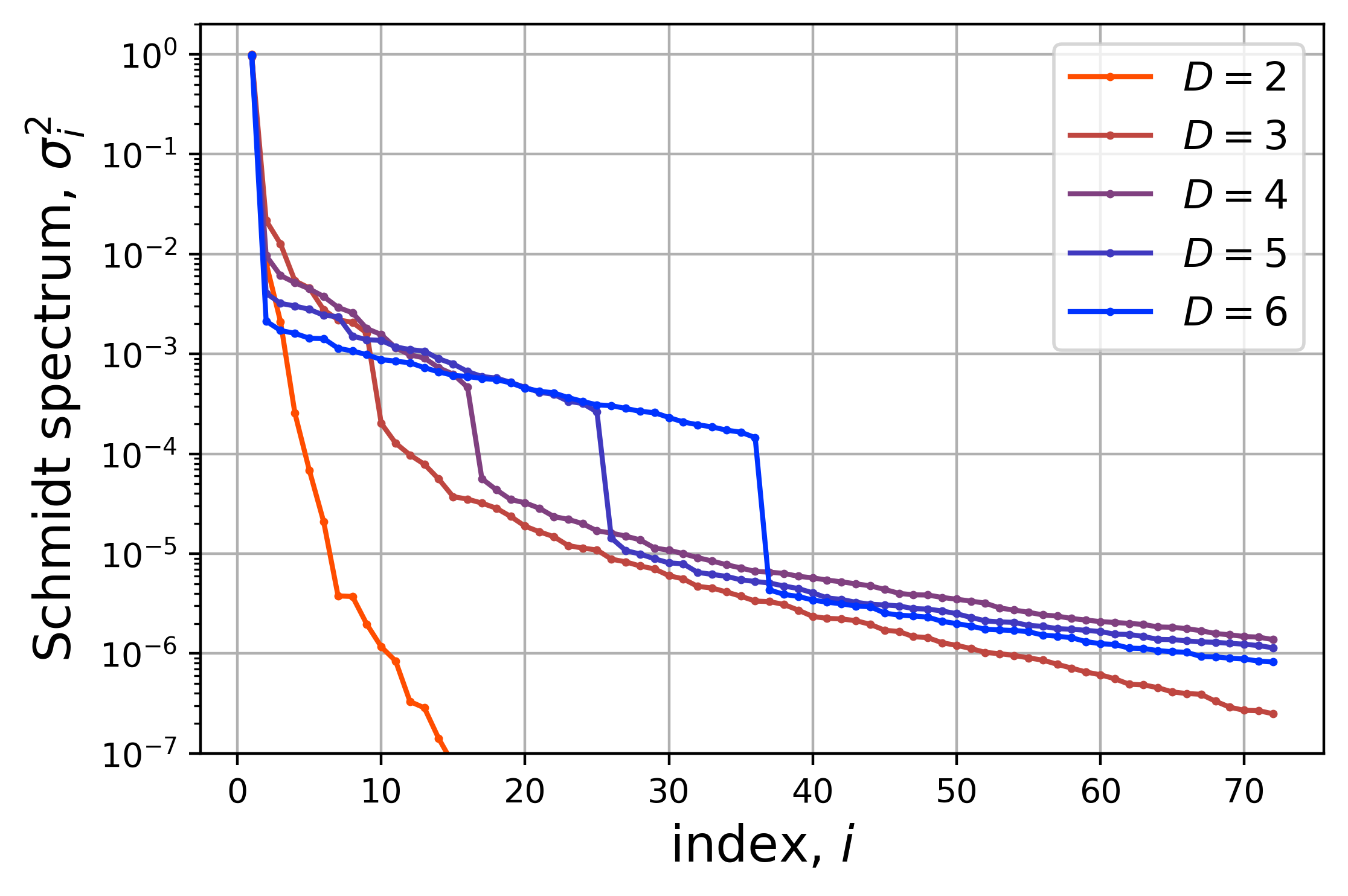}}
  \includegraphics[width=0.67\columnwidth]{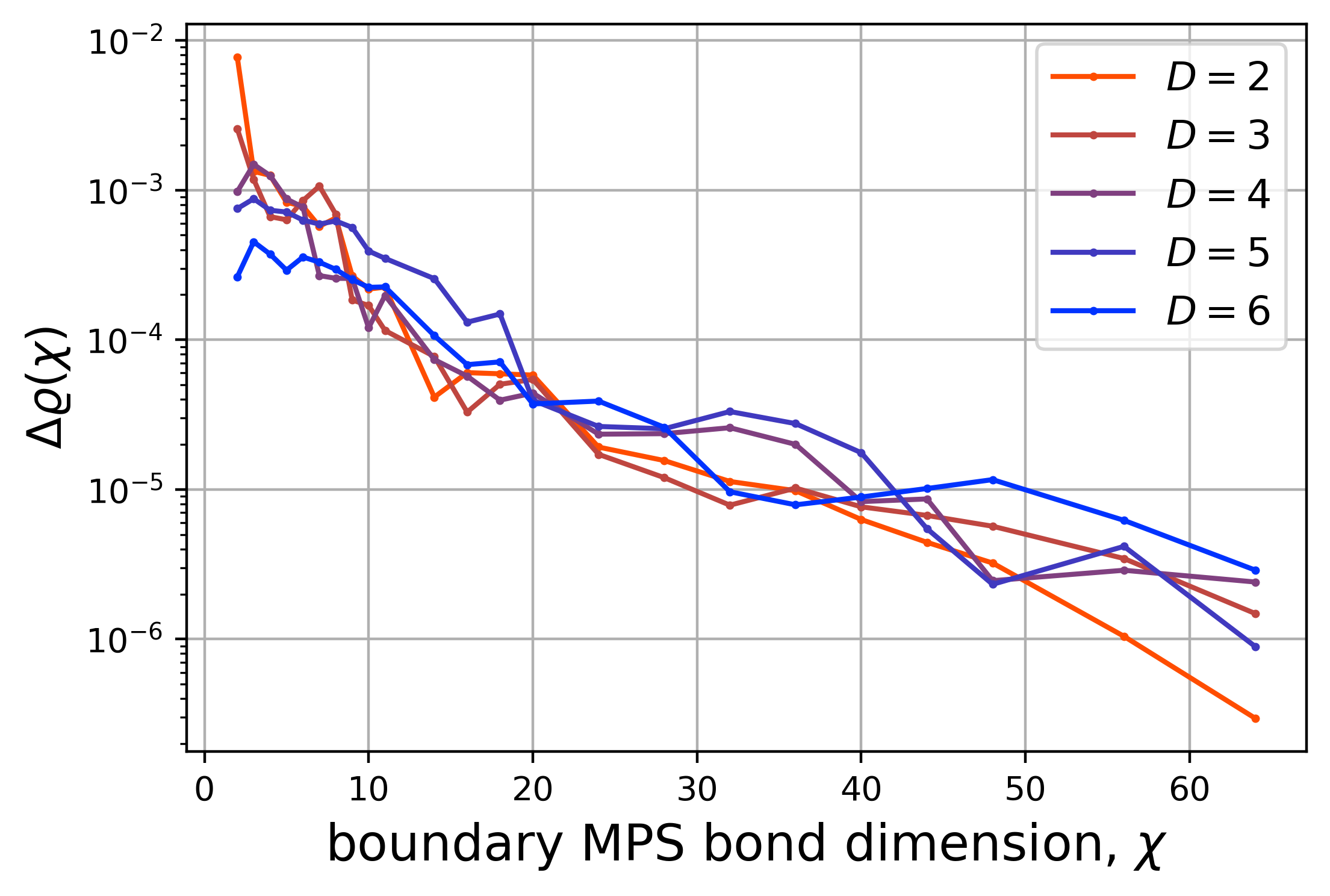}\\
  \hspace{2mm} (a) \hspace{0.66\columnwidth}(b)\hspace{0.66\columnwidth} (c)
  \caption{
  {\bf Boundary MPS contraction of 2d PEPS:} 
  (a) Tensor network representation of the norm $\N=\braket{\Psi|\Psi}$, for a finite $L_x\times L_y$ PEPS $\ket{\Psi}$ (top) and one iteration of the infinite bMPS approximate contraction algorithm (bottom): a row of $TT^*$ tensors is multiplied by the boundary MPS whose bond dimension is then truncated back to $\chi$.
  (b) Schmidt coefficients at a bipartition of the infinite bMPS for $\chi=72$ for a clean random iPEPS for bond dimension $D\in\{2,3,4,5,6\}$.
  (c) Change $\Delta \varrho(\chi)$ in the physical single-site density matrix $\varrho(\chi)$ as a function of $\chi$, where we take $\varrho(\chi_{\rm max}=72)$ as a reference.}
\label{fig:bMPS_diag_and_iPEPS_results}
\end{figure*}

Systematically clarifying the behavior of PEPS and their approximate calculation efficiency is crucial not only for computational physics, but also for refining the targets for quantum computational advantage in materials and chemistry simulation problems~\cite{lee2023}, which are one of the leading prospective applications for quantum computers~\cite{Ma_2020}.
A key challenge is a lack of systematic analytical tools for analyzing the complexity of approximate contraction schemes. To this end, we adopt the philosophy of statistical mechanics and random matrix theory which show that, in many instances~\cite{guhr1998random}, the statistical properties of random ensembles can be systematically understood, even when individual cases cannot be directly analyzed.

With this motivation in mind, we study the ``typical'' contraction complexity of  2d PEPS with \emph{random} tensors. This enables us to systematically address the statistical complexity of random PEPS through complementary analytical and numerical approaches.
 Analytically, we exploit a replica-trick-based mapping between entanglement features of random disordered PEPS and partition functions of classical lattice-``spin" models. We present evidence that standard PEPS contraction schemes can efficiently approximate correlations of a 2d random PEPS, including both of local observables and many non-local observables such as string order parameters.
This contrasts the behavior of computing global properties such as overlaps $\<\Psi'|\Psi\>$ between two PEPS states, which includes individual wave-function amplitudes $\<s_1,s_2,\dots|\Psi\>$,  where one expects a complexity-phase-transition tuned by $D$~\cite{PhysRevB.100.134203}, which has been observed numerically~\cite{2021arXiv210802225L,Yang_2022}.
We emphasize that the arguments presented are not rigorous proofs, but rather physical arguments about the replica statistical mechanics that are grounded in standard paradigms of statistical mechanics and critical phenomena.

To validate the predictions of this analytic derivation, we numerically explore the $D$ dependence of contracting both ``clean" (translation-invariant) infinite PEPS (iPEPS)~\cite{Jordan_2008}, as well as ``clean" and ``disordered" (see below) stabilizer (Clifford) PEPS, which allow for a much larger bond dimension.
The numerical results are consistent with the analytic predictions, including quantitative agreement in the detailed large-$D$ asymptotic behavior. While the analytical findings are limited to disordered PEPS, the numerical results suggest that the predictions also hold for ``clean" translation-invariant PEPS.

It remains an important open question to determine the relevance of these random PEPS results to PEPS representing ground states of Hamiltonians relevant to condensed matter, materials science, and chemistry. 
For example, it is known that despite being highly entangled, at large $D$ the random square-lattice PEPS exhibit only short-range correlations for local observables~\cite{Lancien_2021} (see also Appendix~\ref{app:xi}), whereas the ground-states of physical systems can exhibit relatively long correlation lengths, potentially making them harder to contract.
On the other hand, the statistical mechanics mapping applies more broadly to a large class of network structures, including examples with arbitrarily-long-range correlations (see Appendix~\ref{app:bigxi}).

\paragraph{PEPS and boundary MPS method}
We begin by briefly reviewing standard schemes to approximately contract 2d PEPS.
Consider a 2d square lattice made of $L_x\times L_y \equiv N$ sites, $L_x\leq L_y$, where each site is represented by a $d$-dimensional vector space $\mathbb{C}^d$, and a \textit{bulk} state $\ket{\Psi} \in \bigotimes_{i=1}^{N} \mathbb{C}^d$ of the lattice that accepts a PEPS representation 
of the form $|\Psi\>=\sum_{s_1,\dots, s_{N}=1}^d \mathcal{C}\left[T_{[1]}^{s_1}\dots T_{[N]}^{s_N}\right]|s_1\dots s_N\>$. Here, $T_{[r]}$ denotes the 5-index PEPS tensor on lattice site $r\in \{1,2,\cdots,N\}$, with complex components $\left(T_{[r]}\right)^{s}_{ijkl}$, where $s=1,\cdots,d$ and $i,j,k,l=1,\cdots,D$ are known as \textit{physical} and \textit{bond} indices, respectively, and $\mathcal{C}$ denotes contraction of the bond indices connecting nearest neighbor tensors. 

An archetypal PEPS computation is the contraction of the 2d tensor network in Figure~\ref{fig:bMPS_diag_and_iPEPS_results}a, of size $L_x \times L_y$, which corresponds to the overlap $\mathcal{N} \equiv \braket{\Psi|\Psi}$. Other closely related 2d networks, such as those corresponding to $k$-point correlators of local operators $\<\Psi|O_1O_2\dots O_k|\Psi\>$, are contracted similarly.
An exact contraction incurs a computational cost $O(L_y\exp(L_x))$, but a number of efficient algorithms exist, most notably the boundary MPS (bMPS) method~\cite{Verstraete_2004, Jordan_2008}, which scales linearly both in $L_x$ and $L_y$ at the price of introducing approximations\footnote{The corner transfer matrix (CTM) \cite{Baxter_2007} method is another prominent approximate contraction scheme for the 2d tensor network $\mathcal{N}$. CTM and bMPS are closely related and similar conclusions regarding boundary entanglement, as discussed in this paper, apply to both.}.
The bMPS method considers the \textit{virtual} 1d lattice made of $L_x$ sites, each described by a $D^2$-dimensional vector space $\mathbb{C}^{D^2}$ (corresponding to two PEPS bond indices), obtained from a horizontal cut of the 2d tensor network $\mathcal{N}$, and a so-called \textit{boundary} state $|\psi\rrangle \in \bigotimes_{i=1}^{L_x} \mathbb{C}^{D^2}$ that accepts an MPS representation, $|\psi\rrangle = \sum_{\alpha_1,\cdots, \alpha_{L_x}} \mathcal{C}\left(B_{[1]}^{\alpha_1} \cdots B_{[L_x]}^{\alpha_{L_x}}\right) |\alpha_1 \cdots \alpha_{L_x}\rrangle$. Here the double brackets $|\rrangle$ emphasize that the boundary state is actually a vectorized density operator on the PEPS bond space, $B_{[x]}$ is a 3-index MPS tensor on site $x\in \{1,2,\cdots,L_x\}$ with complex components $\left(B_{[x]}\right)^{\alpha}_{\beta\gamma}$, where $\alpha=1,\cdots,D^2$ and $\beta,\gamma=1,\cdots,\chi$, and $\chi$ is the bMPS bond dimension. In order to approximately contract the 2d tensor network $\mathcal{N}$, which we now regard as made of $L_y$ rows of tensors where each row is labelled by an integer $y=1, \cdots, L_y$, we first represent the boundary state $|\psi(1)\rrangle$ corresponding to the top row ($y=1$) exactly as a bMPS with bond dimension $D^2$, which we then truncate down to $\chi$ (if $D^2 > \chi$). For increasing value of $y=1,2,\cdots$, given a (possibly truncated, approximate) bMPS representation of the boundary state $|\psi(y)\rrangle$, we produce an approximate bMPS for $|\psi(y+1)\rrangle$ by contracting the row $y+1$ with the existing bMPS, see Figure~\ref{fig:bMPS_diag_and_iPEPS_results}a, then truncating the resulting bond dimension $\chi'$ down to $\chi$ (if $\chi' > \chi$). Finally, the value of $\mathcal{N}$ is obtained by contracting the bMPS for $|\psi(L_y-1)\rrangle$ with the bottom row (i.e. $y=L_y$). The leading computational cost of the bMPS method with respect to index dimensions $d, D, \chi$ is $O(D^4\chi^3+dD^6\chi^2)$~\cite{Lubasch_2014} (or $O(D^6\chi^3+dD^8\chi^2)$ in the infinite case \cite{Jordan_2008}, $L_x,L_y\to\infty$). 

The above sequential contraction of the 2d tensor network $\mathcal{N}$ is analogous to a dynamical evolution of the state of a 1d system (represented by the bMPS) by a transfer matrix implementing a completely positive map made of a row of $TT^*$ tensors.
This setting is similar to 1d noisy quantum circuits ~\cite{Noh2020efficientclassical,PhysRevB.107.014307}.

\paragraph{Random PEPS models}
We consider applying the boundary MPS method to  random PEPS, whose tensors $T$ are multi-dimensional arrays of complex entries, with each entry sampled independently and identically distributed (i.i.d.) from the complex normal (Gaussian) distribution with zero mean and unit variance. We consider two different ensembles of random PEPS: i) ``clean", translation-invariant PEPS with a single random instance of $T$ used for each site, and ii) ``disordered" PEPS with a distinct, random $T_{[r]}$ for each site, $r$. 
From the perspective of spatial symmetries, clean and disordered random PEPS are akin to crystalline and disordered materials respectively.

\paragraph{Boundary MPS entanglement and random PEPS contraction} \label{parag:entanglement_contract_entanglement}
The resulting value of $\mathcal{N}= \braket{\Psi|\Psi}$ is approximate due to the errors introduced at each truncation   typically implemented by preserving the $\chi$ largest Schmidt coefficients $\{\sigma_i\}$ (or coefficients $\{\sigma_i^2\}$ in the entanglement spectrum) assigned to each bond of the bMPS (see Appendix \ref{app:methods_iPEPS}). 
Intuitively, if most of the $O(D^{2y})$ Schmidt coefficients (for $y \leq L_x/2$) are `large', then an accurate estimate of $\mathcal{N}$ requires using a bMPS with a bond dimension $\chi$ that grows exponentially in $y$, and thus the method is not efficient. However, if only a small number of Schmidt coefficients are significant, then retaining only a constant value of $\chi$ may suffice for an accurate bMPS representation, thus enabling efficiently approximate contraction of the network. 

Figure \ref{fig:bMPS_diag_and_iPEPS_results}b shows the entanglement spectrum $\{\sigma_i^2\}$ of the \textit{fixed-point} bMPS $\kket{\Tilde{\psi}}$ for clean random iPEPS, where we exploit translation invariance to access the $L_x,L_y\to \infty$ limit, and $\kket{\Tilde{\psi}}$ represents the large-$y$ fixed-point of the bMPS sequence $\{\kket{\psi(y)}\}$ (see Appendix \ref{app:methods_iPEPS} for further details). Notice the exponential decay of $\sigma_i^2$ as a function of $i$, which implies that an accurate approximation of the boundary state can be obtained with a small bond dimension $\chi$. In turn, this results in an accurate approximation $\varrho(\chi)$ to the reduced density matrix $\varrho =\tr_{\bar{R}}|\Psi\rangle\langle \Psi|/\mathcal{N}$ on a local region $R$ of the 2d lattice and thus also to the expectation value $\braket{\Psi|O|\Psi}/\N = \tr (\varrho O)\approx \tr (\varrho(\chi) O)$ for any local observable $O$ supported on $R$. Indeed, consider the distance $\Delta\varrho(\chi)$ between the density matrix $\varrho(\chi)$ and a reference density matrix $\varrho(\chi_{\rm max})$ as given by the largest eigenvalue of $|\varrho(\chi)-\varrho(\chi_{\rm max})|$, where we make the key assumption that $\varrho(\chi_{\rm max})$ is a good approximation to the exact $\varrho = \varrho(\chi\to\infty)$. As shown in Figure \ref{fig:bMPS_diag_and_iPEPS_results}c for a single-site $\varrho$, $\Delta\varrho(\chi)$ decays exponentially with $\chi$, with similar behavior observed when $R$ consists of more than one site (see Figure \ref{fig:2dPESP_red_rho_app}b). We therefore conclude that an accurate approximation to $\braket{\Psi|O|\Psi}/\N$ can be obtained efficiently, for these instances of clean random iPEPS.

In practice, the bMPS entanglement is often characterized using the Renyi entropy of order $n$,
\begin{align} \label{eqRenyi}
    S^{(n)}_A = \frac{1}{1-n} \log \left( \tr \left[\left(\frac{\rho_A}{\tr\rho }\right)^n\right]\right),~~~~
\end{align}
where $\rho \equiv |\psi\rrangle\llangle\psi|$ and $\rho_A = \text{tr}_{\bar{A}}|\psi\rrangle\llangle\psi|$ is the reduced density matrix for a 1d sub-region, $A$, of the evolved bMPS, and the denominator is included to explicitly normalize the state. 
Specifically, for $n\rightarrow 1$, we expect an inefficient contraction when $|\psi(y)\rrangle$ obeys an entanglement \textit{volume law}, where $S^{(n)}_A$ is proportional to the size $|A|$ of region $A$, and an efficient one when it obeys an entanglement \textit{area law}, where $S^{(n)}_A$ is upper bounded by a constant. An advantage of using  a single statistic, such as $S^{(n)}_A$, rather than the full entanglement spectrum is that its ensemble average for $n\geq 1$ can be evaluated using a statistical mechanics model, as we will see below. 

\begin{figure}[t]
\begin{centering}
\includegraphics[width=0.5\textwidth]{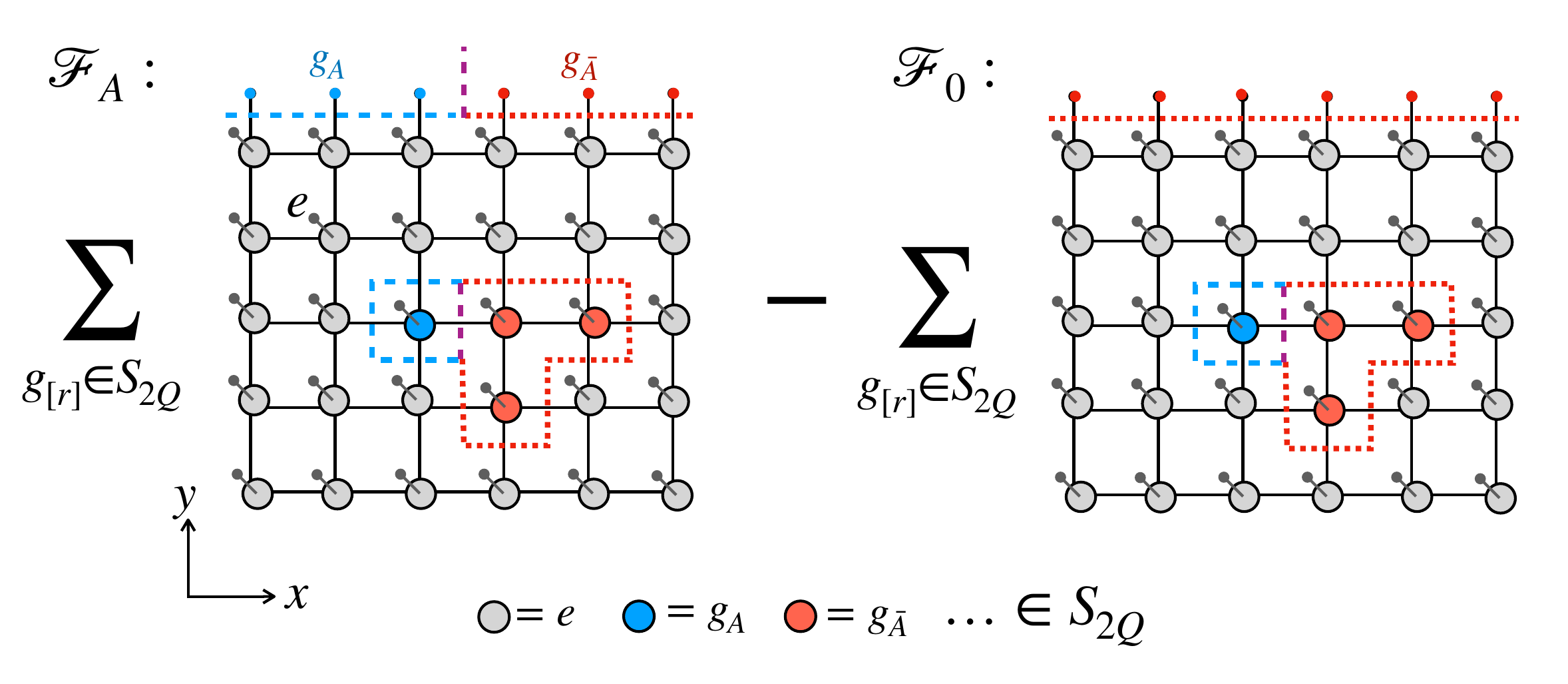}
\end{centering}
\caption{
\textbf{Statistical Mechanics Model (schematic)} -- The bMPS entanglement of an random PEPS maps onto the difference between free-energies $\mathcal{F}_A-\mathcal{F}_0$ that differ in boundary conditions as shown (note in the replica limit free-energy and partition functions coincide). Each site of the random PEPS is replaced by a replica-permutation ``spin" (shown as colored circles; while there are $(2Q)!$ different spin values, we show only the three corresponding to the bulk and boundary fields here). Solid lines along bonds of the random PEPS represent ferromagnetic interactions. Lines ending in a dot denote single-site fields. At large-$D$, the dominant configuration is uniform polarization along $e$ (gray). Fluctuating domains of spins have both a linear boundary tension (dashed lines) due to the interactions, and a surface-area tension due to the bulk $e$ fields.
}
\label{fig:statmech} 
\end{figure}

\paragraph{Statistical mechanics model mapping} \label{par:stat_mech_model}
The entanglement features of random tensor network contractions can be mapped onto free-energy cost of boundary domains in a classical statistical mechanics (stat-mech) model~\cite{HaydenHoloRTN,PhysRevB.100.134203,PhysRevB.102.064202,PRXQuantum.2.010352}. This mapping has been used extensively to study both random holographic tensor networks (tensor networks with physical legs only at the boundary)~\cite{HaydenHoloRTN,PhysRevB.100.134203}, and to explore entanglement growth~\cite{PhysRevX.7.031016,PhysRevX.8.021014,PhysRevB.99.174205}, measurement-induced phase transitions~\cite{PhysRevX.9.031009,PhysRevB.98.205136,PhysRevB.101.104302,PhysRevB.101.104301,PotterVasseurReview2022,doi:10.1146/annurev-conmatphys-031720-030658,PhysRevX.12.021021} in random quantum circuits and channels~\cite{PhysRevB.107.014307}. Here, we adapt it to study the entanglement of the evolved boundary state involved in the contraction of a random, disordered PEPS. In the main text, we merely summarize the key elements of the stat-mech mapping, and refer readers to the supplemental material in the Appendix for details. We focus on the calculation of the PEPS norm, $\mathcal{N} = \<\Psi|\Psi\>$, for a disordered, random, square lattice PEPS $|\Psi\>$, and will later comment to how this is modified for other observables. 
As above, we denote the density matrix of the evolved bMPS in the PEPS contraction procedure described above as $\rho = |\psi\rrangle\llangle\psi|$.

The mapping exploits a replica trick, $\log f = \lim_{m\rightarrow 0} \frac{1}{m}(f^m-1)$, to perform averages of $n^{\rm th}$-Renyi entropies (Equation \ref{eqRenyi}). 
The objects of interest are averages over powers of the reduced density matrix of the boundary MPS for a region, $A$: $\mathbb{E}\[\({\rm tr}\rho_A^n\)^m\]$, where $\mathbb{E}[\dots]$ denotes averaging over the Gaussian-distribution of tensor entries. 
For integer $m,n$, this average can be carried out using Wick's theorem. These quantities contain $Q=nm$ copies of $\rho$. Since each copy of $\rho$ contains $2\times$ $T$ and $2\times $ $T^{*}$ tensors, this results in a sum over all possible Wick pairings of the $2Q=2mn$ replicas of the tensor network. A given Wick pairing is defined by a set of permutation elements $g_r\in S_{2Q}$ where $S_{2Q}$ is the symmetric group (group of permutations) on $2Q$ elements, for each site, $r$.
The resulting entanglement entropy can be written as a free-energy difference~\cite{PhysRevB.100.134203}:
\begin{align} 
    S^{(n)}(A) = \lim_{m\rightarrow 0}\frac{1}{m(n-1)}(\mathcal{F}_A-\mathcal{F}_0),
\end{align}
where $\mathcal{F}_{A,0}=-\log\mathcal{Z}_{A,0}$ are the free energies of a classical statistical mechanics model with partition function $\mathcal{Z} = \sum_{ \lbrace g_r  \in S_{2Q} \rbrace } e^{-H[g]}$, defined by Hamiltonian:
\begin{align}
    \mathcal{H}[g] &= -\sum_{\<rr'\>}
    JC(g_r,g_{r'})- \sum_{r,g\in S_{2Q}} h_{g,r}C(g,g_r). 
    \label{eq:Hsm}
\end{align}
The labels $A$ and $0$ in the free energies $\mathcal{F}_{A,0}$ refer to different boundary fields that will be specified below. 
Here, $\<rr'\>$ denotes neighboring pairs of sites, the function $C(g,g') = C(g^{-1} g')$ counts the number of cycles in the permutation $g^{-1}g'$, and the coupling constant is related to bond dimension: $J=\log D$. Intuitively, $C(g,g')$ measures how similar the permutations $g$ and $g'$ are to each other, with $C(g,g)=2Q$, and minimal value $C\geq 1$. The coupling between neighboring spins is ferromagnetic, favoring aligned spins. This competes with entropic fluctuations of the spins. 

The second term can be thought of as local ``fields" that point in a different ``directions" (in $S_{2Q}$) in the bulk and boundary regions. For $\mathcal{F}_{A}$, the fields are:
\begin{align}
    h_{g,r} = \begin{cases}
        \delta_{g,e}\log d  & r\in {\rm bulk}\\
        \delta_{g,g_A}\log D  & r\in \text{boundary region $A$}\\
        \delta_{g,g_0} \log D  & r \in \text{boundary region $\bar{A}$}
    \end{cases}
    \label{eq:fields}
\end{align}
where $e$ is the identity permutation, and $g_{0,A}$ are permutations corresponding to the trace and cyclic permutation boundary conditions in the entanglement region $A$ and its complement, $\bar{A}$, respectively (see Appendix~\ref{app:statmech} for details). In contrast, the boundary fields in $\mathcal{F}_{0}$ are uniform at the boundary: $h_{g, {\rm boundary}}=\delta_{g,g_0} \log D$ [corresponding to Equation \ref{eq:fields} with only boundary-region $\bar{A}$].

\paragraph{Global observables: Complexity transition}
For computing global properties of a random PEPS, such as individual wave-function components, $\<s_1\dots s_N|\Psi\>$, or overlaps between distinct random PEPS, $\<\Psi'|\Psi\>$, the tensors are not completely positive, and tensor bulk $h_{e}$ fields vanish.
Absent the bulk $h\sim e$ terms, the model has $Q$ copies of each tensor $T$ on each site, and a (bulk) left/right $S_Q\times S_Q$  symmetry. There are two possible phases: a disordered phase with intact symmetry, and an ordered phase where this symmetry is spontaneously broken and the spins polarize towards a spontaneously chosen $g\in S_{Q}$. This model also describes the entanglement of random holographic tensor networks~\cite{PhysRevB.100.134203}, in which physical degrees of freedom arise only at the boundary of a higher-dimensional tensor network with tensors having only virtual bond legs.
In the ordered phase the boundary fields representing the entanglement cut force a $g_A$ to $g_0$ domain wall at the boundary, which costs energy proportional to the length of the domain, corresponding to volume law scaling of entanglement. By contrast, in the disordered phase, domain walls have a vanishing line tension, leading to area-law entanglement scaling.
The universal properties of this complexity transition remain unsolved, but are closely related to those of (forced) measurement-induced entanglement transitions in random unitary circuit dynamics~\cite{PRXQuantum.2.010352}.
Numerical analysis~\cite{2021arXiv210802225L} for square PEPS indicates that the critical bond dimension for this complexity transition is close to $D_c\approx 2$, meaning that essentially all practical calculations are expected to lie on the hard (volume-law) side of the transitions.

\paragraph{Local observables: area-law bMPS}
For local observables such as the wave-function norm, $\<\Psi|\Psi\>$, or $k$-point correlators of local operators: $\<\Psi|O_1O_2\dots O_k|\Psi\>$, the situation is rather different. 
Here, the bulk identity-permutation ($h_e$) fields are present, and  explicitly break the $S_{2 Q}$ symmetry. This erases the distinction between the ordered and disordered phases of the model, and as we will now argue leaves only the area-law phase behind. In fact, we will see that this conclusion holds even for many non-local observables, such as string-order parameters or Wilson loop observables, so long as they lack support on some finite density of sites.

The boundary fields, $g_{A,0}$, associated with the entanglement region $A$, and its complement region $\bar{A}$, compete with the bulk $e$ fields. 
The resulting competition is straightforward to analyze at large $D$, where domain wall configurations are strongly suppressed, and the partition functions are dominated by the lowest energy configuration. We refer to this approximation as ``min-cut".
 When the $L_y$ is large (to approach the thermodynamic limit, we are interested in $L_y\rightarrow \infty$), the number of bulk fields (which scales with system volume) overwhelms that of the boundary fields (which scale with linear dimension), forcing the spins to polarize along $e$. Consequently the linear free-energy per unit length of the boundary regions is proportional to $C(g_{A,0},e)\log D $ in the $A,\bar{A}$ regions respectively. Crucially, while the bulk fields break explicitly break the $S_{2Q}$, there is a residual bulk $S_{Q}$ symmetry that guarantees that $C(g_A,e) = C(g_0,e)$, i.e. that the leading contribution to the boundary-field energies cancels in $\mathcal{F}_A-\mathcal{F}_0$, leading to area-law scaling. 
 
 In fact, in the $D\rightarrow \infty$ limit, this cancellation is exact, and there is strictly zero operator entanglement. 
 Generically, corrections to the min-cut approximation for $S_A$ come from domains that span across the entanglement cut between $A$ and $\bar{A}$. Since these fluctuations are massive, (the probability of getting a large domain of size $\ell$ is expected to be exponentially small in $\ell$), these fluctuations are only sensitive to the local change in boundary fields between $A$ and $\bar{A}$, and hence can give only an area-law contribution to $S_A$.  
 In Appendix~\ref{app:statmechFluct}, we estimate that, beyond the infinite-$D$ limit, the leading corrections to the area-law entanglement come at order $1/D^2$.
 
We remark that these results highlight a counterintuitive feature of random tensor networks. While highly-correlated states require large bond dimension, \emph{typical} large-$D$ PEPS have rather short-range correlations (despite being quite entangled), and the set of highly-correlated large-$D$ PEPS is in fact rare (by the measure of our Gaussian-random ensemble). See Appendix~\ref{secPhysicalvsrandomPEPS} for a more detailed discussion of potential differences between random PEPS and physical groundstates relevant to condensed matter. 
In Appendix~\ref{app:bigxi}, we also discuss different tensor network geometries for which \emph{typical} states exhibit longer range correlations.
 
 We emphasize that, while the calculations described were done at leading order in the large-$D$ limit, the general symmetry principles described above suggest that the prediction of area-law entanglement for the boundary-MPS hold also at any $D$ and that there is no expected singular change (phase transition) in the entanglement as a function of $D$. 

{\bf Correlators and overlaps -- } While we have focused on the computation of the PEPS norm, this is closely related to the calculation of correlation functions of local operators: $\<\Psi|O_1O_2\dots O_k|\Psi\>$. The insertion of local operators amounts to replacing the tensors on those sites with non-positive tensors, which in the statistical mechanics model corresponds to removing the local $e$-field on that site. Yet, so long as a finite-density of sites are unaffected by the operator insertions, there remains a net bulk $e$-field that explicitly breaks the replica symmetry  breaks the replica-permutation symmetry and removes the complexity phase transition. Hence, we expect our ``easiness" result to a very large class of observables including even exotic non-local string correlation functions to used detect topological phases and confinement in gauge theories.
 

\begin{figure}[t]
\begin{centering}
\includegraphics[width=0.5\textwidth]{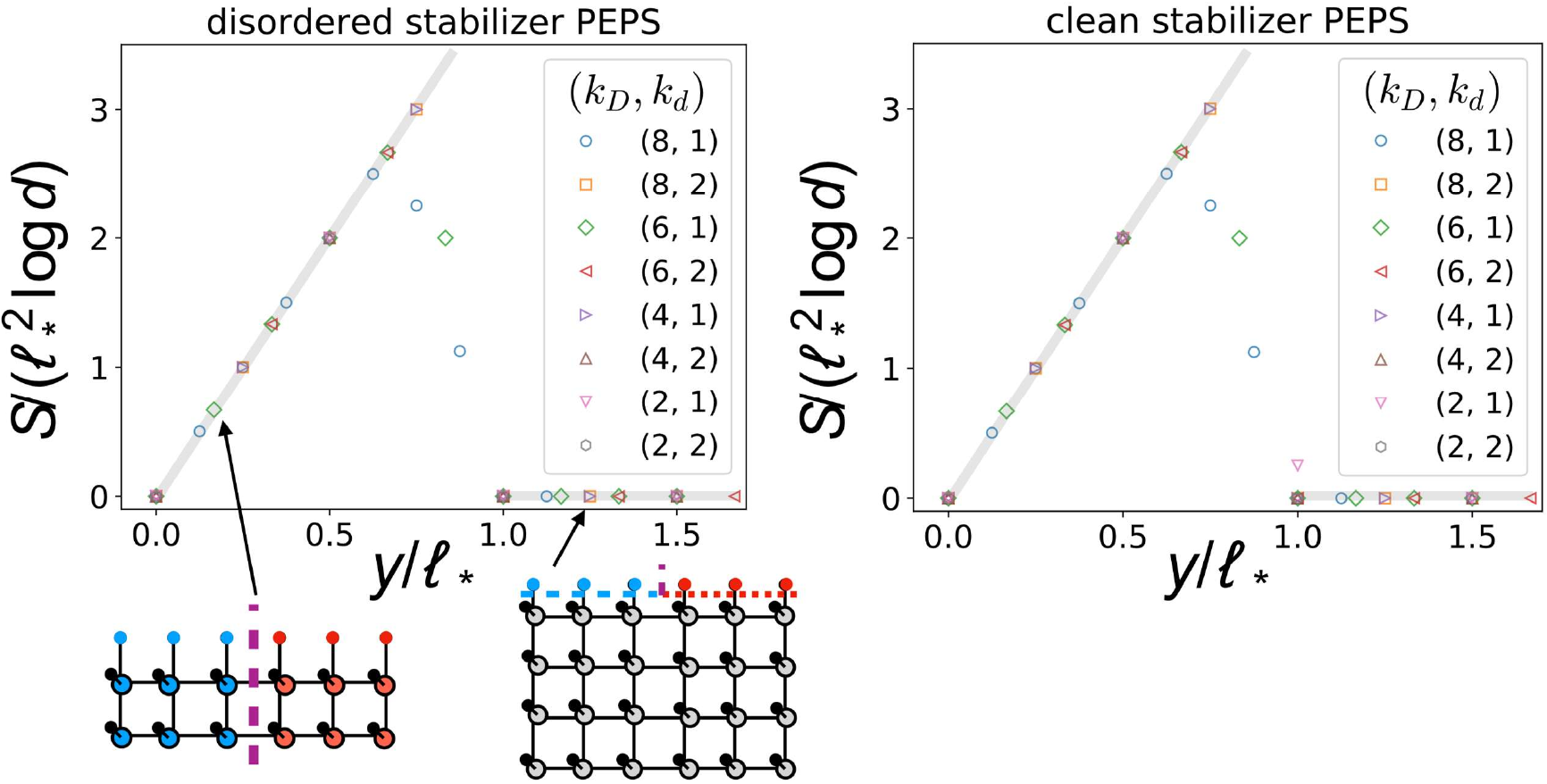}
\end{centering}
\caption{
\textbf{Entanglement barrier } -- Mid-cut entanglement entropy, $S(y)$, of the bMPS $|\psi(y)\rrangle$ for disordered (left) and clean (right) random stabilizer PEPS as a function of layer number $y$, for various choices of $(D,d)=(173^{k_D}, 173^{k_d})$. A schematic representation of the dominant spin configurations at large-$D$ in the corresponding stat-mech model are shown underneath the plot.
$S(y)$ initially increases linearly up to lengthscale $\ell_*\sim \log D/\log d$ (corresponding to a growing vertical domain wall) before dropping to a small constant for $y\gtrsim \ell_*$ (corresponding to the bulk fields expelling the domain wall out to the system's boundary). In both cases, the simulated $S(y)$ agrees well with the stat-mech model in the large-$D$ regime (solid gray lines), with very small instance to instance variance. 
}
\label{fig:entbarrier} 
\end{figure}
\paragraph{Entanglement barrier and random stabilizer PEPS}
 We have so far considered the thermodynamic limit of the boundary-MPS entanglement, where the statistical mechanics model predicts that the evolved boundary MPS eventually reaches an area-law steady state. We can also use the stat-mech mapping to examine the transient dynamics of the boundary MPS at early stages in contraction of a PEPS with open boundaries at $y=0$.
 When $y$ is small, $y\ll \ell^* = \log D/\log d$, for $D>d$, the boundary fields dominate and the spins $g_r$ polarize along the direction of the fields at their closest boundary. For $\mathcal{F}_A$, this adds an extra $g_A$-to-$g_0$ domain will along the y-direction with free energy cost $\sim y$ (see Figure~\ref{fig:entbarrier}). As a result, the statistical mechanics model predicts that entanglement grows as:
 \begin{align}\label{Eq: barrier}
     S^A(y) \sim \begin{cases} 
     y \log D^2 & y\ll \ell^*\\
     {\rm constant} & y \gg \ell^*
     \end{cases}, ~~~~~~\ell^* \equiv \log D/\log d,
 \end{align}
 achieving a maximum of $S_{\rm max}\sim \log^2 D/\log d$ at $y\approx \ell^* = \log D/\log d$, corresponding to a maximum bMPS bond dimension of $\chi_{\rm max}\sim \exp\[\log^2 D/\log d\]$.
We refer to this phenomenon as an ``entanglement barrier". 
We note that a similar entanglement barrier arises in the classical simulation of random quantum circuits in the presence of decoherence~\cite{Noh2020efficientclassical,PhysRevB.107.014307}, which are relevant to ``quantum supremacy" experiments~\cite{QuantumAdvantage}. 
For circuits, the entanglement barrier reflects the initial build up of quantum correlations and entanglement before noise and errors overwhelm the system making it essentially indistinguishable from a maximally-mixed state.
The evolution of the boundary MPS is similar to the dynamics of circuits with decoherence (even though the bMPS evolution is not generally trace-preserving), which provides a physical picture for the low (area law) entanglement of the MPS steady state~\cite{Noh2020efficientclassical,PhysRevB.107.014307}.  In this correspondence, the PEPS physical dimension $d$ represents the strength of noise, and the barrier height decreases with increasing noise strength.

We remark that, in practice, the entanglement barrier may not be an obstacle for extracting bulk properties of a PEPS. As illustrated by the iPEPS numerics above, it may not be necessary to accurately track the transient evolution of the boundary MPS through the entanglement barrier. Instead, one may be able to directly find an accurate representation of the area-law steady state bMPS, with $\chi$ reflecting the much smaller steady-state entropy deep in the bulk. However, for select tasks, such as simulating edge states of topological materials, the entanglement barrier may have practical consequences. We also note that this prediction is consistent with the average-case hardness proof for computing norms of finite-size random PEPS~\cite{Haferkamp_2020} with $D\sim {\rm poly(L)}$: in the stat-mech description this corresponds to an entanglement barrier with $\chi_{\rm max}$ scaling in a (very-weakly) super-polynomial fashion with $L$.

We can also use the entanglement barrier prediction simply as a test for the validity of the min-cut approximation and replica limit for the statistical mechanics model.
To easily access large-$D$, we focus on random stabilizer (Clifford) PEPS, which can be efficiently simulated via the stabilizer formalism with $\sim {\rm poly}(\log D$) complexity (see the Appendix~\ref{app:stabilizer} for the definition of stabilizer PEPS and the algorithm for contracting them). For stabilizer PEPS, each of $T_i$'s bond dimensions needs to be an integer power of some prime number $p$: $d=p^{k_d},~ D=p^{k_D}$. In simulations, we reach the large-$D$ regime by picking $p=173$ (and confirm that qualitatively similar results also hold for $p=2$). 

We performed simulations of both disordered and clean stabilizer PEPS with periodic boundary conditions, and show results in Figure \ref{fig:entbarrier}.
The simulation results agree quantitatively with the theoretical prediction in Equation \ref{Eq: barrier}, suggesting the validity of the statistical mechanics model in capturing the entanglement features of the boundary state. The agreement for the clean case is noteworthy since the statistical mechanics model was derived for disordered PEPS, and suggests that the conclusions of the stat-mech model also apply to translation invariant PEPS.

\paragraph{Discussion}
We have presented numerical and analytical evidence that the complexity of (approximate) 2d random PEPS calculations greatly depends on the nature of the object to be contracted. On the one hand, overlaps between a 2d random PEPS and a product state $\braket{s_1\cdots s_N|\Psi}$ or, more generally, overlaps $\braket{\Psi'|\Psi}$ between two distinct 2d random PEPS, face a complexity phase transition as a function of bond dimension. On the other hand, the overlap $\braket{\Psi|\Psi}$ of a single random PEPS as well as (computationally related) physically relevant properties such as the expectation value $\braket{\Psi|O|\Psi}$ or correlation functions $\braket{\Psi|O_1O_2\cdots O_k|\Psi}$ of local operators, can be efficiently computed.
The statistical mechanics model provides an intuitive picture for this result. 

The chief objective hazard of our analytic analysis is the reliance on the replica trick: away from the infinite-$D$ limit, we are able to analyze properties only for positive integer number of replicas $n=2,3,\dots $. 
By contrast (see Appendix~\ref{app:methods_iPEPS}), errors in observables are best captured by the von-Neumann entanglement entropy (which can be obtained by taking the limit of Renyi index $n\rightarrow 1^+$). We note that stabilizer PEPS have a fine-tuned, completely flat entanglement spectrum and do not constitute an independent check of the replica limit convergence. Therefore, in Appendix~\ref{app:methods_iPEPS}, we numerically explore finite-$\chi$ scaling of the $n$ dependence of bipartite Renyi entropies for clean random iPEPS. We observe a smooth and well-converged Renyi-index dependence up to the largest accessible $D=6$, implying no signs of trouble for the replica limit.

Our statistical mechanics mapping, when generalized to a 3d random PEPS $\ket{\Phi_{3d}}$, predicts again an area-law entangled boundary state for the overlap $\braket{\Phi_{3d}|\Phi_{3d}}$ and expectation value and correlation functions of local observables. This 2d area-law entangled boundary state might again be efficiently approximated with a 2d PEPS $\ket{\Psi_{2d}}$. It is thus plausible that approximate expectation values and correlators of local observables can be efficiently evaluated also for 3d random PEPS. 

An important question is to what extent the present results obtained for random PEPS may also apply to the contraction of PEPS that represent ground states of interest in condensed matter, materials science or quantum chemistry. 
Elucidating such a question would be useful both for the classical simulation of such systems, as well as to determine whether quantum computers may provide an exponential speed-up for such problems~\cite{lee2023}. 
We note that, even if classical TNS methods could efficiently calculate properties of finitely-correlated states in 2d and 3d, many challenging tasks remain ripe for quantum advantage, including simulating highly-entangled states such as metals or phase-transitions, or calculating dynamical responses such as transport coefficients or optical spectra.
Progress in addressing this question will likely require a combination of analytical work and numerical simulations directly targeting ``physically-relevant" systems of interest.\\


    \textit{Acknowledgments.---}%
	We thank Sarang Gopalakrishnan, Andreas Ludwig, Yi-Zhuang You and Yuri D. Lensky for insightful discussions. We acknowledge support from the US Department of Energy, Office of Science, Basic Energy Sciences, under Early Career Award No. DE-SC0019168 (R.V.) and DE-SC0022102 (A.C.P.), as well as the Alfred P. Sloan Foundation through Sloan Research Fellowships (A.C.P. and R.V.). R.V., A.C.P., and G.V. thank the Kavli Institute of Theoretical Physics (KITP) for hospitality. KITP is supported in part by the National Science Foundation under Grant No. NSF PHY-1748958. S.G.G., S.S., T.H. and G.V. acknowledge support by the Perimeter Institute for
    Theoretical Physics (PI), Natural Sciences and Engineering Research Council of Canada (NSERC), and Compute Canada. Research at PI is supported in part by the Government of Canada through the Department of Innovation, Science and Economic Development
    and by the Province of Ontario through the Ministry of Colleges and Universities. G.V. is a CIFAR associate fellow in the Quantum Information Science Program, a Distinguished Invited Professor at the Institute of Photonic Sciences (ICFO), and a Distinguished Visiting Research Chair at Perimeter Institute.

\bibliography{main}
\appendix
\onecolumngrid

\section{Numerical Methods: iPEPS}\label{app:methods_iPEPS}
Consider a 2d square lattice made of $L_x\times L_y \equiv N$ sites, $L_x\leq L_y$, where each site is represented by a $d$-dimensional vector space $\mathbb{C}^d$, and a state $\ket{\Psi}$ on the lattice that can be expressed as a PEPS as introduced in the main text.
The calculation of the norm $\mathcal{N}=\braket{\Psi|\Psi}$ is a common computation in PEPS algorithms, closely related to the computation of more complicated quantities of interest, e.g. expectation values $\braket{\Psi|O|\Psi}$ or correlation functions $\braket{\Psi|O_1O_2\cdots O_k|\Psi}$ of local operators. The exact contraction of the norm of $\mathcal{N}$ has a $O(L_y\exp(L_x))$ computational cost. The boundary MPS method is one of many related approaches~\cite{Nishino_1997,Baxter_2007,Levin_2007,Jordan_2008,Orus_2009, Lubasch_2014} for approximately computing $\mathcal{N}$, while keeping the computational cost to scale linearly in $L_x$ and $L_y$. In this Appendix we consider an infinite PEPS (iPEPS), with $L_x, L_y \to \infty$, by leveraging the translational invariance of the state, and where the computational cost is independent of $L_x$ and $L_y$. 

\begin{figure*}[h!]
  \centering
  {\includegraphics[width=0.75\textwidth]{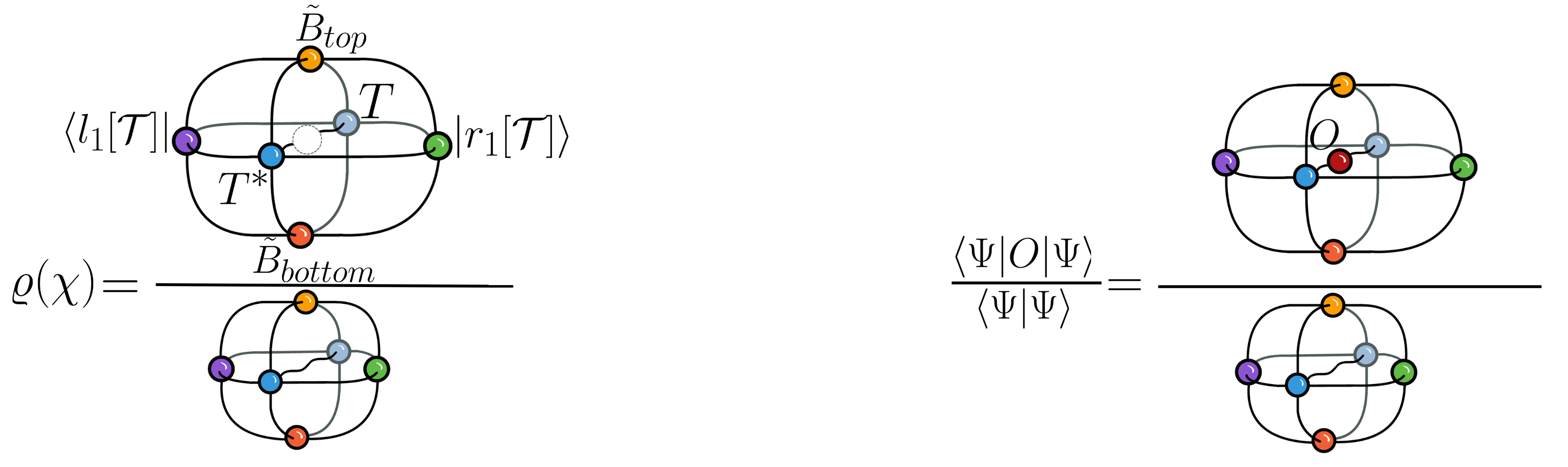}}\\
  \hspace{0.20\textwidth}\hspace{\textwidth}(a) \hspace{0.25\textwidth}(b)\hspace{0.1\textwidth}\\
  \caption{
  {\bf Computation of observable and reduced density matrix:} 
  (a) Tensor network for the computation of observables $\braket{\Psi|O|\Psi}/\braket{\Psi|\Psi}$. (b) Tensor netwok for the computation of normalized $\varrho(\chi)$. More detailed calculation of these objects is described below.}
\label{fig:obvs_and_varrho_norm}
\end{figure*}

\begin{figure*}[h!]
  \centering
  {\includegraphics[width=1\textwidth]{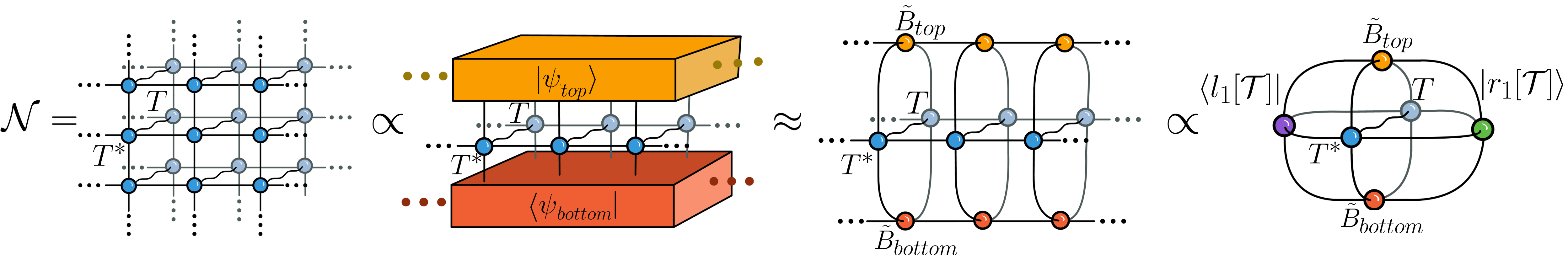}}\\
  \hspace{0.20\textwidth}\hspace{\textwidth}(a) \hspace{0.20\textwidth}(b) \hspace{0.25\textwidth}(c) \hspace{0.2\textwidth}(d)\hspace{0.1\textwidth}\\
  \caption{
  {\bf iPEPS norm computation:} 
  (a) Infinite square lattice tensor network of the norm $\mathcal{N}$ of an iPEPS. (b) Norm boundary states $\ket{\psi_{\text{\tiny \rm top}}}$ and $ \bra{\psi_{\text{\tiny \rm bottom}}}$ with a 1d transfer matrix in between, which is proportional to the iPEPS norm. (c) The bMPS algorithm obtains an approximation to $\ket{\psi_{\text{\tiny \rm top}}}$ and $ \bra{\psi_{\text{\tiny \rm bottom}}}$ with an infinite bMPS. (d) Contracting from the left and from the right of the infinite chain in (c) we obtain the left and right boundary states $\bra{l_1[\mathcal{T}]}, \ket{r_1[\mathcal{T}]}$ (see later for details on their calculation).}
\label{fig:iPEPS_norm_boundaries}
\end{figure*}

The tensor network for $\mathcal{N}$ consists of an infinite square lattice with a pair of tensors $TT^*$ on each site, where $T$ and $T^*$ are connected over their physical index (see Figure~\ref{fig:iPEPS_norm_boundaries}a).  One can show that the norm $\mathcal{N}$ diverges/vanishes with $L_x, L_y$ as $\mathcal{N}=\nu^{L_xL_y}$ where $\nu$ is a non-negative real number known as the \textit{norm-per-site}. Although it is possible to compute $\nu$, in practice this is not needed if we are interested in normalized quantities such as
-$\braket{\Psi|O|\Psi}/\N$ or the reduced density matrix $\varrho$, given by the ratio of two 2d tensor networks that only differ by the insertion of (one or more) local operators (see Figure \ref{fig:obvs_and_varrho_norm}). In this case, we need to compute the fixed point boundary states $\kket{\psi_{\text{\tiny \rm top}}}, \bbra{\psi_{\text{\tiny \rm bottom}}}$ (assumed to be unique) of the 1d transfer matrix given by an infinite row of $TT^*$ (see Figure~\ref{fig:iPEPS_norm_boundaries}b). From now on we drop the double braket notation $\kket{}, \bbra{}$ for boundary states, in favor of single braket notation $\ket{}, \bra{}$. The key idea behind the boundary MPS (bMPS) algorithm is to \textit{approximately} represent the infinite top and bottom boundary states $\ket{\psi_{\text{\tiny \rm top}}}$ and $ \bra{\psi_{\text{\tiny \rm bottom}}}$ with two infinite 1d MPS of bond dimension $\chi$. The algorithm detailed below describes how to obtain the top fixed point boundary MPS (the generalisation to the bottom boundary is straightforward).
\begin{figure*}[h!]
  \centering
  {\includegraphics[width=1\textwidth]{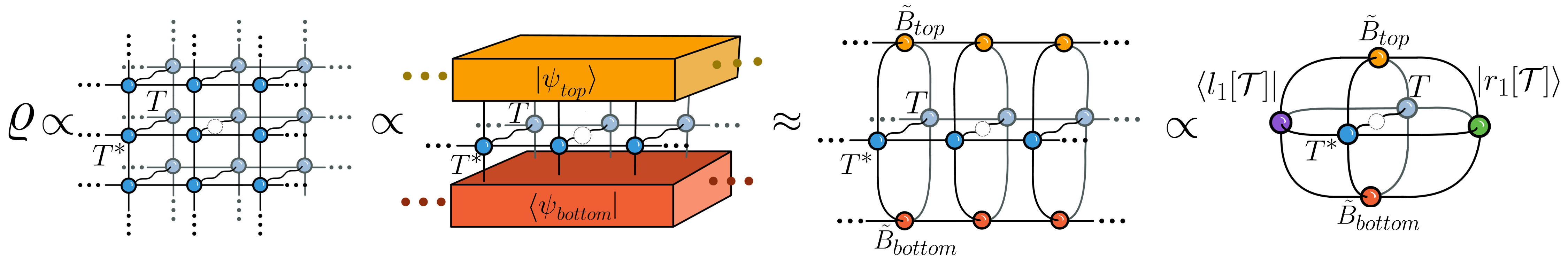}}\\
  \hspace{0.20\textwidth}\hspace{\textwidth}(a) \hspace{0.20\textwidth}(b) \hspace{0.25\textwidth}(c) \hspace{0.2\textwidth}(d)\hspace{0.1\textwidth}\\
  \caption{
  {\bf iPEPS reduced density matrix computation:} Computation of the iPEPS 2d reduced density matrix for one site $\mathcal{\varrho}$. It is closely related to $N$ in Figure \ref{fig:iPEPS_norm_boundaries} but a physical index is left open. Note that to obtain the correct $\varrho(\chi)$ we need to normalize the object in (d) as in Figure \ref{fig:obvs_and_varrho_norm}a.}
\label{fig:iPEPS_varrho_boundaries}
\end{figure*}

\subsection{Boundary MPS: Method}
We build an initial boundary MPS by multiplying pairs of tensors $TT^*$ after removing the vertical upper index (by fixing their value to 1)\footnote{Because we are interested in the fixed point (assumed to be unique) of the 1d transfer matrix, the choice of initial $\kket{\psi(y=0)}$ is not important.}. This translation-invariant bMPS is characterized by a tensor $B$ with 4 indices represented graphically as ``legs": 2 bond legs connecting the tensors with each other, and other 2 legs, one joining $B$ with $T$ and the other with $T^*$. This initial bMPS has bond dimension $D^2$. If $D^2\leq\chi$, we act with the transfer matrix $n$ times until the first truncation is required (namely, when $D^{2n}>\chi$). After this point, we iteratively perform two operations: i) truncation of the boundary MPS bond dimension down to $\chi$ and ii) application of the 1d $TT^*$ transfer matrix which increases the bMPS bond dimension to $\chi D^2$. This algorithm ends once the bMPS has \textit{converged} (see \ref{subsubapp:contract_conv}). The converged bMPS $\ket{\Tilde{\psi}}$, characterized by tensor $\Tilde{B}$, is our approximation to the true fixed point boundary state $\ket{\psi_{\text{\tiny top}}}$. 

\subsubsection{Infinite MPS Bond 
Dimension Truncation} \label{subsubapp:bmps_truncation}
A key component of the algorithm described above is to perform a bond dimension truncation of the infinite bMPS from $\chi'\to\chi$, where $\chi'>\chi$, while trying to keep a high fidelity between the states represented by $B(\chi)$ and the original $B(\chi')$. In order to do this, we consider a bond bipartition into left and right halves of the infinite spin chain represented by the bMPS. We bring this bond into the canonical form~\cite{Vidal_2003}, such that the bond index, $i$, corresponds to the labeling of Schmidt vectors in the Schmidt decomposition of the bMPS representation of the state $|\psi\rangle$ across that index. More explicitly:
\begin{equation}
    |\psi\rangle = \sum_{i=1}^{\chi'}\sigma_{i}\ket{\Phi^L_i}\otimes\ket{\Phi^R_i},
\label{eq:schmidt_decomposition}
\end{equation}
where $\sigma_i$ label the Schmidt coefficients (normalized such that $\sum_i\sigma_i^2 = 1$ and arranged in decreasing value $\sigma_1\geq\sigma_2\geq\cdots\geq\sigma_{\chi'}\geq0$) and $\ket{\Phi^L_i},\ket{\Phi^R_i}$ are the Schmidt vectors, which form an orthonormal set $\braket{\Phi^L_i|\Phi^L_j} = \braket{\Phi^R_i|\Phi^R_j} = \delta_{ij}$. The basis  $\ket{\Phi^L_i},\ket{\Phi^R_i}$ are also the eigenvectors of the reduced density matrices $\rho^A$ and $\rho^B$, with eigenvalues $\sigma_i^2$. Once we perform a gauge transformation to bring the bond into this basis, we can truncate the bond dimension of the bMPS by keeping only the largest $\chi$ Schmidt coefficients. We apply this truncation for every bond in the infinite chain. This truncation procedure will keep truncation errors low as long as the entanglement spectrum eigenvalues $\{\sigma_i^2\}$ decay sufficiently fast with $i$~\cite{Orus_2008,Vidal_2003}.

\begin{figure*}[t!]
  \centering
  {\includegraphics[width=0.5\textwidth]{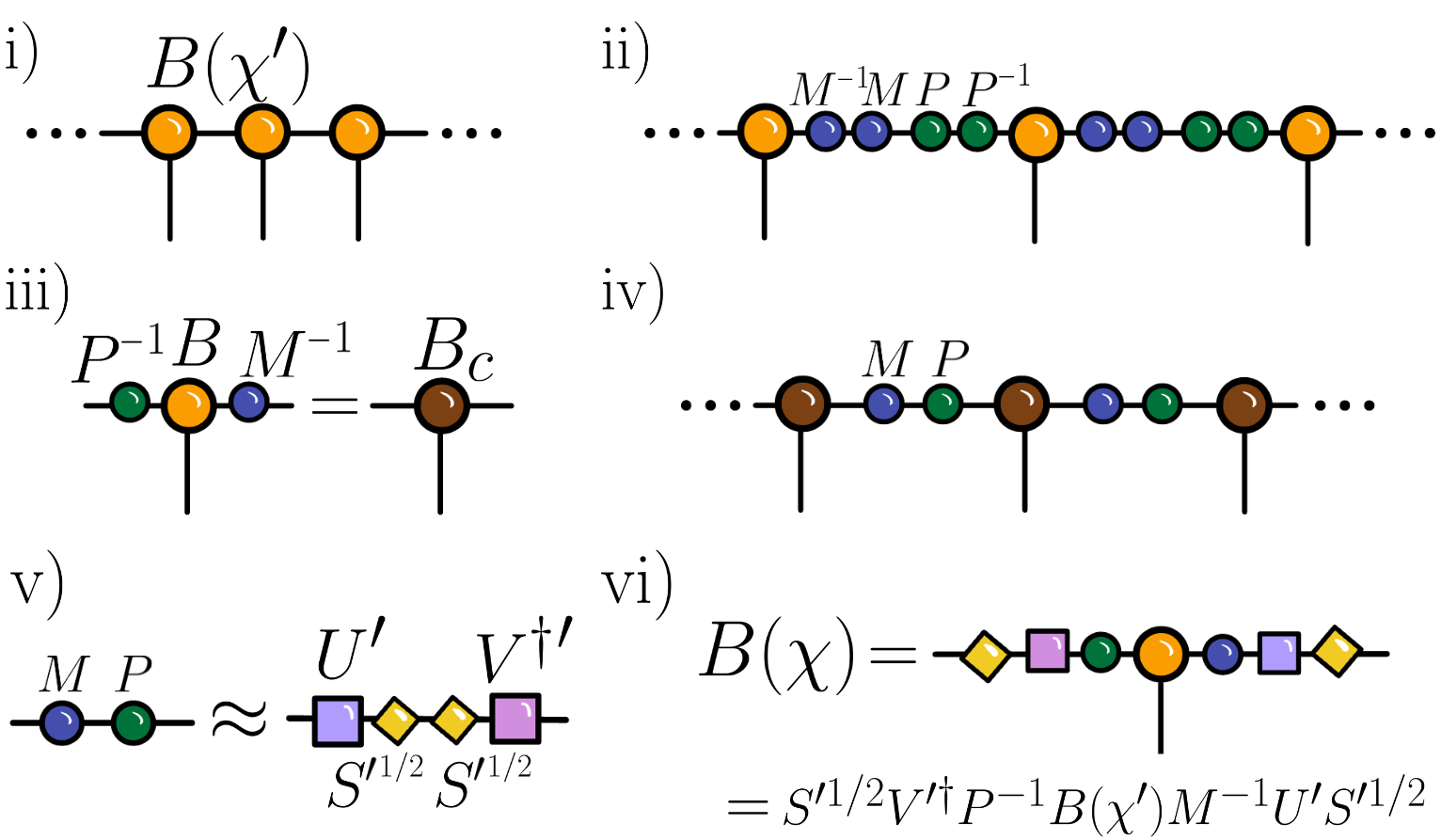}}\hspace{5mm}
  \raisebox{0.15\height}{\includegraphics[width=0.4\textwidth]{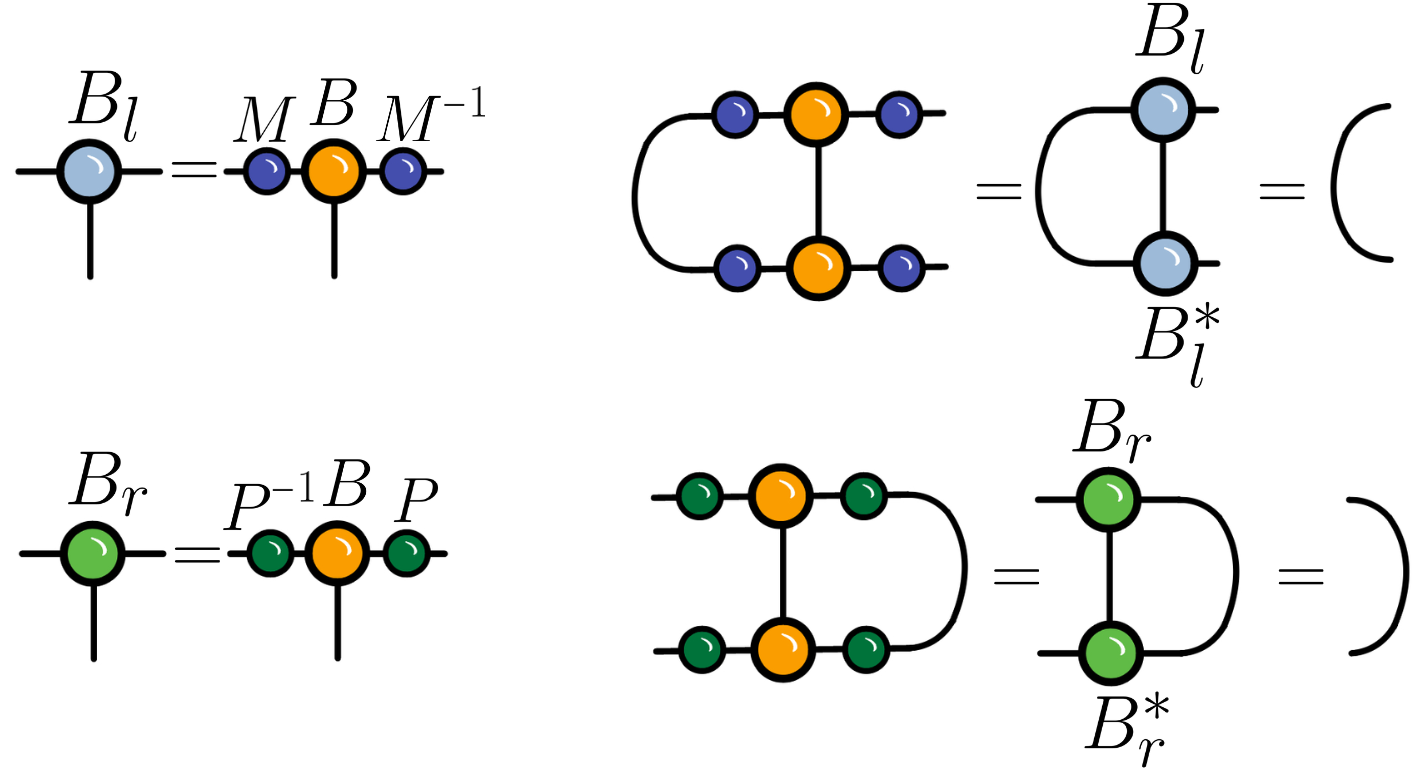}}
  \hspace{\textwidth}(a) \hspace{0.5\textwidth}(b) \hspace{0.35\textwidth}\\
  \caption{
  {\bf Truncation of bMPS bond dimension:} 
  (a) Steps, i)-iv), in the algorithm for truncating the bond dimension of the bMPS.
  (b) Finding left and right canonical form gauge transformation matrices $M$ (top) and $P$ (bottom).}
\label{fig:bMPS_truncation_method}
\end{figure*}

We summarize the truncation algorithm in steps i) to vi) as depicted in Figure \ref{fig:bMPS_truncation_method}a. We need to obtain the invertible matrices $M$ and $P$ that bring the translationally invariant bMPS tensor $B$ into the left and right canonical form $B_l$ and $B_r$ respectively. This gauge requirements are such that the left and right dominant eigenvectors of $B_lB_{l}^*$ and $B_rB_{r}^*$ respectively (contracted over the physical bMPS index as shown in Figure \ref{fig:bMPS_truncation_method}b) should be the identity and its corresponding dominant eigenvalue, $\lambda_1$, should be 1 for normalisation. These conditions are depicted in Figure \ref{fig:bMPS_truncation_method}b. Starting from the large bond dimension infinite bMPS that we wish to truncate (Fig \ref{fig:bMPS_truncation_method}a.i)), we insert two resolutions of the identity $M^{-1}M$ and $PP^{-1}$ at every bond (Fig \ref{fig:bMPS_truncation_method}a.ii)). This yields an infinite chain of $P^{-1}BM^{-1}$ tensors denoted $B_c$ (Fig \ref{fig:bMPS_truncation_method}a.iii)) with a matrix $MP$ on the bonds (Fig \ref{fig:bMPS_truncation_method}a.iv)). We consider the singular value decomposition of the matrix $MP = U S V^{\dagger}$. The diagonal entries of the matrix $S$ and the isometries $U, V^{\dagger}$ correspond to the Schmidt decomposition presented in Equation \ref{eq:schmidt_decomposition} if we had considered only one bond partitioning the infinite chain in half. We truncate these objects and keep only the largest $\chi$ singular values and corresponding vectors therefore approximating the bond with smaller sized matrices $MP\approx U'S'V'^{\dagger}$ (Fig \ref{fig:bMPS_truncation_method}a.v)). Putting everything together, as in Fig \ref{fig:bMPS_truncation_method}a.vi), we obtain a final expression for $B(\chi) = S'^{1/2}V'^{\dagger}P^{-1}B(\chi')M^{-1}U'S'^{1/2}$.

\subsubsection{Convergence of the Boundary MPS to a Fixed Point} \label{subsubapp:contract_conv}
After a sufficient number of iterations, $y_*$, of the 1d $TT^*$ transfer matrix followed by the truncation as described in \ref{subsubapp:bmps_truncation} we observe that our bounday MPS has \textit{converged}, meaning that $B(y_*) \approx B(y_*+1)$ within a set tolerance, in which case we define $\Tilde{B} = B(y_*+1)$. This equivalence can be assessed by looking at the fidelity-per-site (see \ref{subsubapp:fidelity_iMPS} for definition) between $B(y)$ and $B(y+1)$ (which should be approximately 1 if convergence has been reached) or by comparing the Schmidt spectrum of the boundaries in successive iterations (which should be equal). In our work we use both metrics to ensure convergence.

\subsection{Infinite MPS and Infinite PEPS: Computation of Metrics}
\begin{figure*}[h!]
  \centering
  {\includegraphics[width=0.95\textwidth]{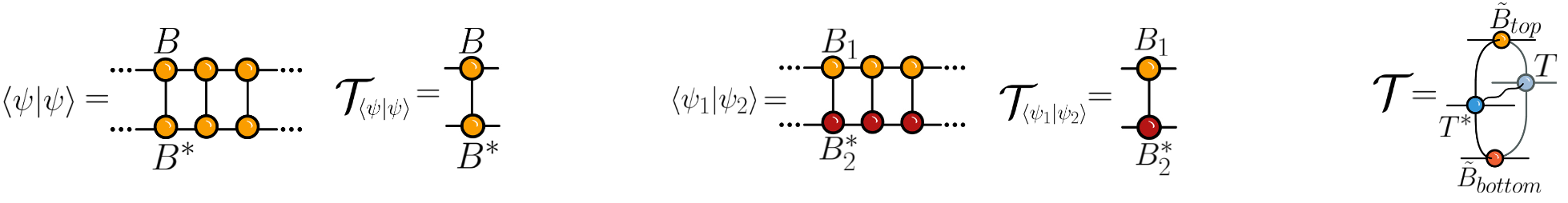}}
  \hspace{0.2\textwidth}(a) \hspace{0.30\textwidth}(b) \hspace{0.35\textwidth}(c) \hspace{0.1\textwidth}\\
  \caption{
  {\bf Transfer matrices:} (a) Norm $\braket{\psi|\psi}$ of an infinite MPS (left) and the transfer matrix $\mathcal{T}_{\braket{\psi|\psi}}$ involved in its calculation. (b) Fidelity  $F =\braket{\psi_1|\psi_2}$ between two infinite MPS and the transfer matrix $\mathcal{T}_{\braket{\psi_1|\psi_2}}$. (c) Transfer matrix $\mathcal{T}$ used for computing $\mathcal{N},\varrho$ and $\braket{\psi|O|\psi}$.}
\label{fig:transfer_matrices}
\end{figure*}
\subsubsection{Infinite MPS: Norm and Correlation Length}
Consider an infinite MPS $\ket{\psi}$ given by tensor $B$ of bond dimension $\chi$. Its transfer matrix $\mathcal{T}_{\braket{\psi|\psi}}$ (see Figure \ref{fig:transfer_matrices}a) has eigenvalue decomposition:
\begin{equation}
    \mathcal{T}_{\braket{\psi|\psi}} = \sum_{i=1}^{\chi^2}\lambda_i\ket{r_i}\bra{l_i},
\end{equation}
where $\{\lambda_i\}$ is the set of eigenvalues, organized such that $|\lambda_1|>|\lambda_2|\geq\cdots\geq|\lambda_{\chi^2}|$, and $\ket{r_i}, \bra{l_i}$ are the right and left eigenvectors respectively. One can show that: i) each eigenvalue $\lambda$ is either real or part of a complex conjugate pair $\lambda, \lambda^*$ and ii) the dominant eigenvalue $\lambda_1$ is real and non-negative. We have also assumed that the dominant eigenvalue $\lambda_1$ is not degenerate. The norm $\braket{\psi|\psi}$ (\ref{fig:transfer_matrices}a) is given by $\lambda_1^{L_x}$, where $L_x$ is the number of tensors in the MPS, and thus $L_x\to\infty$ for an infinite MPS. If $\lambda_1\neq1$, implying that $\braket{\psi|\psi}\neq1$, we may obtain a normalized MPS $\ket{\psi'}$ with tensor $B'=B/\sqrt{\lambda_1}$ so that $\lambda_1' =1$ and $\braket{\psi'|\psi'}=1$.

The correlation length $\xi$ is calculated by considering the dominant and second largest eigenvalue $\lambda_1, \lambda_2$ and it is given by:
\begin{equation}
    \xi = \frac{-1}{\log(|\lambda_2[\mathcal{T}_{\braket{\psi|\psi}}]/\lambda_1[\mathcal{T}_{\braket{\psi|\psi}}]|)},
\end{equation}
where we introduce the notation $\lambda_i[\mathcal{T}_{\braket{\psi|\psi}}]$ to refer to the $i^{\text{th}}$ eigenvalue of the transfer matrix $\mathcal{T}_{\braket{\psi|\psi}}$.

\subsubsection{Infinite MPS: Fidelity} \label{subsubapp:fidelity_iMPS}

Consider two iMPS $\ket{\psi_1}, \ket{\psi_2}$ given by tensors $B_1, B_2$ and bond dimensions $\chi_1, \chi_2$ respectively, normalized such that $\braket{\psi_1|\psi_1}=\braket{\psi_2|\psi_2}=1$, that is $\lambda_1[\mathcal{T}_{\braket{\psi_1|\psi_1}}] = \lambda_1[\mathcal{T}_{\braket{\psi_2|\psi_2}}] = 1$ (see previous section). Their fidelity, $F[\psi_1,\psi_2]=\braket{\psi_1|\psi_2}$ can be expressed as $F=F_s^{L_x}$, where the \textit{fidelity-per-site} $F_s$ is such that $0\leq |F_s|\leq 1$. One can show that $|F_s|=1$ if and only if $B_2$ equals $B_1$ up to a so-called \textit{gauge transformation}, in which case both tensors give rise to the same state.
To calculate $F_s$ consider the transfer matrix given by $B_1B_2^*$, as depicted in Figure \ref{fig:transfer_matrices}b. This transfer matrix can be decomposed in the standard way:
\begin{equation}
    \mathcal{T}_{\braket{\psi_1|\psi_2}} = \sum_{i=1}^{\chi_1\chi_2}\lambda_i\ket{r_i}\bra{l_i},
\end{equation}
where $\{\lambda_i\}$ is the set of eigenvalues, organized such that $|\lambda_1|>|\lambda_2|\geq\cdots\geq|\lambda_{\chi_1\chi_2}|$, and $\ket{r_i}, \bra{l_i}$ the right and left eigenvectors respectively. Here, we have assumed that the dominant eigenvalue $\lambda_1$ is not degenerate. Then, the fidelity-per-site
$F_s$ is simply given by this dominant eigenvalue, $F_s = \lambda_1[\mathcal{T}_{\braket{\psi_1|\psi_2}}]$.

\subsubsection{Infinite PEPS: Reduced Density Matrix} \label{subsubapp:red_rho_2d_peps}

The reduced density matrix for a local region $R$ of the 2d lattice is denoted $\varrho_R=\tr_{\bar{R}}\ket{\Psi}\bra{\Psi}/\N$, where the trace is over all sites except those in $R$. For simplicity, we consider the case where $R$ consists of a single site and refer to this single-site reduced density matrix simply as $\varrho$. In a PEPS calculation with boundary MPS with bond dimension $\chi$, we obtain an approximation $\varrho(\chi)$ to the exact $\varrho$. The calculation of $\varrho(\chi)$ is shown in Figure \ref{fig:obvs_and_varrho_norm}b, where the dominant left and right eigenvectors $\ket{r_1}, \bra{l_1}$ are those of the transfer matrix $\mathcal{T}$, depicted in Figure \ref{fig:transfer_matrices}.

The error in $\varrho(\chi)$ due to finite $\chi$ in the bMPS depends directly on the square of the Schmidt coefficients of the bMPS, as seen in Figure \ref{fig:2dPESP_red_rho_app}a, and therefore on the $n=1$ (von Neumann) Renyi entropy.\\

\subsection{iPEPS and Boundary MPS: Additional Numerical Results}
In this appendix we provide additional numerical results for the contraction of 2d infinite PEPS with the boundary MPS formalism.
\subsubsection{Averaging Over Several Realisations}
\begin{figure}
  \centering{\includegraphics[width=0.35\textwidth]{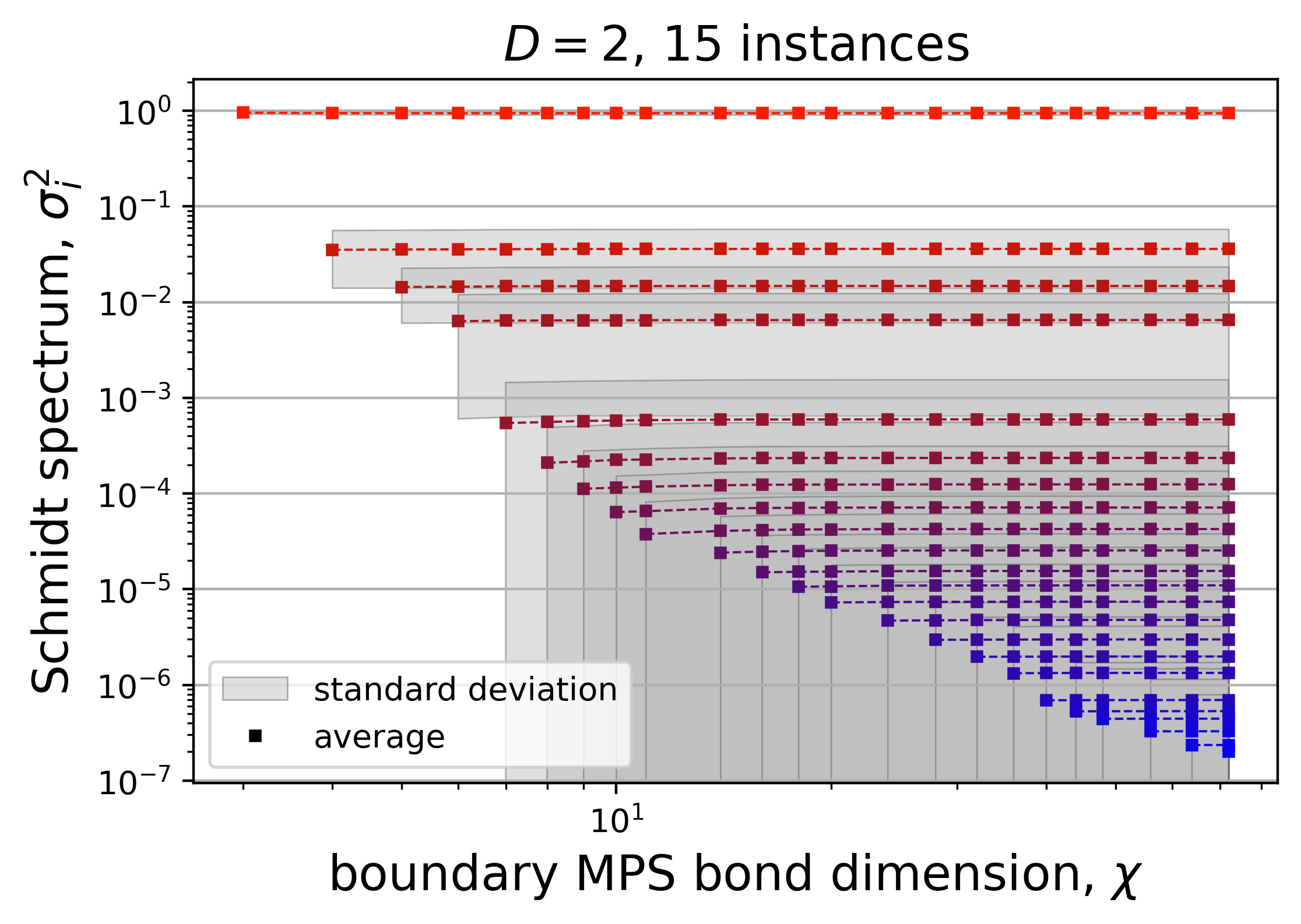}}\hspace{3mm}
  {\includegraphics[width=0.35\textwidth]{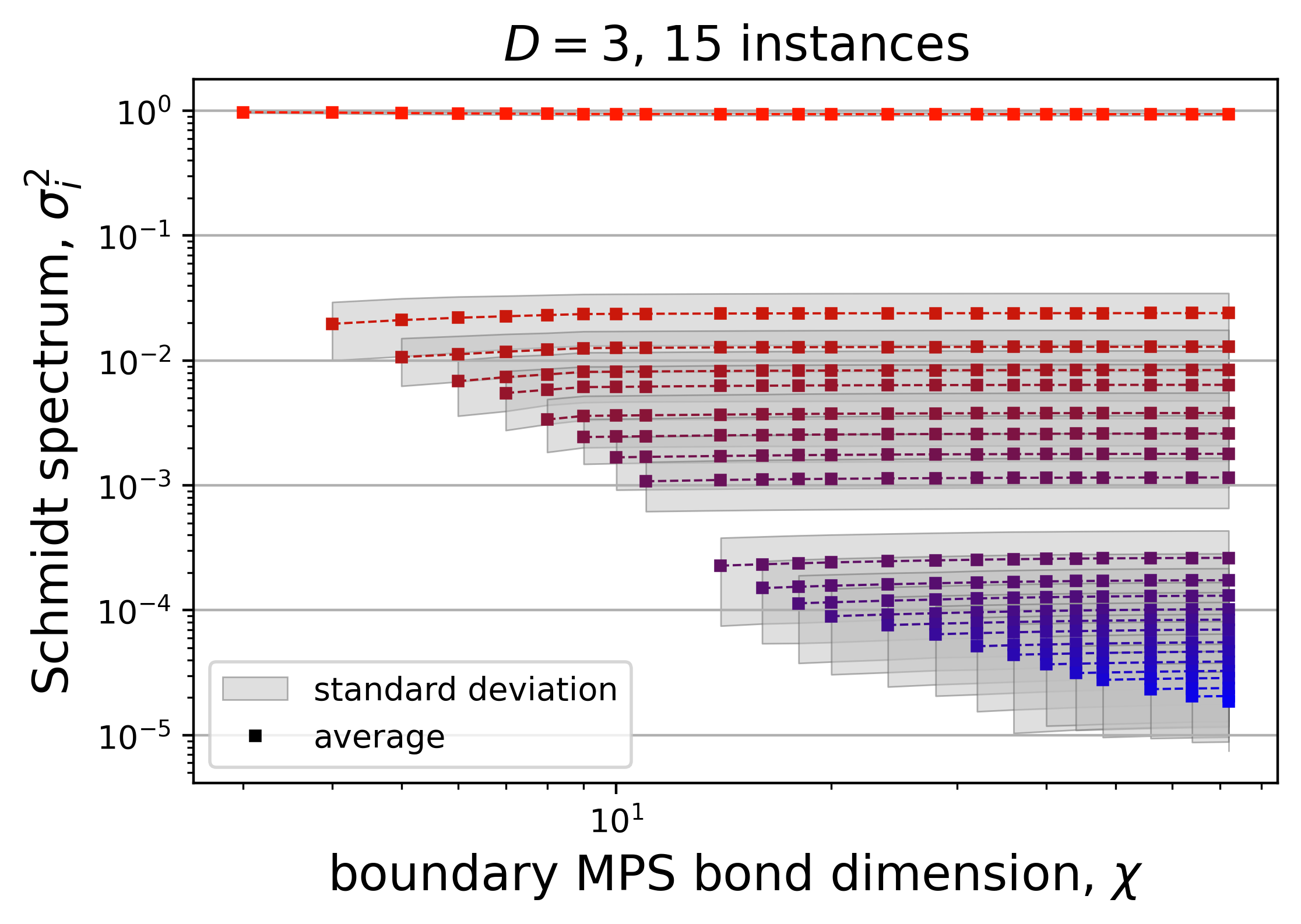}}\\
   \hspace{0.05\textwidth}(a)\hspace{0.35\textwidth}(b)\\
  {\includegraphics[width=0.35\textwidth]{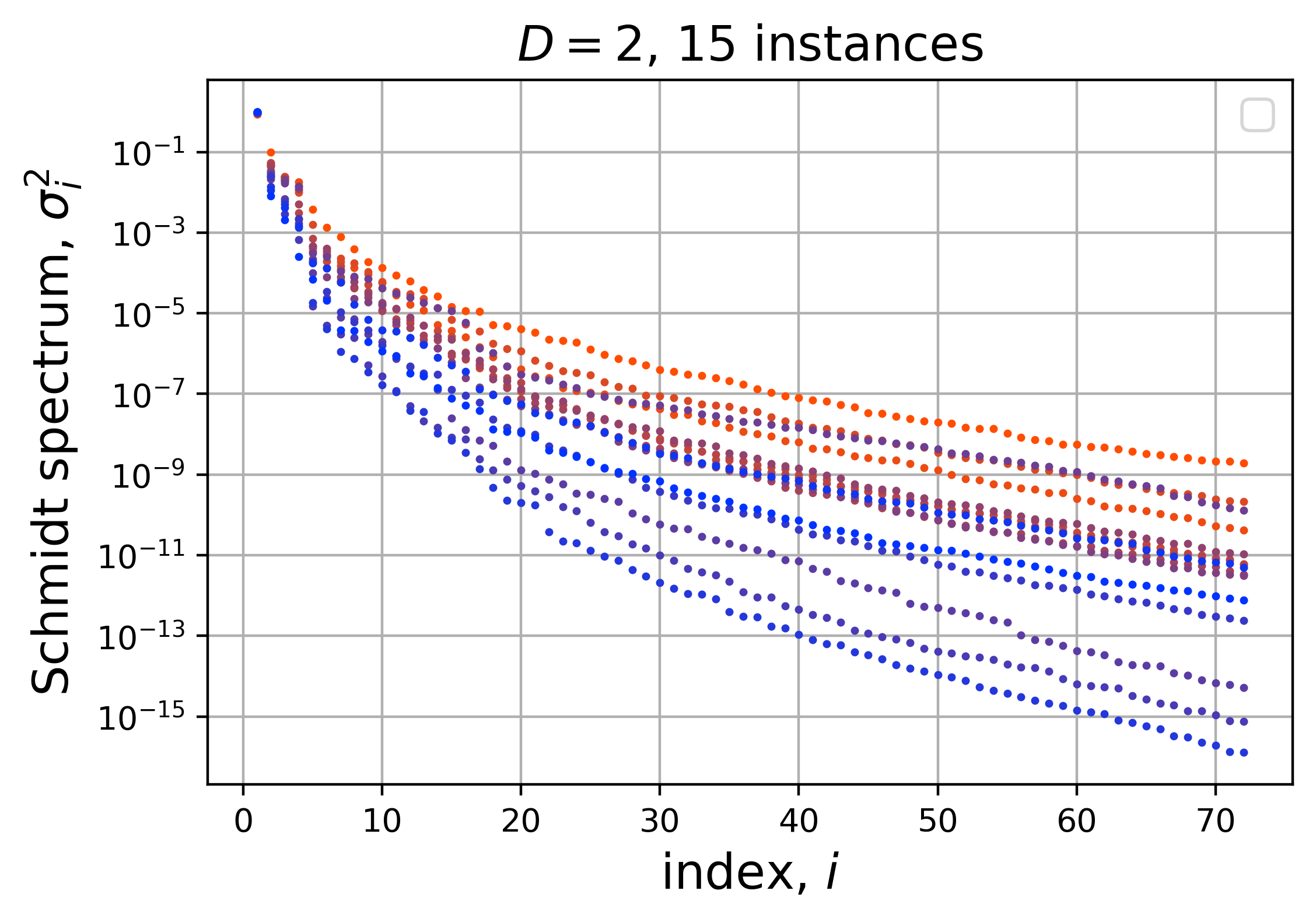}}\hspace{3mm}{\includegraphics[width=0.35\textwidth]{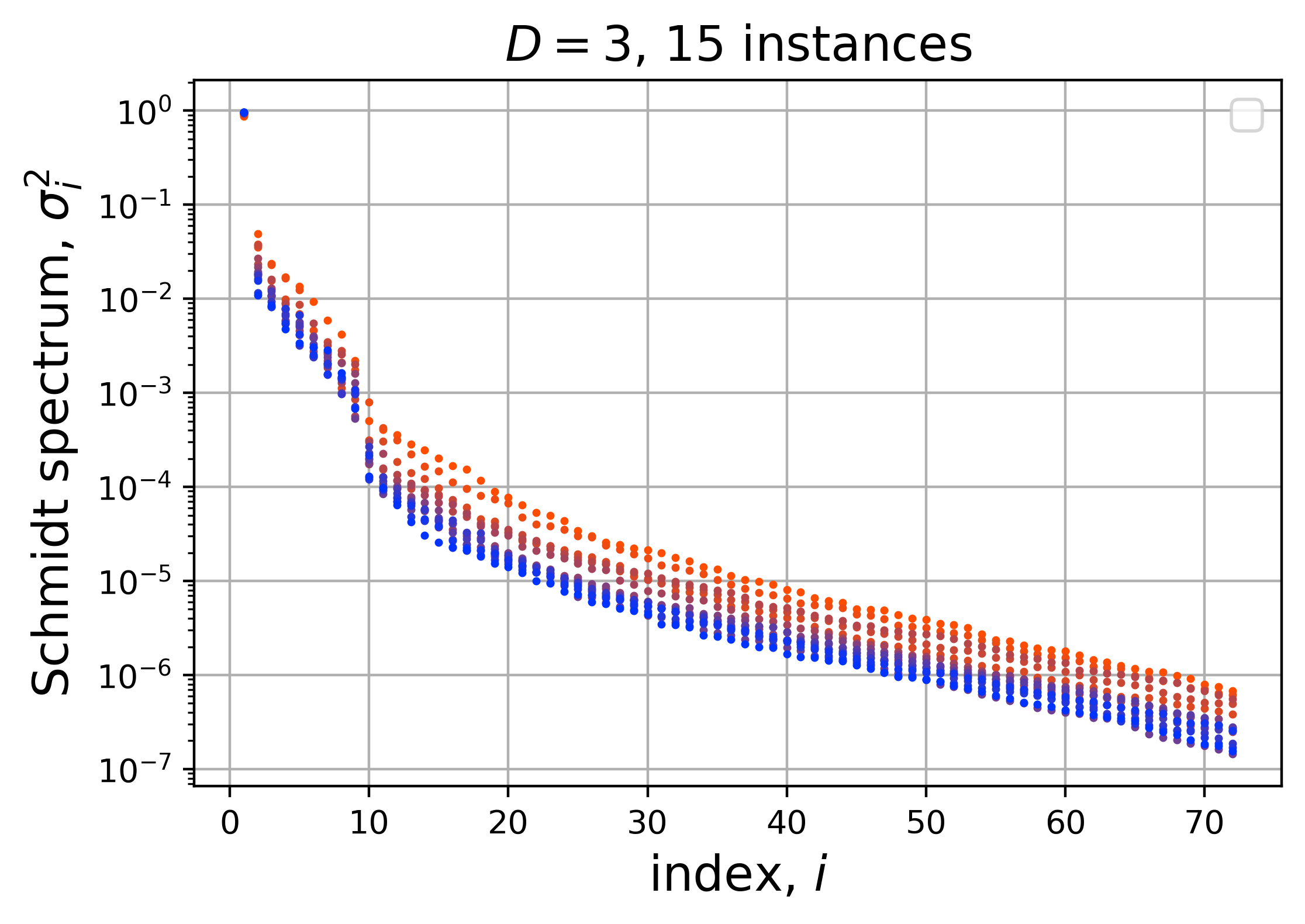}}\\
  \hspace{0.05\textwidth}(c)\hspace{0.35\textwidth}(d)\\{\includegraphics[width=0.35\textwidth]{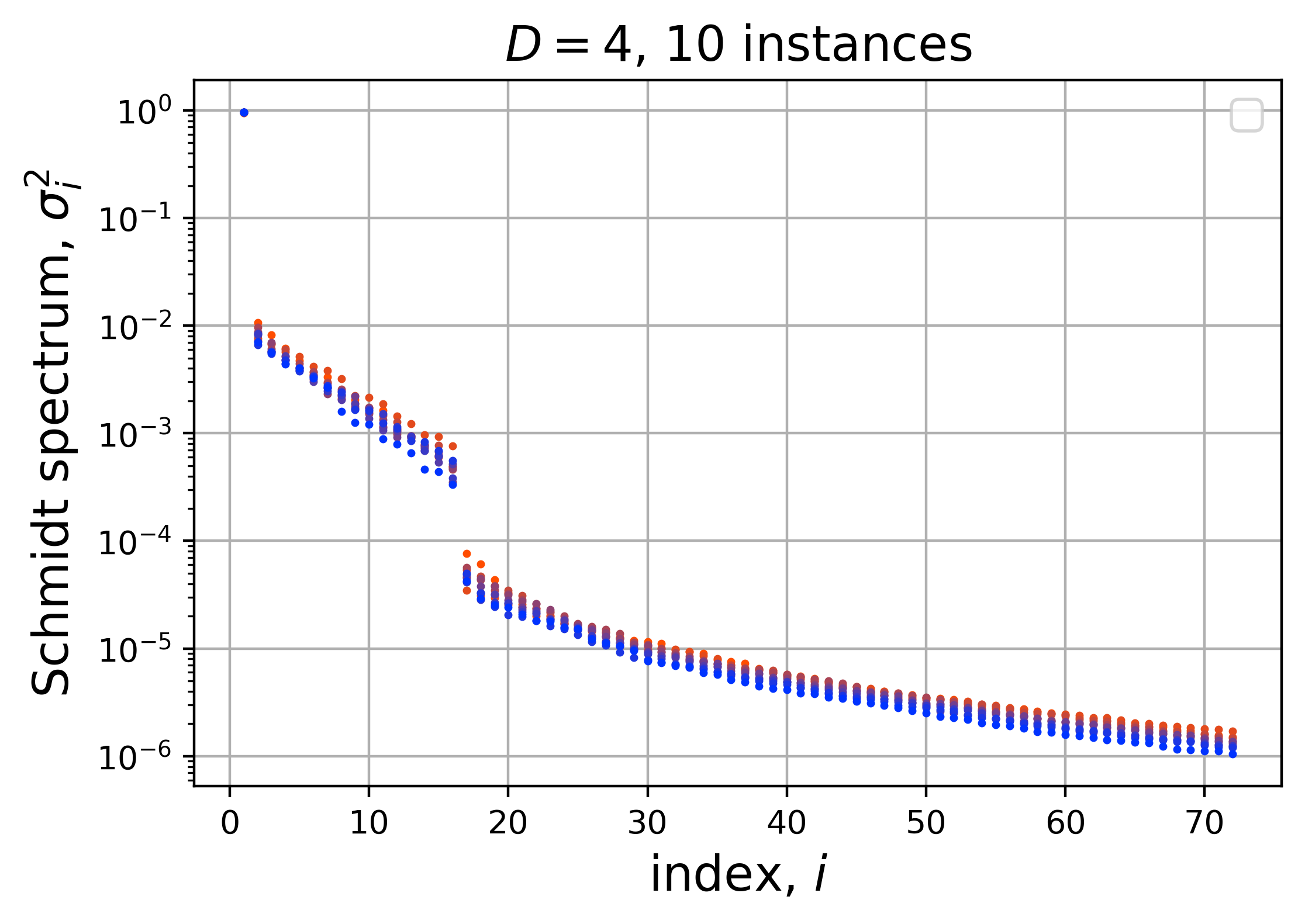}}\hspace{3mm}
  {\includegraphics[width=0.35\textwidth]{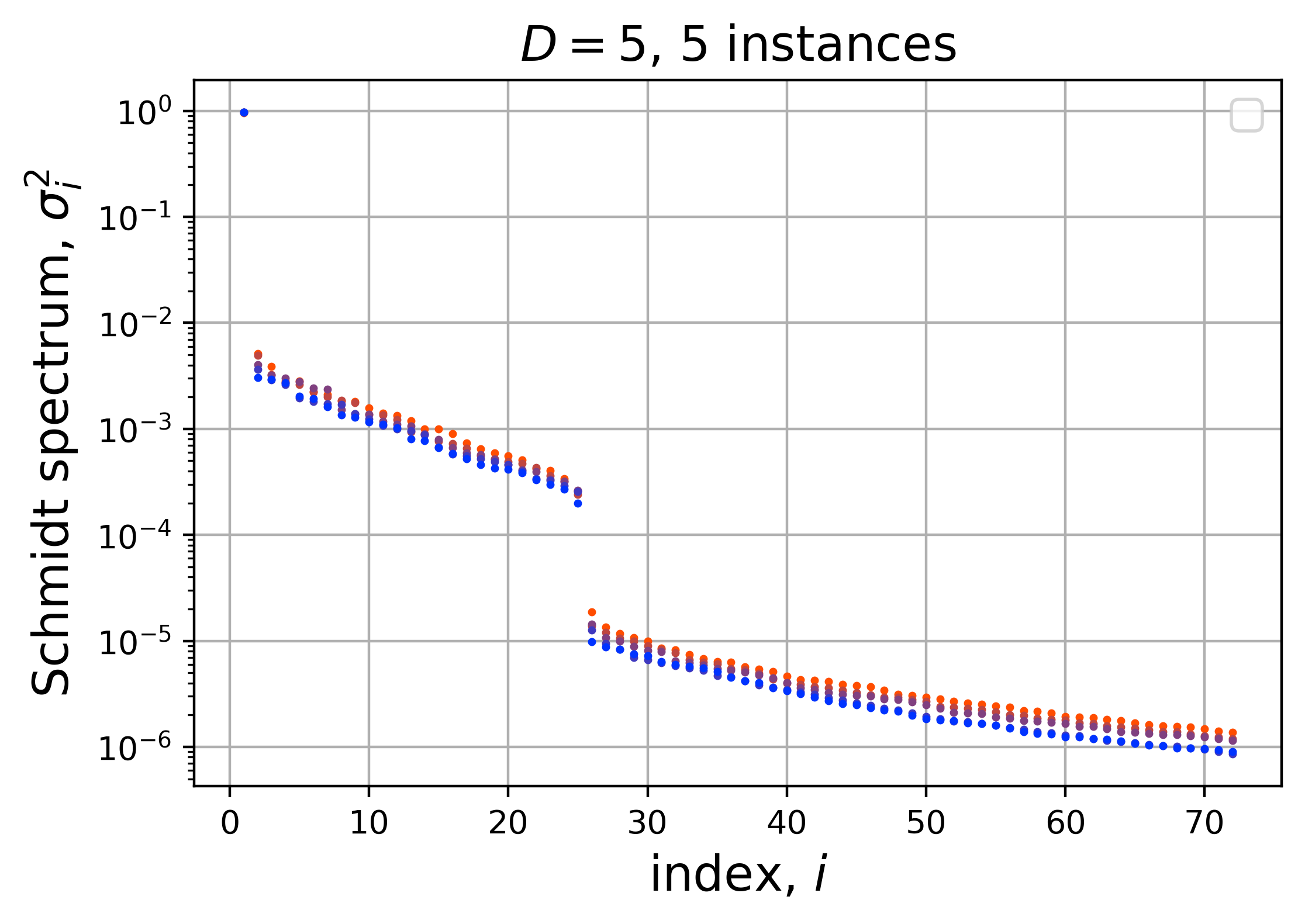}}\\
  \hspace{0.05\textwidth}(e)\hspace{0.35\textwidth}(f)
  \caption{
  {\bf bMPS average Schmidt entanglement spectrum, several 2d PEPS instances $T$:} (a) and (b) $D=2$ and $D=3$ (respectively) of the average and standard deviation (over 15 instances) of the Schmidt spectrum of $\Tilde{B}(\chi)$ for a range of $\chi$ between $\chi_{\rm min}=2, \chi_{\rm max}=72$. The average and standard deviation is calculated for each eigenvalue index. (c)-(f) $D=2,3,4,5$ (respectively) Schmidt spectrum for $\Tilde{B}(\chi_{\rm max}=72)$ for varying number of instances.}
\label{fig:ent_spectrum_averages}
\end{figure}

\begin{figure}[h!]
\centering{\includegraphics[width=0.32\textwidth]{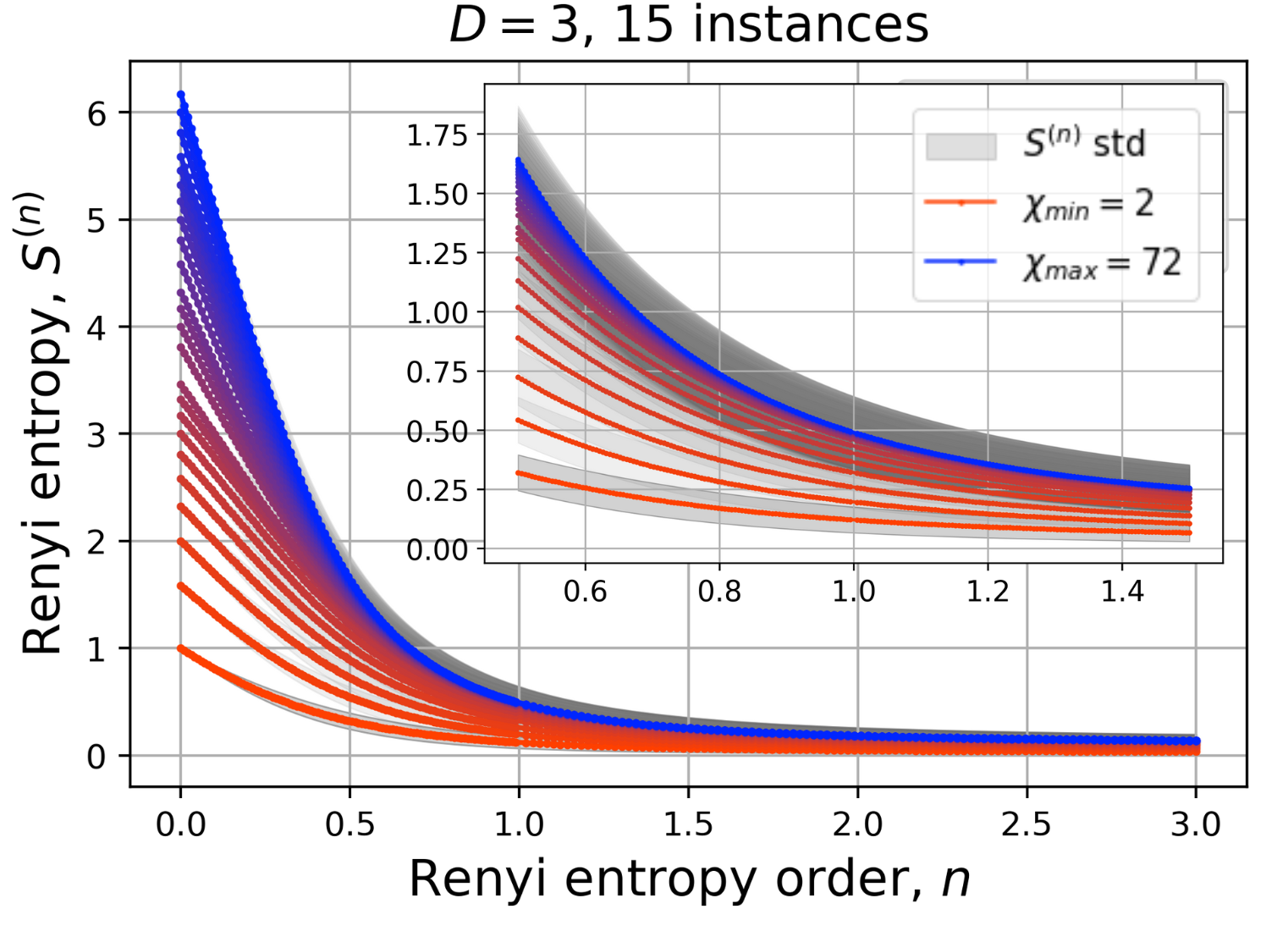}}\hspace{1mm}{\includegraphics[width=0.32\textwidth]{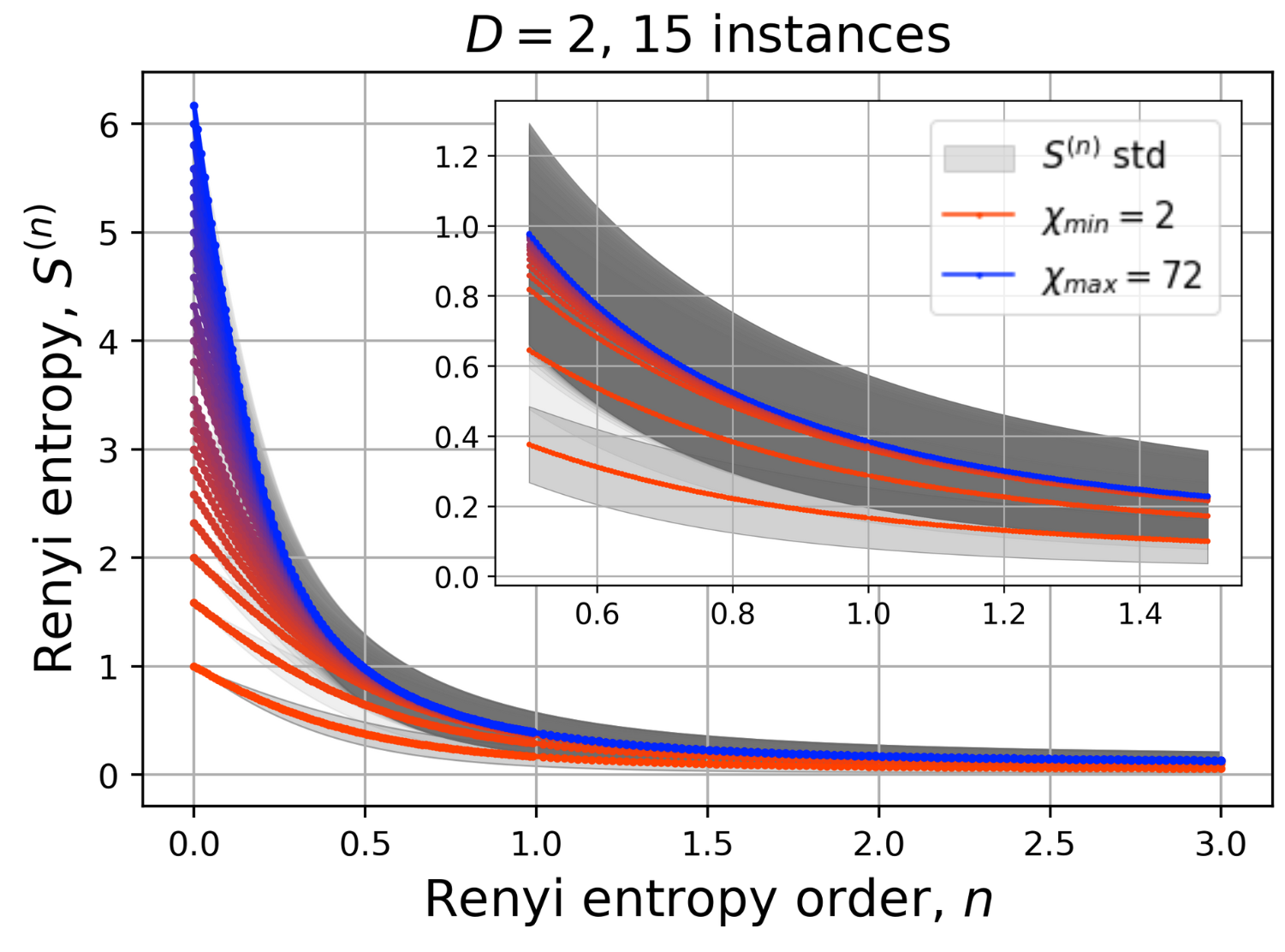}}\hspace{1mm}\raisebox{0.01\height}{\includegraphics[width=0.34\textwidth]{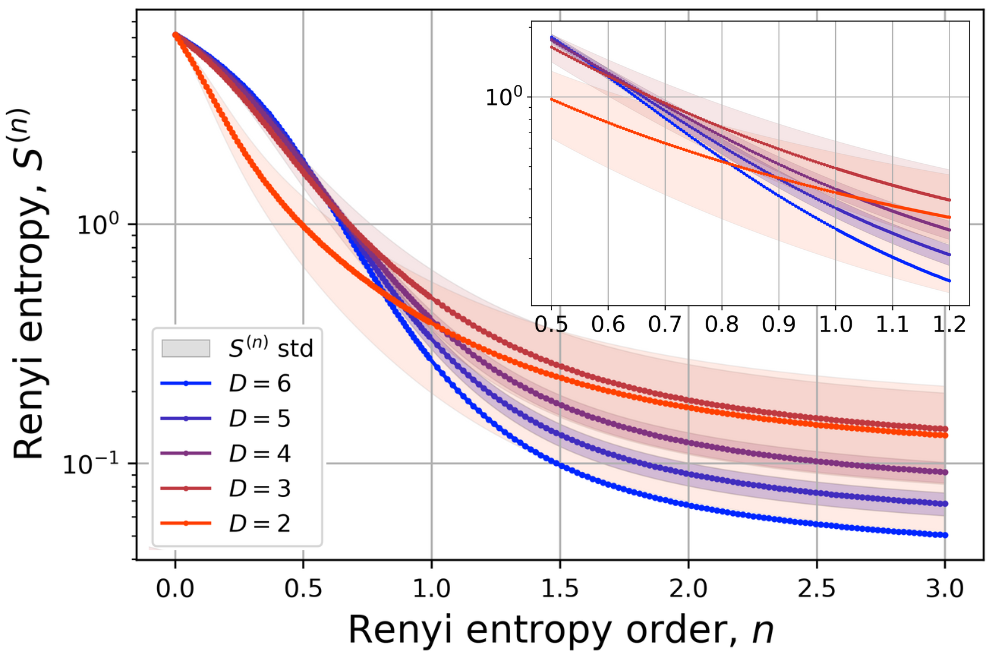}}\\
  (a) \hspace{0.3\textwidth}(b) \hspace{0.3\textwidth}(c)\hspace{0.3\textwidth}\\
  \caption{
  {\bf Renyi entropy order dependence:} (a) and (b) $D=2$ and $D=3$ (respectively) system Renyi entropy as a function of order $n$, for all simulated bMPS bond dimension $2\leq \chi \leq 72$. Average over 15 random initializations of $T$. (b) Renyi entropy for $\Tilde{B}(\chi_{\rm max} = 72)$ as a function of order $n$ for $D=2,3,4,5,6$ averaged over 15,15,10,5,1 iterations respectively.
  We note that the $n=1$ (von Neumann) entropy is expected to control the error of the truncated bMPS approximation, and that the $n=0$ entropy simply counts the number of Schmidt weights and trivially saturates to $\chi_{\rm max}$.}
\label{fig:renyi_ent_order_dep}
\end{figure}

In the main text we have mostly displayed results for a single instance of a random tensor $T$ for each PEPS bond dimension $D$. 
Here, we examine several random instances to build confidence that the reported behavior is typical.
Specifically, we explore average properties of $N_{s}=15,15,10,5,1$ instances of clean, random iPEPS with bond dimension $D=2,3,4,5,6$ respectively. In all cases, we again observe rapid convergence of the entanglement spectrum  (Figure~\ref{fig:ent_spectrum_averages}) and Renyi entropies (Figure~\ref{fig:renyi_ent_order_dep}) with the truncated bMPS bond dimension, $\chi$.
Moreover, the variance in these quantities decreases rapidly with increasing PEPS bond dimension $D$. This suppression qualitatively agrees with the statistical mechanical model, for which fluctuations around the ground state are suppressed at large $D$.

\subsubsection{Fidelities of Converged bMPS}
\begin{figure}[h!]
  \centering
  {\includegraphics[width=0.33\textwidth]{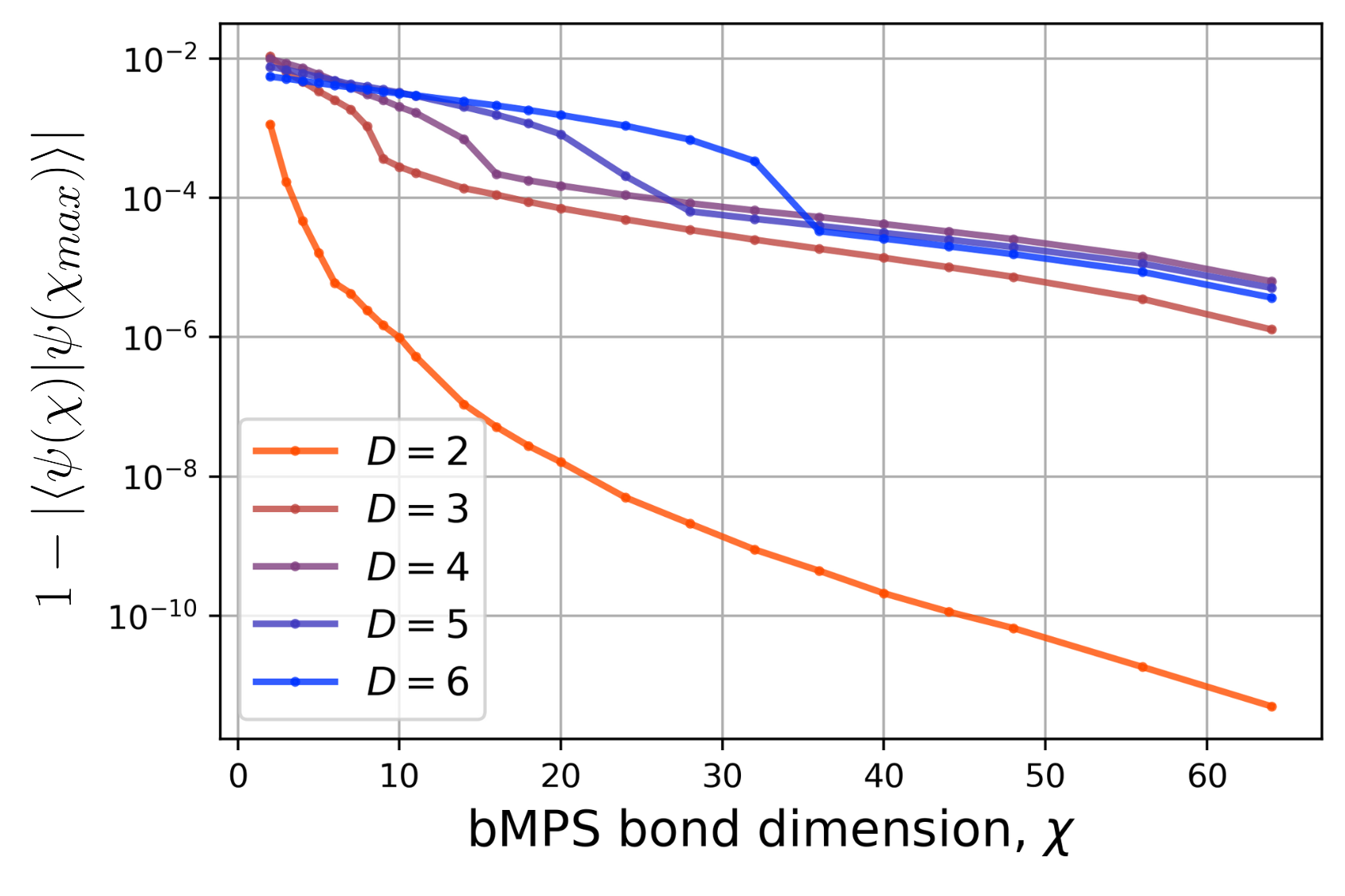}}\hspace{5mm}{\includegraphics[width=0.33\textwidth]{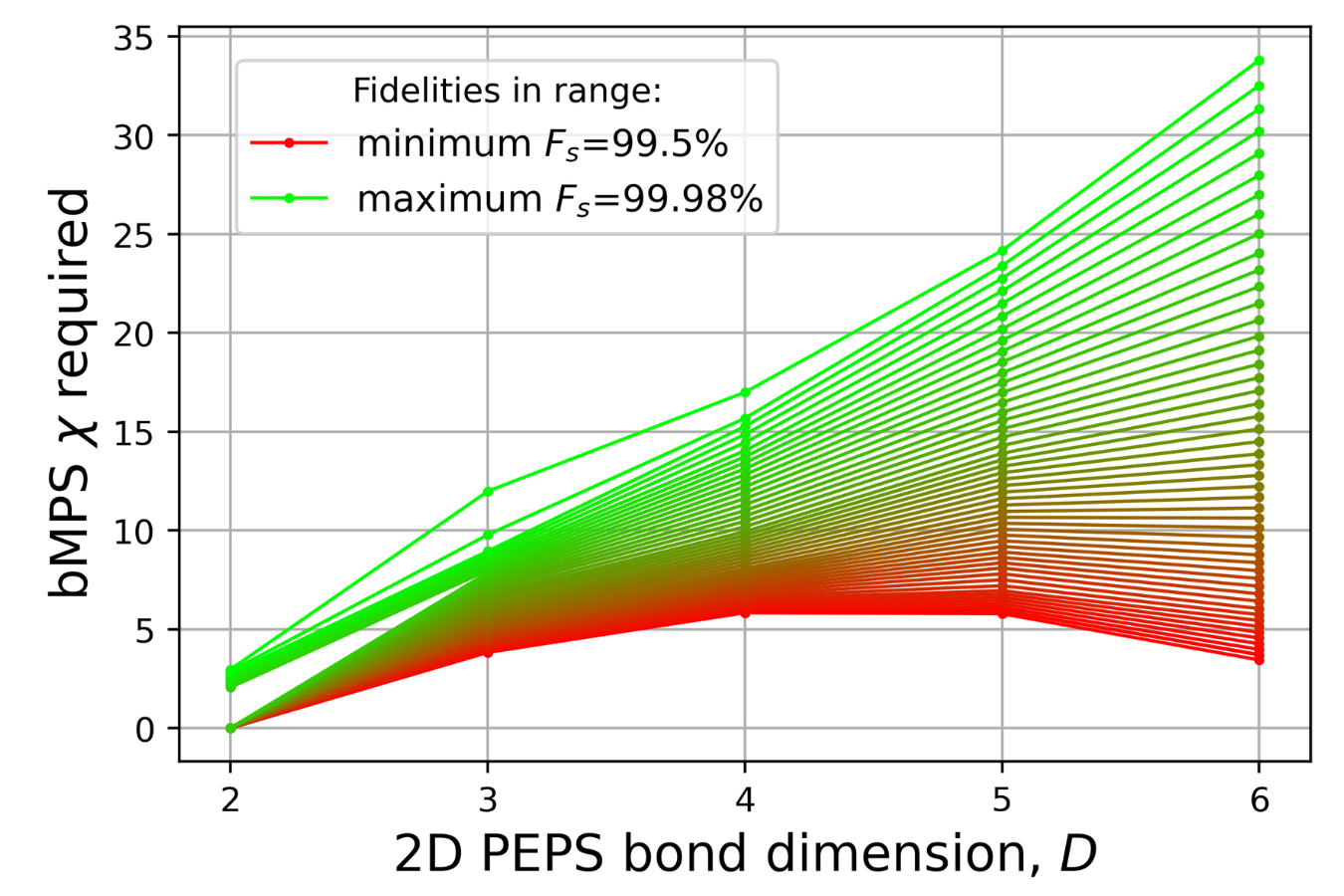}}\\
  (a) \hspace{0.3\textwidth}(b)\\
  \caption{
  {\bf bMPS fidelity numerical results:} (a) Convergence of the fidelity-per-site of the largest simulated bMPS network ($\chi_{\rm max}=72$) with all other smaller systems. 
  (b) Bond dimension required for the bMPS to obtain a certain fidelity-per-site with the largest simulated boundary.}
\label{fig:fidelity_diag_results}
\end{figure}

In order to quantify how close $\ket{\Tilde{\psi}(\chi)}$ is to the exact fixed-point boundary state   we calculate the fidelity-per-site $F_s$ between $\ket{\Tilde{\psi}(\chi)}$  and $\ket{\Tilde{\psi}(\chi_{max})}$, where we use $\ket{\Tilde{\psi}(\chi_{max})}$ as a proxy for $\ket{\Tilde{\psi}}$. We see in Figure \ref{fig:fidelity_diag_results}a that this fidelity is very close to 1 already for a very small value of $\chi$, characteristic of weakly entangled states. This small value of $\chi$ increases slightly as a function of the 2d PEPS bond dimension $D$. The sharp transitions in fidelity match the Schmidt spectrum decay in Figure \ref{fig:bMPS_diag_and_iPEPS_results}b. We also calculate the required $\chi$ to obtain a certain fidelity-per-site as a function of $D$ (Figure \ref{fig:fidelity_diag_results}b) and we notice a $\chi \sim D$ relation. This is in contrast with the $\chi\sim D^2$ relation often quoted in the literature for ground states of local Hamiltonians. This discrepancy is likely due to the weakly entangled character of the random PEPS.

\subsubsection{iPEPS Reduced Density Matrix}
\begin{figure}[h!]
  \centering\raisebox{-0.1\height}
  {\includegraphics[width=0.33\textwidth]{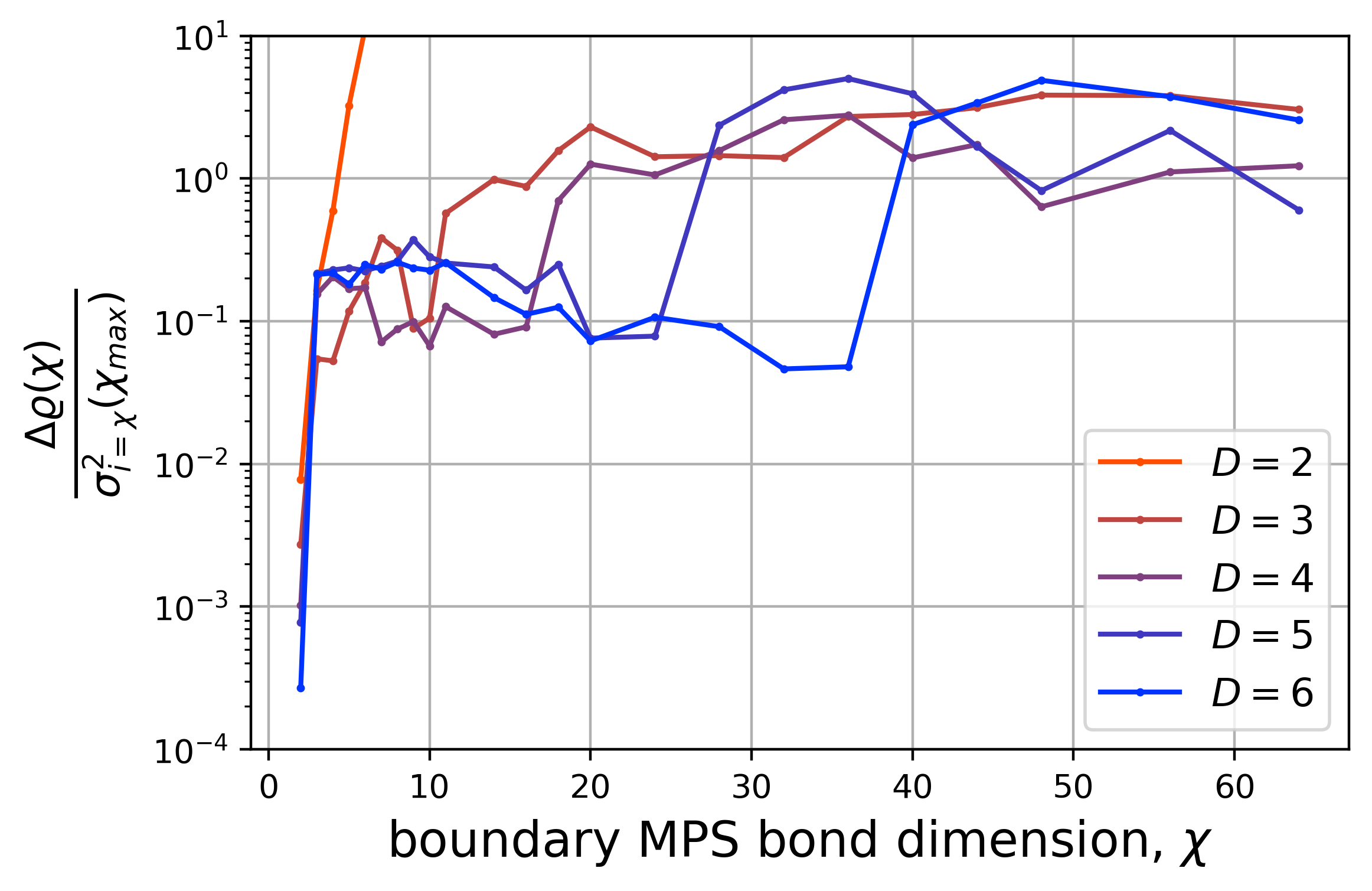}}\hspace{5mm}\raisebox{-0.1\height}{\includegraphics[width=0.33\textwidth]{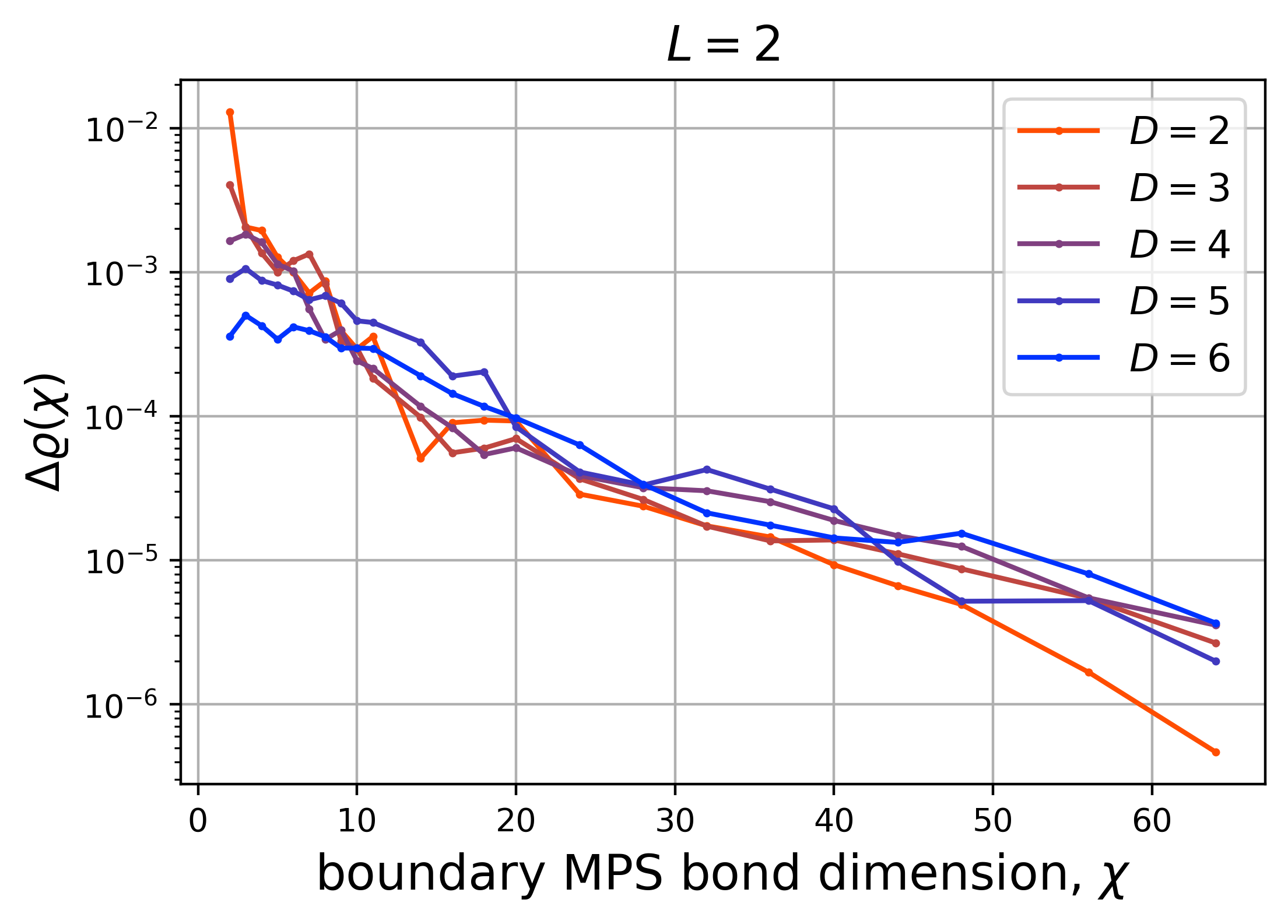}}\\
  (a) \hspace{0.3\textwidth}(b) \hspace{0.3\textwidth}\\
  \caption{
  {\bf 2d PEPS reduced density matrix:} (a) Largest eigenvalue of the difference between the largest and smallest bMPS $\rho(L)$ divided by the squared Schmidt spectrum of the boundary for $\chi_{\rm max}$. This plot shows a constant dependence, indicating that the order-1 Renyi entropy is to be considered when analysing observable accuracy of the 2d PEPS contraction. (b) Change $\Delta \varrho(\chi)$ in the physical $L=2$ density matrix $\varrho(\chi)$ as a function of $\chi$}
\label{fig:2dPESP_red_rho_app}
\end{figure}

In Figure \ref{fig:bMPS_diag_and_iPEPS_results} of the main text we presented the change $\Delta \varrho(\chi)$ in the physical single-site density matrix $\varrho(\chi)$ as a function of $\chi$, where we took $\varrho(\chi_{\rm max}=72)$ as a reference. We observed that $\Delta\varrho(\chi)$ is exponentially suppressed with $\chi$. We expect this result to apply also to the density matrix $\varrho(\chi)$ on a larger local region. As an example, Figure \ref{fig:2dPESP_red_rho_app}b shows $\Delta\varrho(\chi)$ for $\varrho(\chi)$ for a region of two contiguous sites, where indeed we again observe exponential suppression with $\chi$. 

\subsubsection{Correlation Length}
\begin{figure}
  \centering
  {\includegraphics[width=0.33\textwidth]{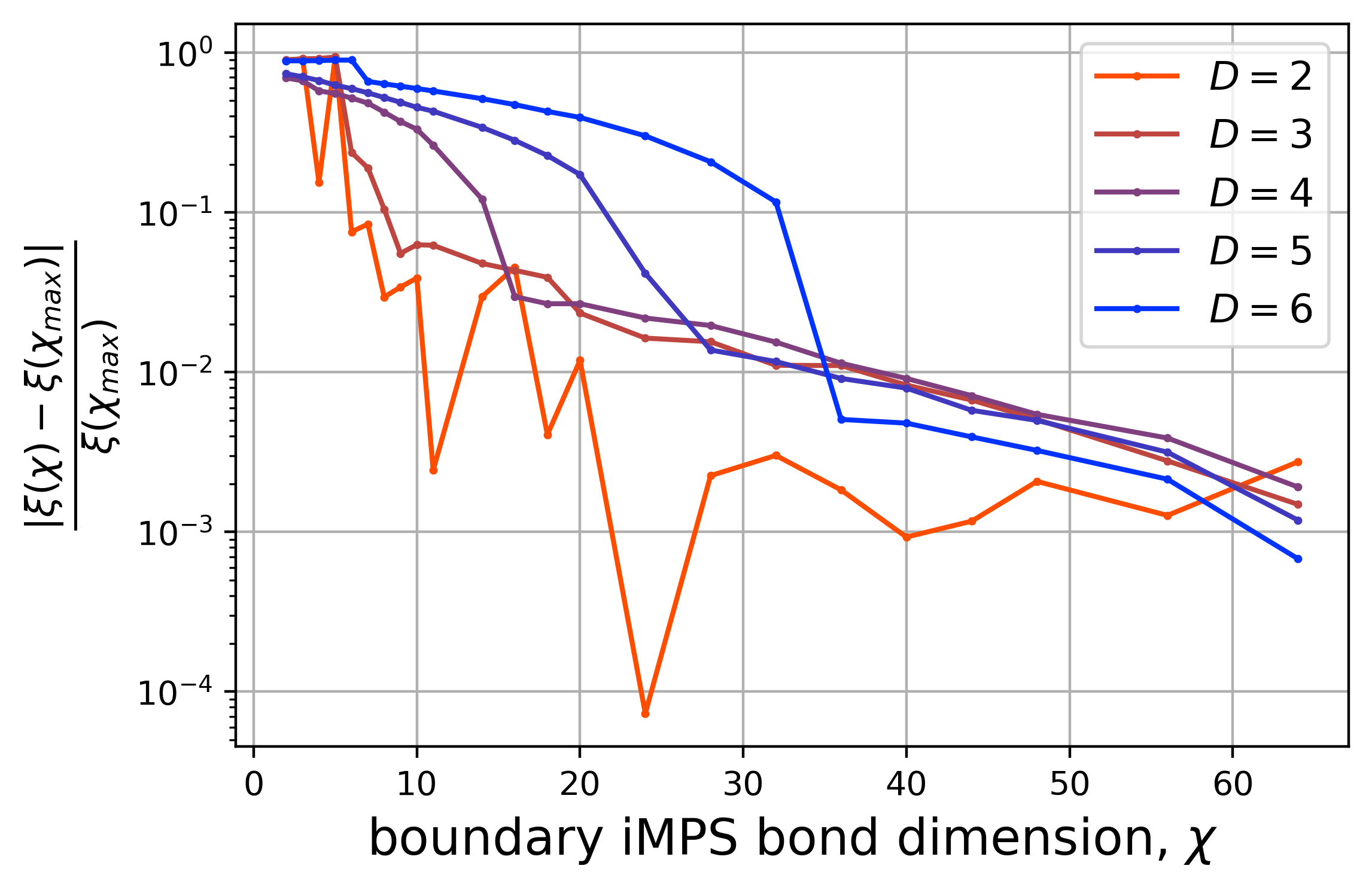}}\hspace{3mm}{\includegraphics[width=0.33\textwidth]{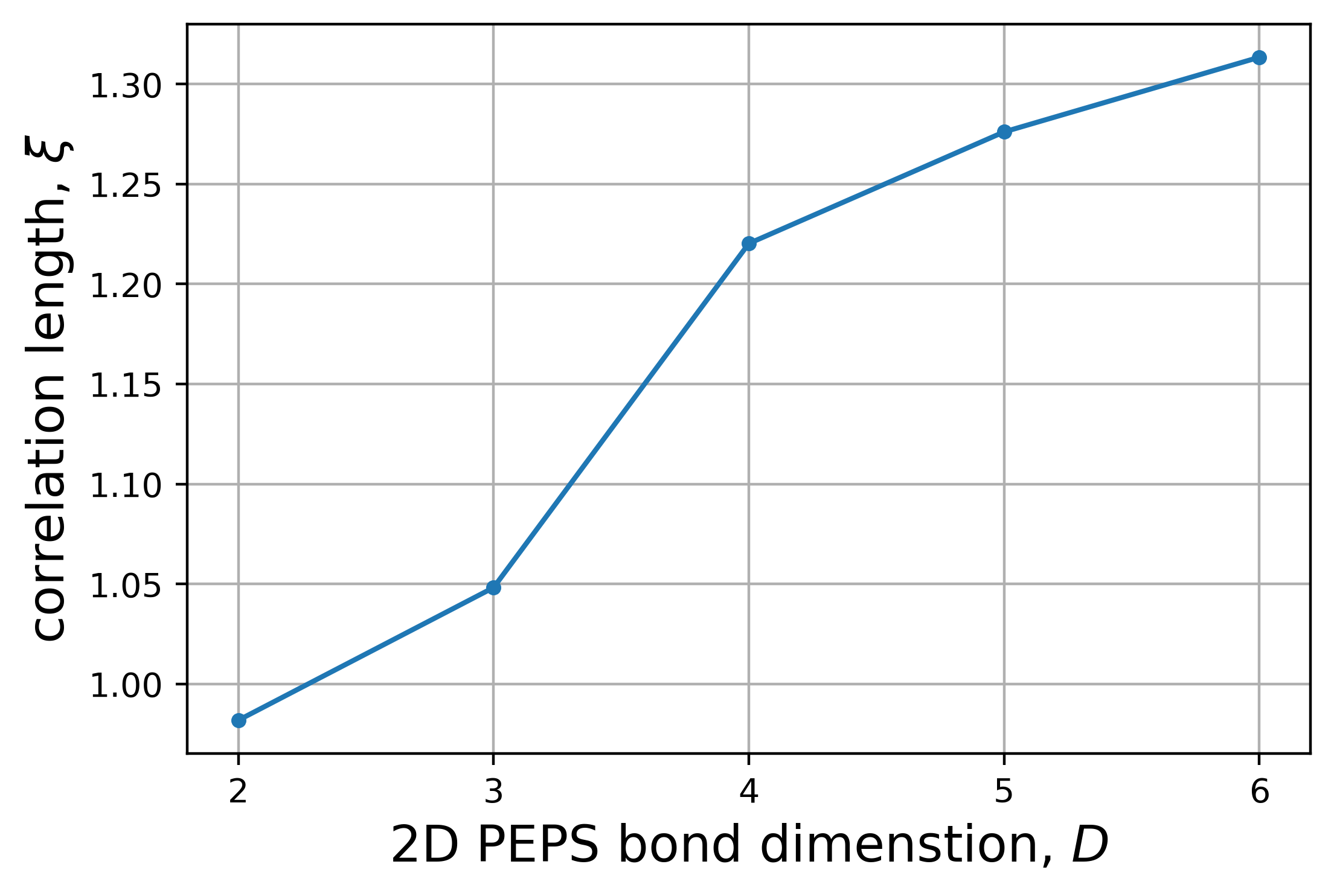}}\hspace{3mm}\\
  (a)\hspace{0.3\textwidth}(b)
  \caption{
  {\bf bMPS correlation length numerical results:} (a) Convergence of the correlation length as a function of bMPS bond dimension $\chi$ and for varying 2d PEPS sizes $D$. (b) Average correlation length for $\Tilde{B}(\chi_{\rm max}=72)$ as a function of $D$ for $D=2,3,4,5,6$, averaged over 15,15,10,5,1 initialisation respectively.}
\label{fig:corr_len_app}
\end{figure}

The correlation length that we obtain from the converged bMPS at maximum bond dimension are low: averaging at order $\sim1$ for all bond dimensions $D$ (Figure \ref{fig:corr_len_app}b). The convergence of the correlation length as a function of bond dimension $\chi$ (Figure \ref{fig:corr_len_app}a) also matches the gaps and exponential decay in the spectrum as per Figure \ref{fig:bMPS_diag_and_iPEPS_results}. This also implies that for relatively low bond dimensions we obtain correlation lengths that are comparable to our best estimate given by $\chi_{\rm max}$.

\subsubsection{bMPS Entanglement Entropy for the 2d Tensor Network Corresponding to the Overlap of two iPEPS}
\begin{figure}[h!]
  \centering
  \includegraphics[width=0.34\textwidth]{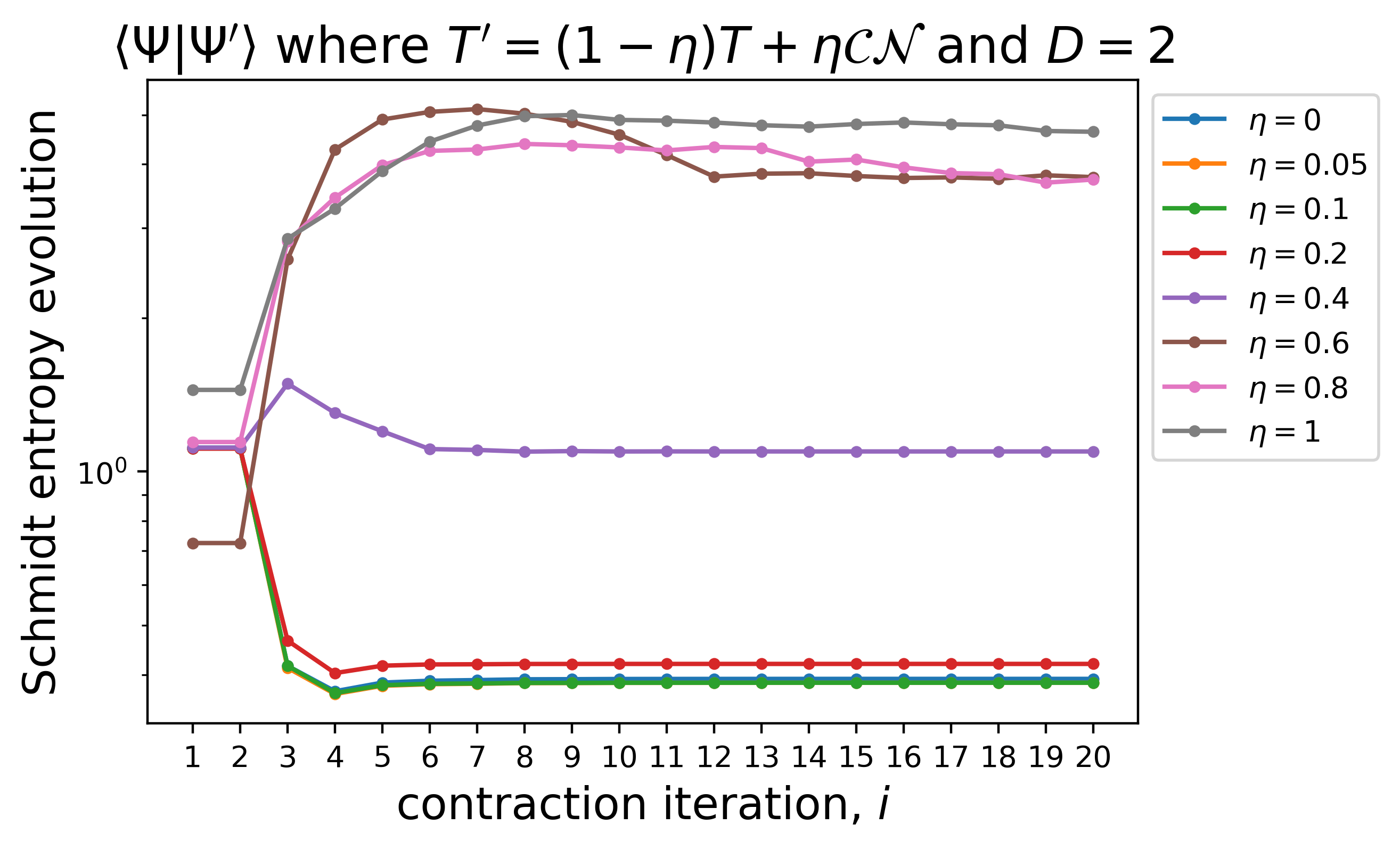}\hspace{3mm}\includegraphics[width=0.34\textwidth]{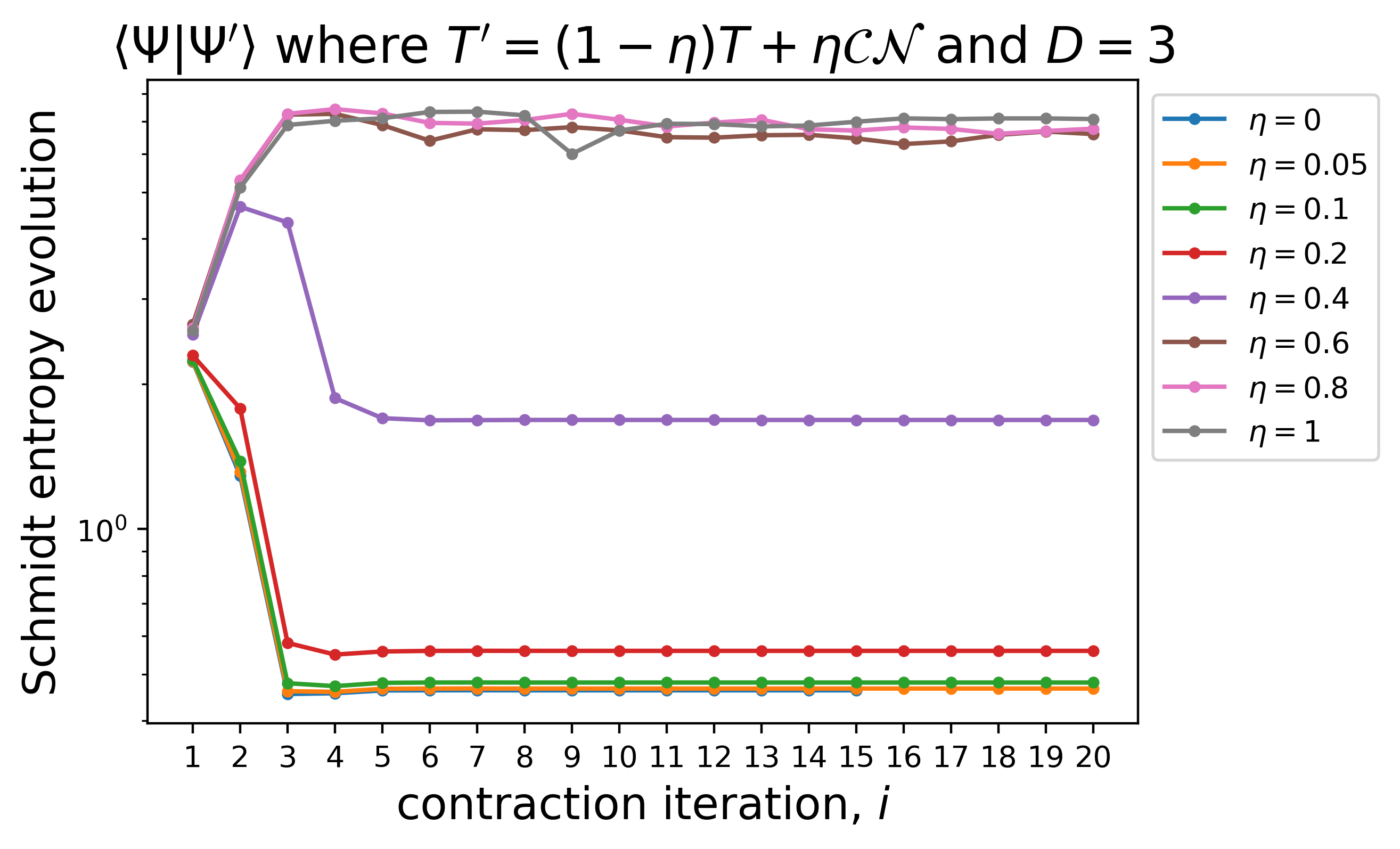}\\
  \hspace{0.2\textwidth} (a) \hspace{0.4\textwidth}(b)\hspace{0.2\textwidth}
  \caption{{\bf Global vs local observables:}
   (a) and (b) show the evolution of the entanglement entropy at the boundary MPS for the overlap $\braket{\Psi|\Psi'}$, where both 2d PEPS have bond dimension $D=2$ and $D=3$ (respectively) and the bMPS has bond dimension $\chi=35$.}
\label{fig:mixed_overlap_calc}
\end{figure}

We have numerically and analytically confirmed that for 2d random PEPS the approximate computation of the norm $\mathcal{N}$ and expectation values of local observables can be performed efficiently. However, other types of observables, such as individual wave-function components, $\<s_1\dots s_N|\Psi\>$, or overlaps between distinct random PEPS, $\<\Psi'|\Psi\>$, are expected to be hard since they involve a 1d transfer matrix map $T{T'}^{*}$, where $T\neq T'$, which is not positive.\\
We consider the computation of the overlap of two distinct 2d random PEPS $\braket{\Psi|\Psi'}$ given by tensors $T$ and $T'$, where we build $T'$ from $T$ with noise tuned by a parameter $\eta \in [0,1]$, such that $T'=(1-\eta)T+\eta\mathcal{C_N}$, where $\mathcal{C_N}$ is the normal complex ditribution with zero mean and standard deviation 1. In the results in Figure \ref{fig:mixed_overlap_calc} we observe an increase in the Schmidt entanglement entropy at the boundary for the contraction evolution of the overlap $\braket{\Psi|\Psi'}$, where, for large $\eta$ the entanglement saturates to a constant value given by the bond dimension $\chi$ of the bMPS.

\section{Stabilizer PEPS formalism and additional numerical results}\label{app:stabilizer}
In this appendix, we provide details about the stabilizer PEPS formalism and its simulation. 

A stabilizer PEPS is a PEPS whose composing tensors $\{T_{\textbf{[r]}}\}$ are stabilizer tensors, \textit{i.e.} each of the states defined through
\begin{equation}
    \ket{T_{\textbf{[r]}}}= \sum_{i, j, k, l, s} \(T_{\textbf{[r]}}\)_{s, ijkl} |s, ijkl\>,
\end{equation}
is a quantum stabilizer state over $p-$qudits, for some prime number $p$. In the expression above, $i,j,k,l$ are virtual bond indices, and $s$ is the physical bond index. Each bond represents a Hilbert space containing an integer number of $p$-qudits, thus both physical and virtual bond dimensions need to be some integer power of $p$: $D = p^{k_D}, d = p^{k_d}$. 

The contraction of two stabilizer tensors can be simulated efficiently by performing several Bell measurements on the contracted bonds, as detailed below. Let us assume $T^1_{ab_1}$ and $T^2_{b_2c}$ are two stabilizer tensors where the two $b$ indices are of the same dimension $p^k$. Let $T^{12}_{ac}$ be the tensor obtained by contacting the two tensors over the $b$ indices:
\begin{equation}
    T^{12}_{ac} := \sum_b T^1_{ab} T^2_{bc}.
\end{equation}
The contraction can be realized by performing several (forced) Bell measurements on qudits within $b_1$ and $b_2$
~\cite{li2021statistical}. If we label the qudits in the $b_{1}$ with $\{1_1,...k_1\}$, and those in the $b_2$ with $\{1_2,...k_2\}$, then:
\begin{equation}
    |T^{12}\> \propto \prod_{i=1}^k\(\mathcal{M}[X_{i_1}X_{i_2}]\mathcal{M}[Z_{i_1}Z^{-1}_{i_2}]\)~ |T^1 \otimes T^2\>,
\end{equation}
where $\mathcal{M}[P]$ is the projector to the $+1$ subspace of the Pauli operator $P$.  Using the Gottesman-Knill theorem~\cite{aaronson2004improved}, one can show that the state $|T^{12}\>$ is still a stabilizer state. Further, its stabilizers can be obtained from those of $T^1$ and $T^2$ with $O(\log^3 D_{\text{total}})$ time complexity, where $D_{\text{total}}=\dim a \cdot \dim c \cdot (\dim b)^2$ is the product of all the open bonds' dimensions. 

It is worth noting that the complexity of contracting stabilizer tensors is independent of the entanglement property of the underlining states. Thus for a given stabilizer PEPS, we are able to compute the boundary state $|\psi(t)\rrangle$'s evolution \textit{exactly} without any truncation and study its entanglement properties. The latter can be computed from the state's stabilizers using the algorithms introduced in \cite{fattal2004entanglement}.

\begin{figure}[h]
\label{fig:app_stabilizer_numerics} 
\begin{centering}
\includegraphics[width=0.28\textwidth]{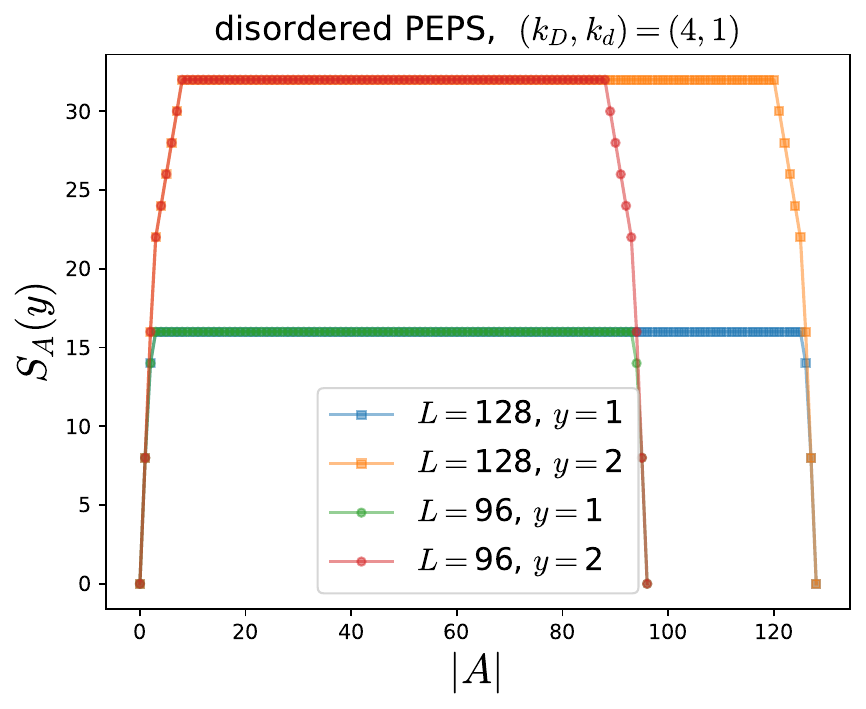}
\includegraphics[width=0.28\textwidth]{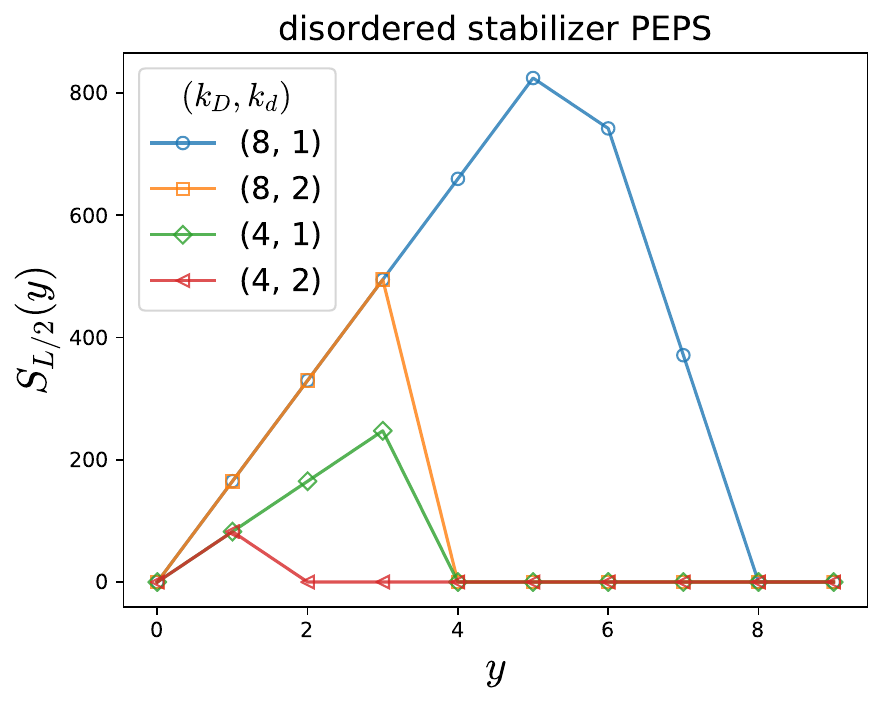}
\includegraphics[width=0.28\textwidth]{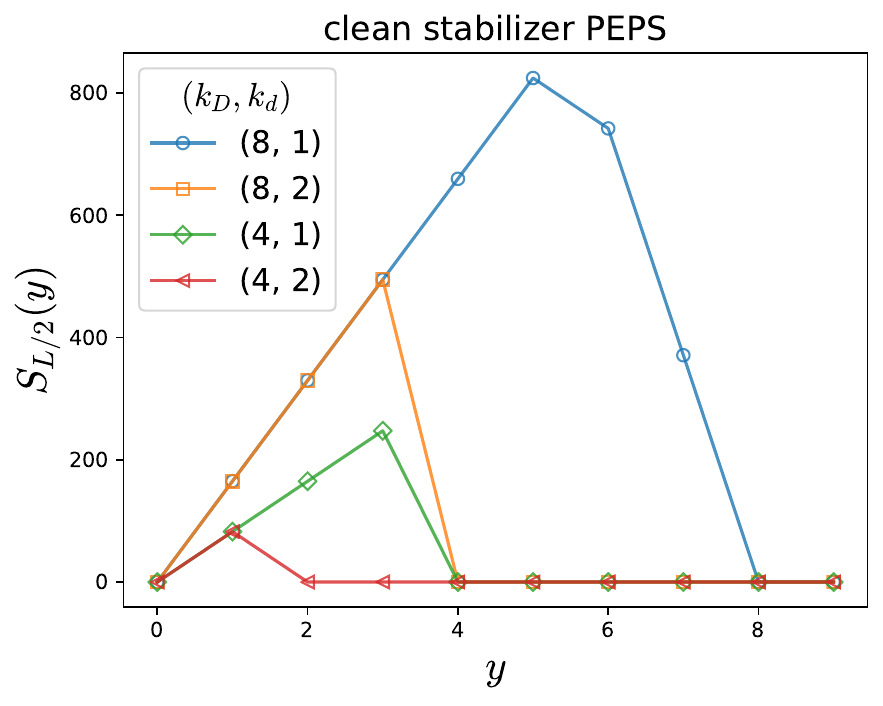}
\caption{
\textbf{Further numerical results for stabilizer PEPS} -- 
(left) Profile of $|\psi(y)\rrangle$'s entanglement entropy in a disordered PEPS. The $L$- and $|A|$- independent plateau suggests area-law entanglement of $|\psi(y)\rrangle$.
(mid) The simulated `entanglement barrier' for various choices of $D$ and $d$ in the disordered PEPS. 
(right) The simulated `entanglement barrier' in the clean PEPS.
}
\end{centering}
\end{figure}

We consider both disordered and clean (\textit{i.e.} translational invariant) stabilizer PEPS. In both cases the PEPS is finite and takes periodic boundary conditions along the $x$ direction, thus the boundary state $|\psi(y)\rrangle$ is also finite and has the periodic boundary condition. To reach the large $D$ regime, we take $p=173$. The simulated von-Neumann entropy $S_A(y)$ \footnote{For a stabilizer state (tensor) and any bi-partition of bonds, the non-zero singular values associated with the bi-partition are all of the same value. Thus the R\'enyi entanglement entropy across the bi-partition is independent of the R\'enyi index $\alpha$, in particular it equals the von Neumann entropy ($\alpha=1$).} of $|\psi(y)\rrangle$ is presented in the Figure \ref{fig:entbarrier} in the main text as well as the Figure \ref{fig:app_stabilizer_numerics}. 

We start by focusing on the disordered case, where each unit stabilizer tensor $T_\textbf{r}$ is sampled independently. The simulation shows that at any given layer number $y$ and when $|A|$ is far from $1$ or $L$, the von Neumann entropy $S_A(y)$ takes a constant value that is independent of $|A|$ or the system size $L$ (Figure \ref{fig:app_stabilizer_numerics}, left). Further, the constant value first increases linearly and then drops with the increase of the number of contracted layers $y$. Both the peak value and the turning point are dependent on the bond dimensions $D$ and $d$ (Figure \ref{fig:app_stabilizer_numerics}, mid). The two plots together suggest that the boundary state $|\psi(t)\rrangle$ is at most area-law entangled at any time (layer number) $y$. Further, the simulated entanglement barrier's dependence on $(d,D)$ matches with the prediction of the stat-mech mapping in Equation \ref{Eq: barrier}, as is shown in Figure ~\ref{fig:entbarrier} in the main text.

Next we come to the clean PEPS case, where all unit tensors $T_{[\vec{r}]}$ are identical and taken to be a randomly sampled stabilizer tensor. The simulation results suggest that the behavior of $S_A(y)$ is almost identical to that in the disordered case. Namely, the $|\psi(y)\rrangle$ is also area-law entangled in the clean case, with an area-law value following the expression Equation~\eqref{Eq: barrier} (Figure \ref{fig:app_stabilizer_numerics}(right) and Figure \ref{fig:entbarrier}(right)).

\section{Details of Statistical-Mechanics Mapping}

\label{app:statmech}
In this appendix, we review the derivation of the stat-mech description of entanglement features of random PEPS.

In the gaussian random PEPS ensemble, the tensor, $T_r$, for each site, $r$, is chosen independently and identically distributed from a Gaussian distribution: 
\begin{align}
\mathbb{E}\[\(T^*_{[r']}\)^{s'}_{i'j'k'l'}\(T_{[r]}\)^s_{ijkl}\] = \delta_{ss'}\delta_{ii'}\delta_{jj'}\delta_{kk'}\delta_{ll'} \delta_{r,r'},
\end{align}
where $\mathbb{E}[\dots]$ denotes averaging over the ensemble, $s=1\dots d$ is the physical index, and $i,j,k,l=1\dots D$ are bond indices, and $r$ label sites of the lattice.
In the following, we drop the indices $s,i,j,k,l$ on the tensors.

\subsection{Mapping RTNs to Replica-Magnets}
Consider the tensor network contraction to compute the norm of the PEPS:  $\mathcal{N}=\<\Psi|\Psi\>$ using the MPS method outlined in Section~\ref{app:methods_iPEPS} above. 
Denote the (unnormalized) density matrix of the evolved boundary state as:  $\rho(y) =  |\psi(y)\rrangle \llangle \psi(y)|$.
Our aim will be to compute the evolution of the ensemble-averaged Renyi entanglement entropy of a region $A$ of the evolved boundary-state, $|\psi(y)\rrangle$. 
%
\begin{equation} \label{Eq: EE}
S_A^{(n)}= \frac{1}{1-n} \log \frac{{\rm tr} \rho_A^n}{({\rm tr} \rho)^n},
\end{equation}
where and $\rho_A$ is the reduced density matrix in some contiguous interval $A$ of size $L_A$ obtained from tracing out the complement of $A$ in $\rho = \kket{\psi}\bbra{\psi}$. 

Since the wavefunction $\kket{\psi}$ is not necessarily normalized, so the denominator in Equation~\eqref{Eq: EE} is crucial to obtain a meaningful entanglement entropy. 
Directly computing the disorder average of this non-linear quantity is challenging.
To avoid this difficulty, we employ a standard replica trick based on the identity:
\begin{align}
\log {\rm tr} \rho_A^n = \lim_{m \to 0} (({\rm tr} \rho_A^n )^m-1)/m.
\end{align}
This allows us to express the disorder average of eq.~\eqref{Eq: EE} as 
\begin{equation} \label{Eq: EEstatmec}
\mathbb{E}\[S_A^{(n)}\]=\frac{1}{n-1}  \lim_{m\to 0}\frac{1}{m} \left( {\cal F}_A-{\cal F}_0 \right),
\end{equation}
with $\mathcal{F}_{A,0} = -\log \mathcal{Z}_{A,0}$ and $\mathcal{Z}_0 \equiv \mathbb{E}\[(\text{tr} \rho^n)^m\]$, $\mathcal{Z}_A \equiv \mathbb{E}\[(\text{tr} \rho_A^n)^m\]$. Using this identity, the calculation of the Renyi entropies reduces to computing $\mathcal{Z}_0$ and $\mathcal{Z}_A$, and to evaluate the replica limit~\eqref{Eq: EEstatmec}. 

When $m$ and $n$ are integers, the averages in $\mathcal{Z}_0$ and $\mathcal{Z}_A$ can be evaluated analytically using Wick's theorem.
One can then express the partition functions $\mathcal{Z}_{A}$ and $\mathcal{Z}_{0}$ in terms of a classical statistical mechanics model, whose degrees of freedom are {\em permutations} labelling different Wick contractions at each vertex of the tensor networks: at each site, each tensor $T_r$ must be paired with a $T_r^*$ possibly belonging to a different replica. Let $Q=nm$ be the number of copies of $\rho$. Then, the partition function $\mathcal{Z}$ involves computing quantities like $\mathbb{E}\[\rho^{\otimes Q}\]$. Note that $|\psi\rrangle$ contains both $T_r$ with $T_r^*$ at each site $R$. Then, in the replicated theory there are $2Q$ copies of $T$, and $2Q$ copies of $T^*$ for each site, which we label, $T^\alpha_r$ with a replica index $\alpha=1\dots 2Q$. We adopt the following ordering for the $2Q$ copies:
\begin{equation}
\{1, \overline{1}, 2, \overline{2}, \dots Q, \overline{Q}\}
\end{equation}
where $k$ labels the state (``ket") in replica $Q$, and $\overline{k}$ denotes the dual state (``bra") in replica $Q$. 
To label permutations we use cycle notation, for example $(124)(35) \in S_6$ denotes the permutation $123456\rightarrow 245136$, i.e. with separate cyclic permutations of elements $(124)$ and of elements $(35)$ [for convenience, we only list the cycles with more than one element].
We will also need to define the cycle counting function
\begin{align}
    C(g,h) \equiv C(g^{-1}h) = \text{\# of cycles in $g^{-1}h$},
\end{align}
where $C(g)$ also includes single-element cycles that are not listed explicitly in our notation e.g. for the above example, $C[(124)(35)] = 3$.

According to Wick's theorem, upon averaging over the Gaussian random tensors, a non-zero contribution is obtained only if each tensor $T^\alpha_r$ is paired with a permuted copy $\(T^{g_{[r]}(\alpha)}_r\)^*$, where $g_{[r]} \in S_{2Q}$ labels a permutation of the replicas, and $S_{2Q}$ is the symmetric group on $2Q$ elements. 

The partition function $\mathcal{Z}$ corresponding to the tensor network contraction, can then be written as a sum over replica-permutation ``spins" for each site:
\begin{align}
\mathcal{Z} = \sum_{\{g_{[r]}\}} W[\{g_{[r]}\}],
\end{align}
where $W$ is the weight of the Wick contraction for the corresponding spin configuration. The weight can be computed analytically for each bond in the tensor network. There are three distinct types of contractions to consider: 
\begin{enumerate}
\item Bulk bonds connecting different nearest-neighbor nodes $i$ and $j$ with permutation ``spins'' $g_{[r]}$ and $g_{r'}$, and bond dimension $D$. The same contraction occurs in the $2Q$ layers of the replicated tensor network. Since Wick contractions force indices to be the same, the resulting weight is equal to $D$:
 \begin{equation}
 D^{C(g_{[r]}, g_{[r']})} = {\rm e}^{(\log D) C(g_{[r]}, g_{[r']}) },
 \end{equation} 
on each link of the square lattice, since the number of independent bond indices is equal to $C(g)$, the number of cycles in the permutation $g$. This is most easily seen by a graphical representation: each cycle in $g_{[r]}^{-1} g_{[r']}$ leads to a ``loop'' where indices have to be the same, with weight $\sum_{\alpha=1}^D \delta_{\alpha \alpha} = D$. Interpreting this positive weight as a Boltzmann weight, this terms leads to a ferromagnetic interaction (favoring $g_{[r]}=g_{r'}$ for neighboring $i,j$ to maximize the number of cycles to $2Q$) with interaction strength $\log D$. This Boltzmann weight has a left/right symmetry $\(S_{2Q} \times S_{2Q} \)\rtimes \mathbb{Z}_2$ (where the extra $\mathbb{Z}_2$ symmetry corresponds to $g \to g^{-1}$). 

\item Bulk contraction of $T_{r}$ with $T_{r}^*$ along the physical leg with dimension $d$. This contraction can be implemented by adding a site with fixed permutation equal to identity $e=()$: in each replica $k$, we pair $k$ with itself (corresponding to gluing $T$ with $T^*$ in the ket), and $\overline{k}$ with itself  (corresponding to gluing $T$ with $T^*$ in the bra). This leads to a factor
 \begin{equation}
 d^{C(e g_{[r]})} = {\rm e}^{(\log d) C(g_{[r]}) },
 \end{equation} 
on each site. This can be seen as a $S_{2Q} \times S_{2Q} \to S_{2Q}$ symmetry-breaking field favoring the identity permutation. This bulk field on every site prevents any phase transition by creating an energy costs for domains of spins with $g_{[r]}\neq e$ that scales as the volume of the domain.

\item Boundary contractions at the top layer to compute  $\mathcal{Z}_0 \equiv \overline{(\text{tr} \rho^n)^m}$, $\mathcal{Z}_A \equiv \overline{(\text{tr} \rho_A^n)^m}$. At the top layers we have dangling legs, that should be contracted to implement the trace and partial trace operations to compute entanglement. In $\mathcal{Z}_0$, we want to compute $\text{tr} \rho$ in each replica. This means that in each replica (and at each site at the boundary), we want to contract $T_{r}$ (resp. $T_{r}^*$) in the ket with  $T^*_i$ (resp. $T_{r}$) in the bra. In our language this corresponds to the permutation
 \begin{equation}
g_0 = (1 \overline{1}) (2 \overline{2}) \dots (Q \overline{Q}),
 \end{equation} 
Note that this permutation is {\em{not}} identity, it has $Q$ cycles while $e=()$ has $2Q$ cycles. At the end of each leg, we fixed the permutation to $g_0$:
 \begin{equation}
 D^{C(g_0^{-1} g_{[r]})} = {\rm e}^{(\log D) C(g_0 g_{[r]}) },
 \end{equation} 
for $i=1,\dots, L$ at the top boundary. To implement the partial trace in $\mathcal{Z}_A \equiv \overline{(\text{tr} \rho_A^n)^m}$, we fixed the permutation to $g_0$ if $i$ is in $\overline{A}$, and to $g_A$ is $r \in A$, with 
 \begin{equation}
g_A = \left( \left( 1 \overline{2} \right)  \left( 2 \overline{3} \right) \dots  \left( n \overline{1} \right)  \right)^{\otimes m}.
 \end{equation} 
\end{enumerate}
Assembling these ingredients results in the effective Hamiltonian of the main text.

\subsection{Comparison to related stat-mech models}

A nearly identical replica-spin model was derived in~\cite{PhysRevB.100.134203} for holographic random tensor network states (rTNS), i.e. whose tensors had physical legs only at the boundary, and only virtual bond legs in the bulk. In that work, the key difference was that the holographic rTNS did not have positive tensors. As a result the permutation spins were $S_Q$- rather than $S_{2Q}$- valued, and the bulk retained the $S_Q$ symmetry since there was no field along $e$. This crucial difference led to two possible phases of the stat-mech model: a disordered (paramagnetic) phase at weak coupling (low-$D$) in which the permutation spins are short-range correlated, and an ordered (ferromagnetic) phase at strong coupling (large-$D$) in which the $S_Q$ symmetry was spontaneously broken and the permutation spins have long range order. In the ordered phase, domain walls had a non-vanishing surface-tension, resulting in an extensive free-energy for the boundary twist in the entanglement region size, resulting in volume-law entanglement scaling. By contrast, in the disordered phase, there is only a local free-energy cost at the edge of the boundary-domains, resulting in area-law entanglement. 

Coming back to the stat-mech model for the PEPS norm computation: the key difference is that the tensors are completely positive, i.e. are composite tensors made up of $T$ and $T^*$ with physical legs contracted. This results in a bulk field along the identity ($e$) permutation that explicitly breaks the $S_{2Q}$ symmetry. A similar statistical mechanics model emerges in the context of random quantum channels~\cite{PhysRevB.107.014307}. In that language, bond dimension corresponds to physical dimension in the channel, and our physical dimension $d$ maps to the strength of channel (effectively the number of Kraus operators). 

 Intuitively, this explicit symmetry breaking destroys the ordering transition, such that the entire phase diagram is effectively ordered (in the sense that domain walls have a non-zero surface tension). Naively, one might expect this to result in a volume-law entanglement throughout for any $D$. However, as we show next, there is an exact cancellation of the volume-law contribution to the free-energy with twisted boundary conditions, generically resulting in area-law entanglement for the evolved boundary state.

\begin{figure}
 \label{fig:permutations} 
\begin{centering}
\includegraphics[width=0.5\textwidth]{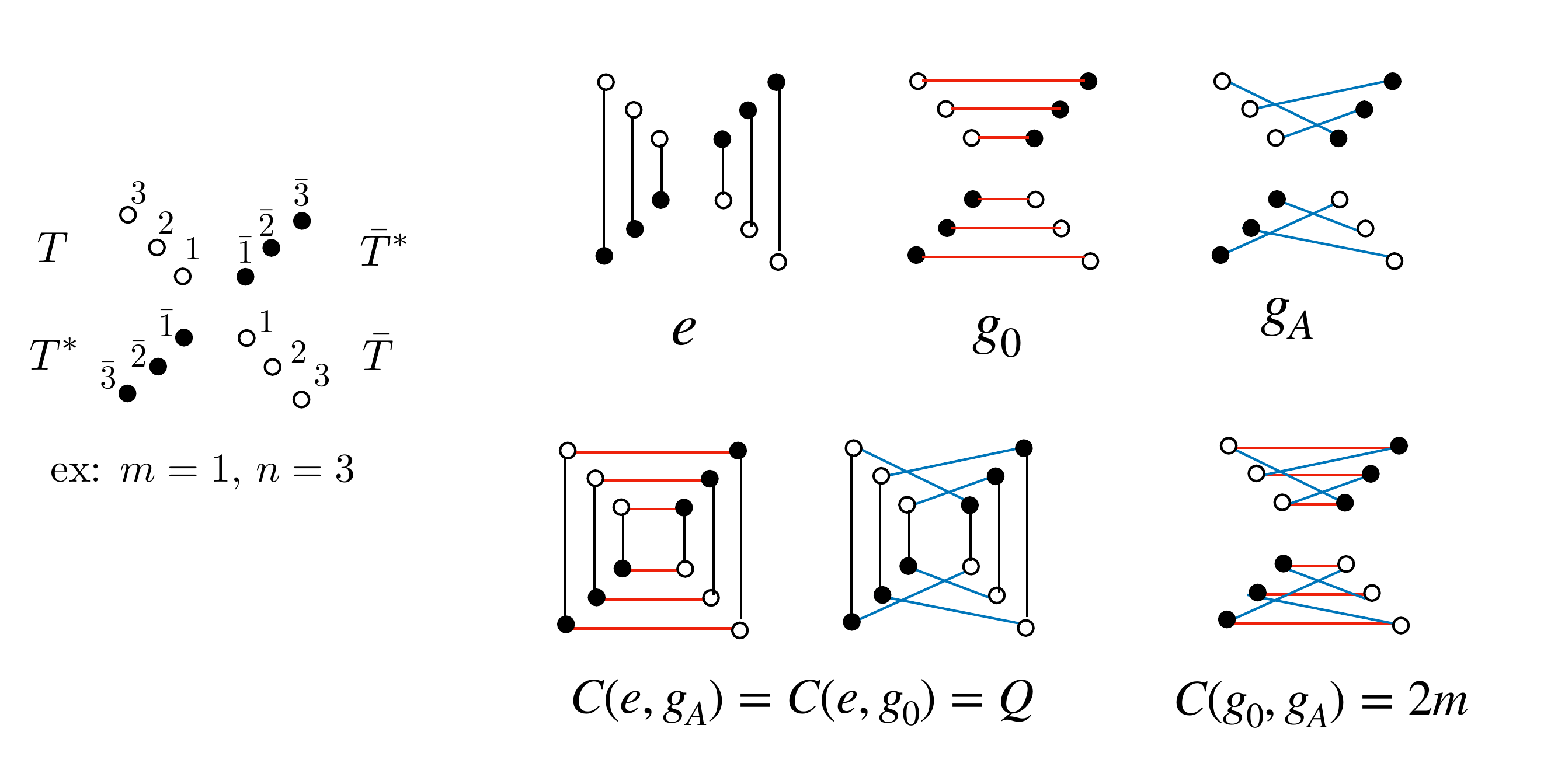}
\end{centering}
	\caption{ {\bf Graphical notation for permutations and cycle counting -- }
	(Left) In each of the $Q=mn$ replicas of $\rho=\ket{\psi}\bra{\psi}$, there are $2Q$ $T$ tensors, and $2Q$ $T^*$ tensors. We label $T,T^*$ tensors from the bra $\bra{\psi}$ with an over-bar. It is convenient to draw the $T$ and $T^*$'s in $m$ groups of $n$ (shown here for $m=1$). Averaging over tensors forces a wick contraction between $T_{i}$ and $T^*_{g(i)}$ where $g \in S_{2Q}$. (Top right) The three permutations corresponding to the bulk ($e$ = identity) and boundary ($g_A,g_0$) fields. (Bottom right) Graphical calculation of cycle counting for various fields, $C(\sigma)$ is given by counting the number of independent loops in the picture (note, for $m>1$, not shown, there would simply be $m$ independent copies of this picture).}            
\end{figure}

\subsection{Minimum cut picture of random PEPS contraction}
At large-$D$, the permtuation spins are strongly locked to each other by their ferromagnetic interactions, and pinned to the bulk $e$ fields. Here, domain walls have a non-zero line tension, and we can approximately compute the domain wall free-energy for the entanglement entropy, by minimizing this line-tension.

Let us focus on the thermodynamic limit, $L, y \to \infty$ and on half-system entanglement. The statistical mechanics model has a bulk symmetry-breaking field that prevents an entanglement phase transition which would be associated with a spontaneous breaking of $S_{2Q}$ symmetry. Specifically, the $e$ field produces an energy cost for domains with $g_{[r]}=e$ that scales like the volume of the domain. This field favors the identity permutation $e$ in the bulk (with fluctuations suppressed if $D \gg 1$), while the boundary fields favor $g_0$ or $g_A$. However, the energy cost of the domain walls between those permutations and $e$ are the same, since $g_A$ and $g_0$ each have $Q=nm$ cycles. Therefore, ${\cal F}_A$ and ${\cal F}_0$ each contain an extensive term $L \log D $, but importantly this extensive term is the same for both ${\cal F}_{A,0}$ and cancels in the difference. This cancellation can be traced back to the $S_Q\times S_Q$ symmetry of the $e$-field in the bulk, and that $g_0$ and $g_A$ differ by a transformation in this symmetry group so that the two types of boundary conditions are locally equivalent. Consequently, the only difference between $A$ and $0$ arises from local energy cost associated with the domain wall between $g_0$ and $g_A$ BCs, which in 2d has constant size independent of $L$. In general, we thus have:
\begin{equation} 
S_A^{(n)} \sim \text{constant}.
\end{equation}
That is, the top boundary is always area law. In fact, as $D \to \infty$, both partition functions are dominated by the ground-state configuration where all spins are $g_{[r]}=e$, and we have ${\cal F}_A-{\cal F}_0=0$ corresponding to a disentangled state.
We note, in passing, that an identical calculation for a 3D PEPS shows that the operator entanglement would scale linearly in the system size. It is plausible that a boundary 2d PEPS, which can account for such linear scaling of operator entanglement, would again enable an efficient contraction.

\subsection{Fluctuation Corrections}
\label{app:statmechFluct}

As we now briefly discuss, fluctuation corrections to the $D=\infty$ limit can be viewed as an expansion in dilute gas of spin flips on top of the $e$-polarized ground state.

For a finite number of replicas, the minimal-energy excitations are single spin flips (1SF's), where we replace $g_{[r]}:e\rightarrow \sigma\neq e$ at some site $r$.
Denoting $J=\log D$ and $h=\log d$, and $c(g,g')= C(g,g')-C(e,e)$, the 1SF costs energy:
\begin{align}
    E_1(g) = 
    \begin{cases} 
    (4J+h)c(g,e) & {\rm bulk}
    \\
    (3J+h)c(g,e)+h\[c(g,g_{A/0})-c(e,g_{A,0)}\] & \text{r $\in A/\bar{A}$ boundary}
    \end{cases}
\end{align}
The lowest-energy spin-flips correspond to transposing a single pair of replicas: $g=(ab)$, which have bulk energy: $E_1\((ab)\)_{\rm bulk} = 4J+h$.
There are $D_1^{\rm bulk}) = Q(2Q-1)$ different single-transposes, leading to a corresponding degeneracy of the single SF excitations in the bulk.
Near an $\bar{A}$ boundary, the cheapest spin flip is is of the form $(k\bar{k})$ for some $k=1,\dots 2Q$, and costs energy $E_1\((k\bar{k})\)_{\bar{A}-{\rm bdry}} = 2J+h$, and degeneracy $D_1^{\rm edge} = Q$.
The minimal-energy spin flips and corresponding degeneracy near $A$ boundary are related to those near the $\bar{A}$ boundary by the symmetry generator: $g_Ag_0^{-1}$, which commutes with the bulk $e$ fields. At large $D$, we can use these excitations to approximate the free-energy by a dilute gas of spin-flip excitations. This expansion is however subtle in the replica limit, as permutations with a number of cycles proportional to $m \to 0$ become dominant in the replica limit. While this caveat prevents us from systematically computing the free energy in a controllable way, this simple counting of low energy excitations predicts that the coefficient of the area-law coefficient scales as 
\begin{align}
\mathcal{F}_A-\mathcal{F}_0 = -\log \frac{\calZ_A}{\calZ_0} \approx D^{-2}d^{-2} + \mathcal{O}(D^{-3}).
\label{eq:area_law_coeff}
\end{align}
Though we are unable to explore a large enough range of $D$ with sufficient precision in the iPEPS numerics to test this asympotic prediction in detail, we note that the large-$D$ suppression in sample-to-sample variance of entanglement features observed in the iPEPS numerics is in qualitative agreement with the suppression of fluctuation contributions to the stat-mech model at large-$D$.

\subsection{Correlation length of random PEPS \label{app:xi}}
The stat-mech mapping also enables one to estimate the correlation length-scale for observables in random PEPS. Namely, consider computing the typical amplitude of a correlation function:
\begin{align}
\log C_{\rm typ} 
\equiv {\mathbb E}\[\log\[\frac{\<\Psi|O_1O_2|\Psi\>\<\Psi|\Psi\>}{\<\Psi|O_1|\Psi\>\<\Psi|O_2|\Psi\>}\]\],
\end{align}
where $O_{1,2}$ are local observables on sites $1,2$, and we have explicitly normalized the wave-function, and also divided by the (normalized) one-point correlators: ($\frac{\<\Psi|O_{1,2} \|\Psi\>}{\<\Psi|\Psi\>}$) to remove dependence on the operator norm of $O_{1,2}$.

For concreteness, and without loss of generality, let us specialize to the case where $O_{[r]} = |s\>_{[r]}\<s|$ is a projector onto physical state $|s\>$ at site $r$ (and identity elsewhere). Generic observables can be written as linear combinations of such projectors (up to a basis transformation that can be absorbed into the randomly-drawn tensor on site $[r]$).

One can evaluate the average of the log in $C_{\rm typ}$ via a replica trick as outlined above for the bMPS entanglement. 
The result is that:
\begin{align}
C_{\rm typ} = \exp\[-\(F_{O_1,O_2}-F_{O_1}-F_{O_2}+F_0\)\] = \frac{\mathcal{Z}_{O_1O_2} \mathcal{Z}_0}{\mathcal{Z}_{O_1}\mathcal{Z}_{O_2}},
\label{eq:ctyp}
\end{align}
where $F_0$ is the free-energy associated with the stat-mech Hamiltonian (\ref{eq:Hsm}), $F_{O_1,O_2,\dots O_k}$ is the same free-energy except with the projectors inserted at sites $1,2,\dots k$, and $\mathcal{Z} = e^{-F}$ is the associated partition function. The projectors restrict the sum over the physical index values to $s$, which is equivalent to removing the replica symmetry-breaking: $e$-field on that site.
Equivalently, $F_{O_1,O_2,\dots O_k}$ is given by the free-energy of the Hamiltonian discussed in the main text, but perturbed by a term: $\Delta H = +\log d \sum_{[r]=1\dots k} ~C(e,g_r)$.

At large-$D$, we can estimate the leading contribution to $C_{\rm typ}$ as follows. The leading contribution to the stat-mech partition function is from a uniformly $e$-polarized replica-spin configuration. Fluctuations about this come in the form of small domains of non-$e$ polarized spins. By inspection, only domains that include both sites $1,2$ make a non-cancelling contribution to the ratio in (\ref{eq:ctyp}). At large-$D$, this contribution is dominated by the smallest such spanning domain, which is a line of flipped replica-spins, $g_i\neq e$, along a short path connecting points $1$ and $2$. This domain has a line tension $\mathcal{F} \approx \(\log D^4d\) r_{12}$, where $r_{12}$ is the length of the shortest path through the network connecting points $1,2$. This contributes  exponentially decaying correlations:
\begin{align}
C_{\rm typ} \approx e^{-r_{12}/\xi},
\end{align}
with characteristic correlation length: $\xi \approx \(\log dD^4\)^{-1}$. Notice that the correlation length decreases with increasing $D$. 
However, note that random PEPS states at large $D$ \emph{are not} close to product states, but, in fact saturate the maximal entanglement allowed for the given bond-dimension PEPS.

This shows that large-$D$ random square lattice PEPS actually have rather short range correlations, in accordance with previous studies~\cite{Lancien_2021}, and our numerical observations for clean random iPEPS.

\section{Random PEPS vs physically relevant ground states}
\label{secPhysicalvsrandomPEPS}

An important question is to what extent the results presenting in this paper, strongly indicating that random PEPS can be efficiently approximately contracted, can be extended to PEPS representing ground states of physically relevant Hamiltonians, e.g. in the context of condensed matter, materials science and quantum chemistry. That is, can our results shed some light into the performance of PEPS algorithms when simulating such systems? Here we restrict our considerations to two-dimensional ground states that obey the entanglement area law (2d ground states that violate the entanglement area law, such as ground states in the presence of a 1d Fermi surface, are expected to be harder to contract).

We have seen below that random PEPS have a very short correlation length $\xi$ on the order of one lattice site or less. In contrast, the correlation length $\xi$ in a physically relevant ground state can be arbitrarily large (for instance, the correlation length diverges as we approach a quantum critical point). Relatedly, we have numerically seen that the boundary MPS for a random PEPS has very limited amount of entanglement whereas, as PEPS practitioners have learned over the last 15 years, the entanglement in the boundary MPS for a physically relevant 2d ground state can again be arbitrarily large (even in those cases where the boundary MPS obeys an area law).

We have therefore identified two structural differences between random PEPS and physically relevant ground states, namely differences in correlation lengths $\xi$ and in the amount of boundary MPS entanglement. How fundamental are these structural differences? Based on experience with renormalization group, random-circuit dynamics, and random matrix theory it is tempting to conjecture that the large-$D$ random PEPS might represent a sort of coarse-grained ``fixed-point" representation of physically relevant ground states. However, below we will see that while coarse-graining the PEPS for a physically relevant ground state would indeed effectively remove the difference in correlation length, it would not change the difference in boundary MPS entanglement. Since boundary MPS entanglement determines the computation cost in approximate PEPS contractions, we cannot conclude that our results for random PEPS apply to such PEPS.

That is not to say that the stat-mech approach used in this paper to successfully characterize the boundary MPS entanglement for random PEPS is restricted to studying states with a short correlation length $\xi$. On the contrary, as discussed below, we will see that the same approach can be used for 2d random tensor network states (which are not 2d PEPS) with arbitrarily large correlation length $\xi$.



\subsection{Coarse-graining by blocking tensors}
First, note that, any finitely correlated PEPS, i.e. with finite correlation length, $\xi$, can be transformed into a PEPS with shorter correlation length $\xi'\sim 1$ by ``blocking" together sites in $\xi\times \xi$ blocks of physical sites. This blocking adds constant overhead to the bond-dimension of each tensor, $D' \sim D^\xi$. While this cost may be severe in practice, from an asymptotic complexity standpoint, it is merely a constant overhead. This argument suggests that one can perhaps think of a random PEPS as reflecting a block-spin renormalization group (RG) style coarse-graining of a physical PEPS with longer-range correlations. 

\begin{figure*}[t!]
  \centering
  {\includegraphics[width=0.9\textwidth]{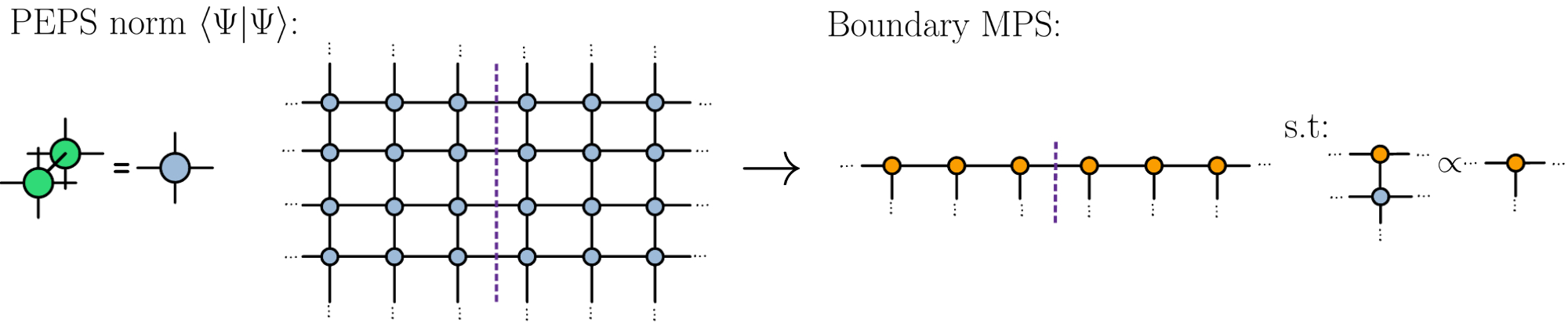}}
  \hspace{\textwidth}(a) \hspace{0.5\textwidth}(b) \hspace{0.35\textwidth}\\
  \caption{
  (a) Square lattice of the 2D PEPS norm.
  (b) Boundary MPS obtained from the contraction of the norm in (a).}
\label{fig:app_converge_norm_cut}
\end{figure*}

\begin{figure*}[t!]
  \centering
  {\includegraphics[width=0.7\textwidth]{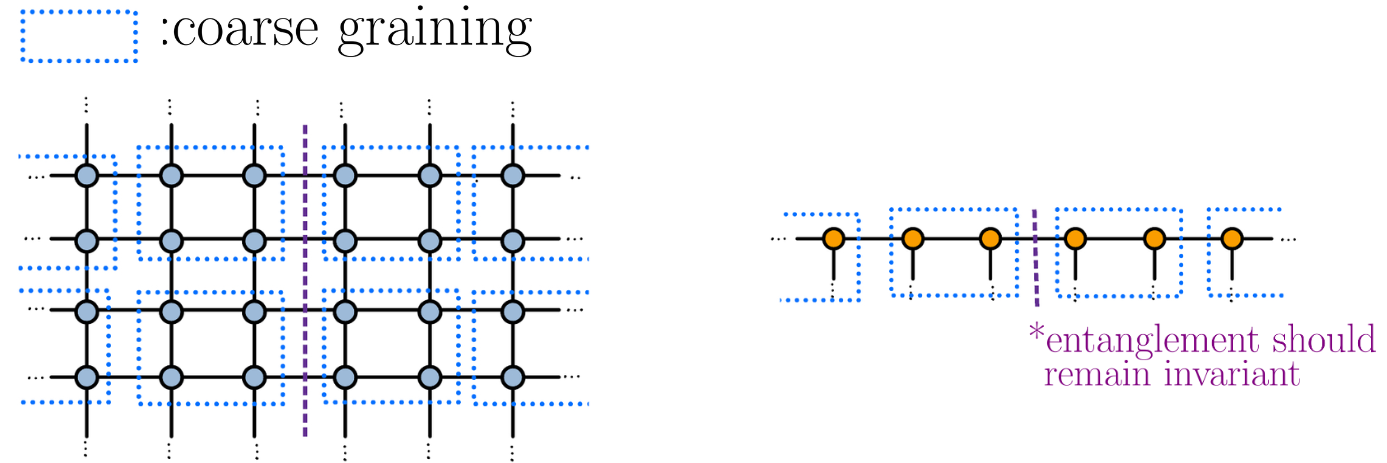}}
  \hspace{\textwidth}(a) \hspace{0.4\textwidth}(b) \hspace{0.35\textwidth}\\
  \caption{
  (a) Coarse grained square lattice of the 2D PEPS norm.
  (b) Boundary MPS obtained from the contraction of the coarse grained norm.}
\label{fig:app_converge_norm_cut_CG}
\end{figure*}

However, the following observation suggests that there is no connection between a coarse-grained PEPS for physical models with large $\xi$, and a random PEPS with bond-dimension $D'$. 
Years of numerical experience~\cite{Verstraete_2004, Corboz_2014, Corboz_2014_2, Niesen_2017, Zheng_2017, Ponsioen_2019, Chen_2020} show that ground-states with large correlation length have corresponding large entanglement both for the physical state, and the bMPS for its norm and correlation functions.
While blocking reduces the correlation length, it does not change the entanglement spectrum of the bMPS (see Fig.~\ref{fig:app_converge_norm_cut},\ref{fig:app_converge_norm_cut_CG}).
Hence, for physical states, the bMPS entanglement should grow with $\xi$, whereas for random PEPS with bond dimension $D'=D^\xi$, the method of Appendix~\ref{app:statmechFluct} above predict bMPS entanglement decreasing exponentially with $\xi$ as $\sim 1/D^{2\xi}$.
This argument shows that large-$D$ random PEPS do not have the correct entanglement structure to capture block coarse-grainings of long-range correlated states encountered in simulation of physical systems.

However, in standard renormalization group approaches, coarse-graining does not simply involve merely blocking sites together, but of hierarchically decomposing the state via layers of coarse-graining steps that act on different distance scales. Inspired by this, in the next section, we construct a class of hierarchical random tensor network states that have arbitrarily-long correlation length, $\xi$, and which have bMPS entanglement that grows with $\xi$ in a manner qualitatively consistent with that found in simulations of physical systems (though it remains an open question whether variants of such hierarchical tensors networks reflect all the important structure found in physical states).

\subsection{Random Tensors Networks with large correlation lengths\label{app:bigxi}}
In this section, we construct an ensemble of random 2d tensor networks that:
\begin{enumerate}
\item  have arbitrarily long correlation length, $\xi$, 
\item can be viewed as PEPS with effective bond-dimension $D_{\rm eff}\sim {\rm poly}(\xi)$, and 
\item can be reliably analyzed by stat-mech mapping in a large-$D$ limit, which predicts that their physical properties can be efficiently computed via an area-law bMPS.
\end{enumerate}

\begin{figure}
 \label{fig:statmech} 
\begin{centering}
	\includegraphics[width=0.6\textwidth]{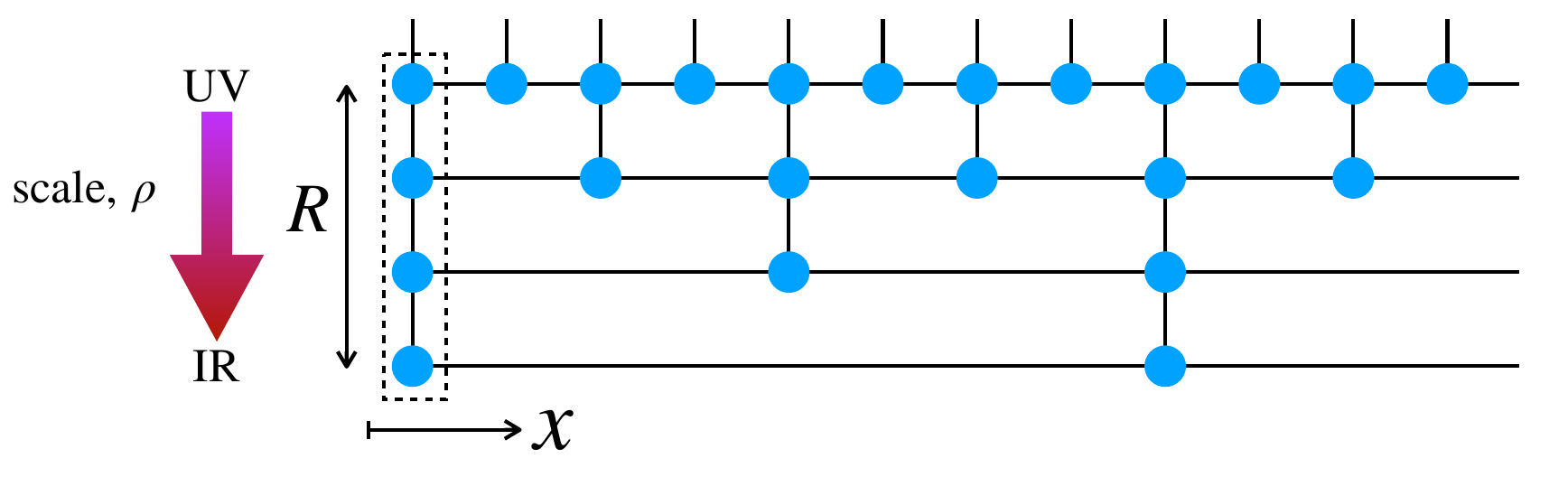}
\end{centering}
	\caption{ {\bf Random Tensor Networks with Arbitrarily Long Correlation Length -- }
	Schematic of $1d$ gMERA (with obvious generalization to higher-dimensional versions) which has arbitrarily-long correlation length $\xi\sim 2^R$ even in the large-$D$ limit, showing that correlation length of random tensor networks is not necessarily short. Blue dots are tensors with Gaussian-random entries. Internal bonds have dimension $D$, physical bonds (sticking up at the top) have dimension $d$. The holographic ``scale" dimension has size $R$. The correlation length is $\xi \sim 2^R$. The $2d$ version of this network can be contracted by a boundary MPS (dashed box) with effective bond dimension is $D_{\rm eff}\sim D^{R}$, which scales polynomially with $\xi$.
 \label{fig:bigxi}
    }             
\end{figure}

Specifically, inspired by expectation that an RG coarse-graining can reduce a PEPS with any finite correlation length $\xi$ to one with $\xi\lesssim 1$, we consider a shallow generalized multi-scale entanglement renormalization ansatz (gMERA) architecture introduced in~\cite{anand2022holographic}. 
A $1d$ version of this structure is shown in Fig.~\ref{fig:bigxi}, with obvious generalizations to higher-d. It consists of a depth, $R$, layers of tensors, in which at layer $1\leq j\leq R$, the tensors are connected only at distance $2^j$~\footnote{We note that, while this gMERA structure was originally introduced in the context of quantum circuit tensor networks, and considered unitary or isometric tensors, the isometry constraint will have little impact on the stat-mech description, and we drop it for simplicity.}.
This shallow gMERA geometry allows one to neatly interpolate between finitely-correlated states ($R$-finite) and critical states ($R \rightarrow \infty$). 
Heuristically, we can view this as a discrete version of the holographic AdS/CFT correspondence, where physical legs live only at the boundary of a (short) extra ``scale" dimension, which runs from short-distance (UV) at the physical boundary, to longer-distance (IR).
Here, we consider the case where each tensor in this network has Gaussian random entries, and all internal legs have bond-dimension $D$, and physical legs have dimension $d$.
Adapting the discussion of typical two-point correlation functions above to this network, one again concludes that the correlations decay exponentially with the size of the smallest domain that includes both sites $1,2$. 
In this network, the smallest domain will run along the IR edge of the network, resulting in:
\begin{align}
C_{\rm typ}(r) \sim 
\begin{cases}
1/r^{p}~~{\rm w/}~~p\sim \log_2 D  & ;r < 2^R\\
e^{-r/\xi}~~\rm{w/}~~\xi \sim 2^R &; r \geq 2^R
\end{cases}
\end{align}
where $\xi \sim 2^{R}$.
We note that, the functional forms listed merely reflect an overall asymptotic decay of the envelop of correlations. In addition, there is a complicated fractal/self-similar modulation inherited from the geometry of the network.

From this expression, we see that the correlation length, $\xi$ can be made as large as desired by controlling the depth of the shallow gMERA.
At the same time, this shallow gMERA can be viewed as a PEPS with effective bond dimension $D_{\rm eff} = D^R = \xi^{\log_2 D}$, which scales polynomially with the correlation length.

The stat-mech mapping of the bMPS entanglement for these shallow gMERA proceeds similarly to that for the 2d square PEPS, except that the physial legs arise only at the UV layers. Therefore, for contracting networks representing norms and correlations, the replica-symmetry breaking $e$-field is only present in the UV. 
Nevertheless, this is still sufficient to explicitly break the replica symmetry, and give an area-law for the bMPS for any $D_{\rm eff}$, i.e. for any correlation length, $\xi$. 

The area-law coefficient may be estimated in the large-$D$ expansion as outlined above for the square-lattice PEPS. 
The leading contribution again comes form two-site domains of flipped replica spins ($g\neq e$) that straddle the entanglement cut.
A new feature is that the straddling domain can occur at any layer in the (shallow) holographic dimension without effecting its free-energy cost, giving rise to an entropic factor that scales as $\sim R$.
The resulting bMPS entanglement in the large-$D,R$ limit is then:
\begin{align}
    S\sim \frac{1}{D^4}R \sim \frac{1}{D^4}\log \xi.
    \label{eq:SgMERA}
\end{align}
We note that the physical dimension, $d$ does not appear in this expression because the bulk tensors have only virtual legs.
At large $\xi$, the bMPS entanglement in (\ref{eq:SgMERA}) scales like that of a nearly-critical $1d$ system with effective ``central charge" $\sim 1/D^4$, with critical scaling cut off by a finite correlation length $\xi$. In particular, 

This example shows that short correlation length is not an intrinsic limitation of random large-$D$ tensor networks. However, it remains unclear whether this example fully capture the structure relevant to physical ground-states.

\end{document}